\documentclass[%
 reprint,
nobibnotes,
 amsmath,amssymb,
aip,
jcp,
citeautoscript,
]{revtex4-1}

\usepackage{siunitx}
\usepackage{bm}%
\usepackage{xcolor} %
\usepackage{graphicx}
\usepackage[utf8]{inputenc}
\usepackage{epstopdf}
\usepackage{afterpage}
\usepackage{setspace}
\usepackage{booktabs}
\usepackage{subcaption}
\usepackage{hyperref}
\usepackage{multirow}

\expandafter\ifx\csname package@font\endcsname\relax\else
 \expandafter\expandafter
 \expandafter\usepackage
 \expandafter\expandafter
 \expandafter{\csname package@font\endcsname}%
\fi
\begin{document}

\title{Impact of the damping function in dispersion-corrected density functional theory on the properties of liquid water}

\author{K. Nikolas Lausch}
\affiliation{Lehrstuhl f\"ur Theoretische Chemie II, Ruhr-Universit\"at Bochum, 44780 Bochum, Germany}
\affiliation{Research Center Chemical Sciences and Sustainability, Research Alliance Ruhr, 44780 Bochum, Germany}
\author{Redouan El Haouari}
\affiliation{Lehrstuhl f\"ur Theoretische Chemie II, Ruhr-Universit\"at Bochum, 44780 Bochum, Germany}
\affiliation{Research Center Chemical Sciences and Sustainability, Research Alliance Ruhr, 44780 Bochum, Germany}
\author{Daniel Trzewik}
\affiliation{Lehrstuhl f\"ur Theoretische Chemie II, Ruhr-Universit\"at Bochum, 44780 Bochum, Germany}
\affiliation{Research Center Chemical Sciences and Sustainability, Research Alliance Ruhr, 44780 Bochum, Germany}
\author{J\"{o}rg Behler}
\email{joerg.behler@rub.de}
\affiliation{Lehrstuhl f\"ur Theoretische Chemie II, Ruhr-Universit\"at Bochum, 44780 Bochum, Germany}
\affiliation{Research Center Chemical Sciences and Sustainability, Research Alliance Ruhr, 44780 Bochum, Germany}

\date{12 June 2025}

\begin{abstract}
Accounting for dispersion interactions is essential in approximate density functional theory (DFT). Often, a correction potential based on the London formula is added, which is damped at short distances to avoid divergence and double counting of interactions treated locally by the exchange-correlation functional. Most commonly, two forms of damping, known as zero- and Becke-Johnson (BJ)-damping, are employed and it is generally assumed that the choice has only a minor impact on performance even though the resulting correction potentials differ quite dramatically. Recent studies have cast doubt on this assumption pointing to a significant effect of damping for liquid water, but the underlying reasons have not yet been investigated. Here, we analyze this effect in detail for the widely used Tkatchenko-Scheffler and DFT-D3 dispersion models. We demonstrate that, regardless of the dispersion model, both types of damping perform equally well for interaction energies of water clusters, but find that for the two investigated functionals zero-damping outperforms BJ-damping in dynamic simulations of liquid water. Compared to BJ-damping, zero-damping provides, e.g., an improved structural description, self-diffusion, and density of liquid water. This can be explained by the repulsive gradient at small distances resulting from damping to zero that artificially destabilizes water's tetrahedral hydrogen-bonding network. Therefore, zero-damping can compensate for deficiencies often observed for generalized gradient functionals, which is not possible for strictly attractive BJ-damping. Consequently, the improvement that can be achieved by applying a dispersion correction strongly depends on the employed damping function suggesting that the role of damping in dispersion-corrected DFT needs to be generally reevaluated.

\end{abstract}

\maketitle

\section{Introduction}\label{sec:introduction}

Despite its simple chemical composition water exhibits a complex phase diagram and a variety of interesting properties such as a density maximum at 277\,K, which leads to ice floating on liquid water.\cite{mallamace_anomalous_2007, pettersson_watermost_2016, morawietz_how_2016, chen_ab_2017, omranpour_perspective_2024} This peculiar behavior of water arises from a delicate balance of covalent and non-covalent interactions (NCIs) such as polarization, hydrogen bonding and dispersion interactions.\cite{del_ben_probing_2015, gillan_perspective_2016, pettersson_watermost_2016, chen_ab_2017} 
\\
Understanding how these different interactions determine the unique properties of water is a formidable challenge and modeling the structure and dynamics of water faithfully in computer simulations requires a reliable \textit{ab initio} description of the atomic interactions.\cite{del_ben_probing_2015} High-level electronic structure methods such as second-order Møller–Plesset perturbation theory (MP2)\cite{del_ben_bulk_2013, del_ben_probing_2015}, the random phase approximation (RPA)\cite{del_ben_probing_2015}, and coupled cluster theory using singles, doubles and perturbative triples (CCSD(T))\cite{daru_coupled_2022, yu_q-aqua_2022, chen_data-efficient_2023} can provide such a description and results based on these methods are in excellent agreement with experimental data. However, these high-level methods are computationally prohibitively expensive for more complex systems like solid-liquid interfaces or molecular chemistry in solution, e.g. in fields such as electrochemistry, catalysis or biochemistry. For these systems, density functional theory (DFT) is the method of choice due to its reasonable accuracy and modest computational costs.\cite{burke_perspective_2012, becke_perspective_2014, gros_ab_2022, montero_de_hijes_density_2024}
\\
It is well-known that the employed exchange-correlation (xc) functional strongly affects the quality of the DFT description of water.\cite{hassanali_aqueous_2014, gillan_perspective_2016} Due to their high efficiency, generalized gradient approximation (GGA) functionals are among the most popular functionals for condensed systems.\cite{gillan_perspective_2016} However, in particular for water and aqueous systems often the results provided by GGA functionals do not compare favorably to experiment.\cite{gillan_perspective_2016} Typical observations include an over-structuring of the liquid, suppression of self-diffusion, underestimation of the equilibrium density and the absence of a density maximum.\cite{gillan_perspective_2016, morawietz_how_2016, wang_density_2011, kuhne_static_2009}
\\
While different origins for these shortcomings have been proposed, a concensus is that GGA functionals are unable to accurately describe the different interactions, which are relevant to model water. The main errors are related to the GGA parametrization\cite{sprik_ab_1996, gillan_perspective_2016, hassanali_aqueous_2014, tonigold_dispersive_2012}, the lack of exact exchange to remedy the self-interaction error (SIE) of approximate density functionals,\cite{gillan_perspective_2016, hassanali_aqueous_2014} and the inability to describe long-range electronic correlation.\cite{grimme_dispersion-corrected_2016, gillan_perspective_2016, hassanali_aqueous_2014} For each of these shortcomings, corrections are available and the extent to which they improve the properties of liquid water has been studied extensively.\cite{todorova_molecular_2006, guidon_ab_2008, lin_importance_2009, yoo_communication_2011, tonigold_dispersive_2012, bankura_structure_2014, distasio_individual_2014, gillan_perspective_2016, morawietz_how_2016}  While systematically addressing all known shortcomings should in principle result in an improved description of water, some results suggest that important error cancellations may break down if only specific improvements are considered.\cite{marsalek_quantum_2017}
\\
Today, a broad consensus has been reached that accounting for non-local correlation effects, especially attractive dispersion interactions, is necessary when applying approximate DFT to large molecular or condensed systems.\cite{grimme_dispersion-corrected_2016, hermann_first-principles_2017} In simulations of liquid water, many studies have reported a significant effect on the structure and properties upon including long-range dispersion interactions.\cite{lin_importance_2009, yoo_communication_2011, bankura_structure_2014, distasio_individual_2014, gillan_perspective_2016, morawietz_how_2016} While there are different approaches to account for dispersion interactions such as nonlocal van der Waals (vdW) functionals\cite{dion_van_2004, klimes_chemical_2009, vydrov_nonlocal_2010} or effective one-electron potentials\cite{von_lilienfeld_optimization_2004}, the most frequently employed correction method is to apply a pair-wise additive potential based on the London formula\cite{eisenschitz_uber_1930} after electronic convergence.\cite{grimme_dispersion-corrected_2016} This approach was popularized by Grimme with the DFT-D family of methods\cite{grimme_accurate_2004, grimme_semiempirical_2006, grimme_consistent_2010, grimme_effect_2011, caldeweyher_generally_2019} and also includes the Tkatchenko Scheffler (TS) method\cite{tkatchenko_accurate_2009} and the exchange-dipole model\cite{becke_exchange-hole_2005, becke_exchange-hole_2007} (XDM) by Becke and Johnson. 
\\
These pairwise methods include an empirical, functional-dependent damping term, which controls the behavior of the correction from small to intermediate interatomic distances.\cite{grimme_dispersion-corrected_2016} This damping term is necessary to avoid the diverging behavior of the London formula at short distances and takes into account that xc functionals differ in the range and extent to which they describe dispersion locally. Therefore, the behavior in the respective region needs to be adapted to combine any of the pairwise methods with a given xc functional and to avoid double counting of short-range dispersion interactions. This is typically achieved by fitting the damping function parameters to reference interaction energies obtained from a higher level of theory, which is able to describe long-range correlation correctly, i.e., CCSD(T). However, when parametrizing a damping function for a particular xc functional it is assumed that the only error in the produced interaction energies is missing long-range correlation even though this not necessarily correct.\cite{hermann_electronic_2018} Therefore, there is the risk of the damping function overcompensating for other ill-described NCIs contained in the benchmark set. As a consequence, the correction has a limited transferability\cite{hermann_electronic_2018} and can lead to unintended behavior when applied to systems beyond the benchmark set.
\\
There are two common approaches to damping. One seeks to minimize double counting of short-range interactions treated locally by the xc functional by employing a function that approaches zero at small interatomic distances. This approach is often referred to as ``zero-damping'' and is used in DFT-D3\cite{grimme_consistent_2010} and the TS model\cite{tkatchenko_accurate_2009}. A downside to this form of damping is that it leads to a negative gradient of the dispersion energy at small to medium distances,\cite{grimme_effect_2011} i.e., repulsive forces. This is inconsistent with the general understanding of dispersion interactions being attractive at any range. The other approach avoids this behavior and approaches a constant finite value at small interatomic distances instead.\cite{grimme_effect_2011} This form of damping represents the physical behavior of the dispersion energy at small distances\cite{koide_new_1976} and was first proposed by Becke and Johnson\cite{becke_density-functional_2005,johnson_post-hartreefock_2005,johnson_post-hartree-fock_2006,grimme_effect_2011}. It is employed in XDM,\cite{becke_exchange-hole_2005, becke_exchange-hole_2007} was later implemented into the framework of DFT-D3\cite{grimme_effect_2011} and is exclusively used in DFT-D4.\cite{caldeweyher_generally_2019} In the case of DFT-D3, the performance of the two types of damping has been compared by Grimme et al.\cite{grimme_effect_2011} for a large database of molecular interaction energies and only a small influence of the choice of damping has been observed. For liquid water however, there have been a few reports featuring a direct comparison between the two forms of damping, which showed differences in the produced structure \cite{sakong_structure_2016, dodia_structure_2019} and density isobar\cite{montero_de_hijes_density_2024} of water depending on the employed damping function, which indicates that the choice of damping might have a more significant effect than concluded in Ref. \onlinecite{grimme_effect_2011}. Still, the reason for the different behavior has not been explored in detail yet.
\\
Given these observations and the general uncertainty regarding the transferability of the damping function parameters, we feel it is necessary to revisit and reevaluate the effect of the damping function in dispersion-corrected DFT, especially for liquid water, due to its central importance for many different fields of research.
\\
To thoroughly assess the effect of damping, it is necessary to differentiate between effects resulting from the parametrization and functional form of the damping function, and from the choice of xc functional. Therefore, we start this investigation by parametrizing both a zero-type and BJ-damping function in the framework of the TS model for the PBE\cite{perdew_generalized_1996} and RPBE\cite{hammer_improved_1999} GGA functionals. We chose the TS model over a DFT-D method for the ease of implementation, featuring fewer adjustable parameters and relying less on precomputed data. However, we will compare our results to DFT-D3, where both types of damping are implemented, using the parameters provided by Grimme and coworkers\cite{grimme_consistent_2010, grimme_effect_2011} throughout this work.
\\
To assess the effect of the reference data, we use two benchmark sets that are commonly employed for this purpose\cite{tkatchenko_accurate_2009, klimes_chemical_2009, schroder_reformulation_2015, caro_parametrization_2017} - the S22\cite{jurecka_benchmark_2006} and the extended S66\cite{rezac_s66_2011, rezac_erratum_2014} benchmark sets - which include CCSD(T) interaction energies of model complexes  representing different NCIs. Afterwards, we test the obtained parameters on water cluster interaction energies contained in the Benchmark Energy and Geometry DataBase (BEGDB) benchmark set\cite{rezac_quantum_2008, temelso_benchmark_2011} to determine the performance and relative interaction strengths. 
\\
Using the obtained dispersion models, we explore the effect of damping on results obtained from molecular dynamics (MD) simulations of liquid water. Here, we focus on radial distribution functions (RDFs), the self-diffusion coefficient and density isobar of water. To be able to converge the properties with respect to system size and simulation time, we employ efficient and accurate high-dimensional neural network potentials\cite{behler_generalized_2007, behler_constructing_2015, behler_first_2017, behler_four_2021} (HDNNPs), trained on bulk water DFT reference data obtained from the different dispersion models, to run the simulations. 
We show, that dispersion models can perform similarly in static benchmark sets, and yet differ in MD simulations, where the parametrization and functional form of the damping function influences the produced results significantly for both investigated xc functionals. This demonstrates that the quality of a dispersion model cannot be reliably assessed on static geometries only, underlining the importance of also assessing the dynamical properties resulting from dispersion corrections.

\section{Methods}\label{sec:methods}

\subsection{Dispersion models}
\subsubsection{Overview}

The DFT-D3\cite{grimme_consistent_2010, grimme_effect_2011} and TS models~\cite{tkatchenko_accurate_2009} determine the dispersion energy $E_\mathrm{disp}$ using pair-wise additive, damped potentials based on the well-known London equation~\cite{eisenschitz_uber_1930},
\begin{equation}
    E_\mathrm{disp} = -\frac{1}{2}\sum^N_\mathrm{AB,\textbf{L}}{'}\sum_{n} s_n\frac{C_{n,\mathrm{AB}}}{{r_{\mathrm{AB},\textbf{L}}^n}}f_{\mathrm{damp},n}(r_{\mathrm{AB},\textbf{L}}),
    \label{eq:Edisp(2)}
\end{equation}
where $r_{\mathrm{AB},\textbf{L}} = |\textbf{R}_{A}-(\textbf{R}_{B}+\textbf{L})|$ is the distance between atoms A and B, which can be of the same or different elements.\cite{grimme_consistent_2010, bucko_tkatchenko-scheffler_2013} $\textbf{L}$ is a translation vector of the unit cell, which is either the zero vector for interactions inside the cell or a vector pointing to any replicated neighboring cell. The sum is over all $N$ atom pairs up to a large cutoff radius and the prime indicates that atoms do not interact with themselves inside the unit cell, i.e., $\mathrm{A}\neq \mathrm{B}$ for $\textbf{L} = \textbf{0}$. The dispersion coefficients $C_{n,\mathrm{AB}}$ of atom pair AB, which are of orders $n=6$ and $n=8$ in the case of DFT-D3 and $n=6$ for TS, describe the interaction strength of the pair. DFT-D3 scales the potentials using an additional xc functional-dependent factor $s_n$.  For GGA functionals the $n=6$ parameter is fixed to one while the $n=8$ parameter is optimized for the specific xc functional.\cite{grimme_consistent_2010, grimme_effect_2011} The TS model does not scale the interactions, i.e., $s_n=1$. 
\\
Integral to both models is the Casimir-Polder expression of the dispersion coefficient\cite{casimir_influence_1948, grimme_dispersion-corrected_2016},
\begin{equation}
C_{6,\mathrm{AB}} = \frac{3}{\pi} \int^\infty_0 \alpha_{\mathrm{A}}(i\omega)\alpha_{\mathrm{B}}(i\omega)d\omega,
\label{eq:Casimir}
\end{equation}
where $\alpha_\mathrm{A}(i\omega)$ and $\alpha_\mathrm{B}(i\omega)$ are the frequency-dependent polarizabilities of the respective atoms, evaluated at the imaginary frequency $\omega$. 
The main difference between the two models\cite{otero-de-la-roza_application_2020} is the way the $C_{6,\mathrm{AB}}$ coefficient is approximated for a given atom pair AB, which will be discussed in Sections \ref{sec:d3} and \ref{sec:ts}. 
\\
The $C_{n,\mathrm{AB}} / {r_{\mathrm{AB},\textbf{L}}^n}$ term in Eq.\:\ref{eq:Edisp(2)} diverges at short interatomic distances. Moreover, local correlation is typically included to some extent in xc functionals. Consequently, the pair-wise potentials must be damped by a function $f_\mathrm{damp}(r_{\mathrm{AB},\textbf{L}})$, which controls the short-range behavior and adapts the dispersion correction to a given xc functional.\cite{klimes_perspective_2012} The details of this damping will be further discussed in Sec.\:\ref{sec:damping}.
\\
We note that the simple pair-wise approach can be extended by considering three-body interactions via the Axilrod-Teller potential\cite{axilrod_interaction_1943} regardless of the dispersion model. However, three-body contributions to $E_\mathrm{disp}$ are rather small ($<5$--$10\%$)\cite{grimme_consistent_2010} and they are only rarely considered when using the original TS model. Therefore, to enable a consistent comparison of dispersion interactions in both models, three-body interactions will not be taken into account in the present work. 

\subsubsection{DFT-D3 model}\label{sec:d3}

In the DFT-D3 method the environment-dependent $C_{6,\mathrm{AB}}(\mathrm{CN_A},\mathrm{CN_B})$ coefficients depend on the fractional coordination numbers $\mathrm{CN_{A}}$ and $\mathrm{CN_{B}}$ and are determined from reference coefficients of atom pairs in different well-defined coordination environments. For this purpose, the fractional coordination number $\mathrm{CN_{A}}$ is given by 
\begin{equation}
    \mathrm{CN_{A}}=\sum^N_\mathrm{B\neq A} \frac{1}{1+e^{-k_1(k_2(R_\mathrm{A,cov}+R_\mathrm{B,cov})/r_{\mathrm{AB},\textbf{L}}-1)}},
    \label{eq:cn_d3}
\end{equation}
where the sum goes over all $N$ atoms  irrespective of the element. $R_\mathrm{A,cov}$ and $R_\mathrm{B,cov}$ are scaled, element-dependent covalent radii\cite{pyykko_molecular_2009}, and $k_1 = 16$ and $k_2 = 4/3$ are empirically determined scaling factors.\cite{grimme_consistent_2010}
\\
The reference environments of the atoms are realized by forming hydrides with different stoichiometries $\mathrm{A_mH_n}$ for each element in the periodic table up to $Z=94$.\cite{grimme_consistent_2010} For the majority of considered hydrides $\mathrm{m}=1$, but for elements such as carbon, where the value of $C_{6,\mathrm{AB}}$ changes significantly depending on the hybridization, i.e., $sp^3$ and $sp^2$ in ethane ($\mathrm{C_2H_6}$) and ethylene ($\mathrm{C_2H_4}$), respectively, m can be larger than one.\cite{grimme_consistent_2010} Additionally, the free atom is considered as well. For each reference compound, the frequency-dependent polarizability $\alpha_{\mathrm{A_mH_n}}(i\omega)$ (or $\alpha_{\mathrm{B_kH_l}}(i\omega)$) 
is obtained from time-dependent DFT (TDDFT).\cite{grimme_consistent_2010}
\\
Using Eq.\:\ref{eq:Casimir} and the approximate additivity of polarizablilities\cite{miller_additivity_1990, caldeweyher_generally_2019}, the reference $C^\mathrm{ref}_{6,\mathrm{AB}}$ coefficient of a pair in the associated coordination environment is determined according to
\begin{equation}
    \begin{split}
           C^\mathrm{ref}_{6,\mathrm{AB}} =& \frac{3}{\pi} \int^\infty_0 \frac{1}{m}\left[\alpha_{\mathrm{A_mH_n}}(i\omega)-\frac{n}{2}\alpha_{\mathrm{H_2}}(i\omega)\right]\\
        &\cdot \frac{1}{k} \left[\alpha_{\mathrm{B_kH_l}}(i\omega)-\frac{l}{2}\alpha_{\mathrm{H_2}}(i\omega)\right]d\omega,
    \end{split}
    \label{eq:D3_Casimir}
\end{equation}
where $\alpha_\mathrm{H_2}(i\omega)$ is the frequency-dependent polarizability of a hydrogen molecule.
These $C^\mathrm{ref}_{6,\mathrm{AB}}$ coefficients and the CNs of atom A and B in their respective reference hydrides or free atom forms act as supporting points for a two-dimensional interpolation procedure.\cite{grimme_consistent_2010}
The $C_{6,\mathrm{AB}}(\mathrm{CN_A},\mathrm{CN_B})$ coefficient for a given pair is then approximated by determining $\mathrm{CN_{A}}$ and $\mathrm{CN_{B}}$ using Eq.\:\ref{eq:cn_d3} and interpolating between these available supporting points.\cite{grimme_consistent_2010} Higher order coefficients are calculated recursively from the $C_{6,\mathrm{AB}}$ values using element pair-dependent scaling coefficients.\cite{grimme_consistent_2010, schroder_reformulation_2015}

\subsubsection{Tkatchenko-Scheffler model}\label{sec:ts}

The TS model takes advantage of the  relation between the polarizability $\alpha$ and volume $V$\cite{brinck_polarizability_1993} of an atom to approximate the $C_{6,\mathrm{AB}}$ coefficients from free atom reference data. The ratio $v_\mathrm{A}$ of the volume $V_\mathrm{A}$ of atom A in a molecule to the volume of the respective free atom  $V^\mathrm{free}_\mathrm{A}$ is determined using a Hirshfeld partitioning\cite{hirshfeld_bonded-atom_1977} of the electron density $n(\mathbf{r})$,
\begin{align}
\begin{split}
v_\mathrm{A} &= \frac{V_\mathrm{A}}{V^\mathrm{free}_\mathrm{A}},\\
\frac{V_\mathrm{A}}{V^\mathrm{free}_\mathrm{A}} &= \frac{\int r^3w_\mathrm{A}(\mathbf{r})n(\mathbf{r})d^3\mathbf{r}}{\int r^3(\mathbf{r})n_\mathrm{A}^\mathrm{free}(\mathbf{r})d^3\mathbf{r}},\\
 w_\mathrm{A}(\mathbf{r}) &= \frac{n_\mathrm{A}^\mathrm{free}(\mathbf{r})}{\sum_\mathrm{B}n_\mathrm{B}^\mathrm{free}(\mathbf{r})}.
\end{split}
\end{align}
Using the determined ratios $v_\mathrm{A}$, the reference free atom polarizabilities $\alpha_{0,\mathrm{A}}^\mathrm{free}$ and the $C^\mathrm{free}_{6,\mathrm{AA}}$ coefficients of neutral atom pairs of the same element in vacuum published by Chu and Dalgarno\cite{chu_linear_2004} are scaled according to
\begin{align}
    C_{6,\mathrm{AA}} &= {v_\mathrm{A}^2}C_{6,\mathrm{AA}}^\mathrm{free}, \\
    \alpha_{0,\mathrm{A}} &= v_{\mathrm{A}}\alpha^{\mathrm{free}}_{0,\mathrm{A}},
\label{eq:scale_c6}
\end{align}
to give the $C_{6,\mathrm{AA}}$ coefficient and polarizability $\alpha_{0,\mathrm{A}}$ of the atom in the molecule. Using these scaled values, the $C_{6,\mathrm{AB}}$ coefficient for an atom pair AB is determined as
\begin{equation}
C_{6,\mathrm{AB}} = \frac{2C_{6,\mathrm{AA}}C_{6,\mathrm{BB}}}{\left[\frac{\alpha_{0,\mathrm{B}}}{\alpha_{0,\mathrm{A}}}C_{6,\mathrm{AA}}+\frac{\alpha_{0,\mathrm{A}}}{\alpha_{0,\mathrm{B}}}C_{6,\mathrm{BB}}\right]}.
\label{eq:TS_C6}
\end{equation}
This equation is the result of a series expansion for $\alpha_{\mathrm{A/B}}(i\omega)$ and a rearrangement of Eq.\:\ref{eq:Casimir}~\cite{tkatchenko_accurate_2009}.
Higher order dispersion coefficients are not determined in the TS model.

\subsubsection{Damping Functions}\label{sec:damping}

\begin{figure}
    \centering
    \includegraphics[scale=0.8]{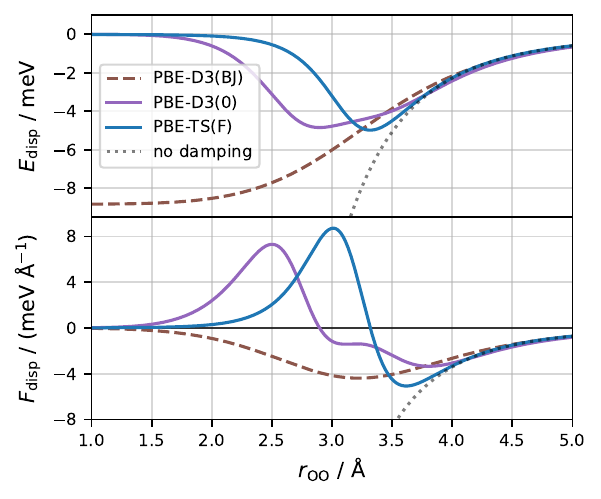}
    \caption{Dispersion energy $E_\mathrm{disp}$ and resulting dispersion forces $F_\mathrm{disp}$ for an oxygen atom pair (OO) as a function of the pair distance $r_{\mathrm{OO}}$. For each curve, the respective parameters for the PBE functional and $C_{6,\mathrm{OO}}$ and $C_{8,\mathrm{OO}}$ coefficients of a free oxygen pair were employed (no environment dependence). In the case of TS $v_\mathrm{A/B}$ was fixed to one. Additionally, the $-C^\mathrm{free}_{6,\mathrm{OO}} \cdot r_\mathrm{OO}^{-6}$ curve without damping, with $C^\mathrm{free}_{6,\mathrm{OO}} = 15.6\,E_\mathrm{h}\,{a_0^6}$ taken from Ref.\:\onlinecite{chu_linear_2004} is shown for comparison.}
    \label{fig:damping-comparison}
\end{figure}

The damping function $f_\mathrm{damp}$ determines the behavior of the dispersion correction at short to intermediate interatomic distances.\cite{grimme_effect_2011} As shown in Fig.\:\ref{fig:damping-comparison}, the shape of the dispersion correction potential is quite drastically affected by this choice of the damping function. 

There are two different ``damping-philosophies''.\cite{grimme_effect_2011} One argues that $E_\mathrm{disp}$ should approach zero in the $r_{\mathrm{AB},\textbf{L}} \rightarrow 0$ limit to avoid double-counting of short-range correlation interactions because they are treated locally by the xc functional. Damping functions, which show this behavior, include the so called ``zero-damping function'' proposed by Chai and Head-Gordon\cite{chai_long-range_2008},
\begin{equation}
f^0_\mathrm{damp}(r_{\mathrm{AB},\textbf{L}}) = \frac{1}{1+6\left(\frac{r_{\mathrm{AB},\textbf{L}}}{s_{r,n}R_{0,\mathrm{AB}}}\right)^{-\gamma_n}},
\label{eq:zero_damping}
\end{equation}
and the Fermi-type damping function proposed by Wu and Yang\cite{wu_empirical_2002},
\begin{equation}
    f^\mathrm{F}_\mathrm{damp}(r_{\mathrm{AB},\textbf{L}}) = \frac{1}{1+\exp{\left[-d\left(\frac{r_{\mathrm{AB},\textbf{L}}}{s_{r,n}R_{0,\mathrm{AB}}}-1\right)\right]}}.
    \label{eq:fermi_damping}
\end{equation}
Both are functions of the pair distance $r_{\mathrm{AB},\textbf{L}}$ and include an element-pair-dependent cutoff radius $R_\mathrm{0,AB}$, which determines the range of the damping and defines the region in which correlation is described by the xc functional.\cite{grimme_consistent_2010}  Since the range can differ for different xc functionals, a scaling factor $s_{r,n}$ is used to scale the cutoff radii that needs to be optimized for a given xc functional. Additionally, $\gamma_n$ in Eq.\:\ref{eq:zero_damping} and $d$ in Eq.\:\ref{eq:fermi_damping} determine the slope of the damping function.\cite{grimme_effect_2011, tkatchenko_accurate_2009}
\\
Eq.\:\ref{eq:zero_damping} was chosen as damping function in the original DFT-D3 publication\cite{grimme_consistent_2010} with $\gamma_6 = 14$ and $\gamma_8 = 16$. Only the $s_{r,6}$ factor is typically optimized while the $s_{r,8}$ is commonly set to one for GGA functionals.\cite{grimme_consistent_2010} DFT-D3 employing the zero-damping function Eq.\:\ref{eq:zero_damping} will be referred to as DFT-D3(0) in the following. For DFT-D3(0) individual cutoff radii were determined for each element pair by finding the distance at which the first-order DFT interaction energy of a given element combination reaches a predefined threshold.\cite{grimme_consistent_2010}
\\
The Fermi-type damping function Eq.\:\ref{eq:fermi_damping} is usually employed for damping within the framework of the TS model.\cite{tkatchenko_accurate_2009} We use the term ``zero-damping'' also to refer to the behavior of Eq.\:\ref{eq:fermi_damping}, but use the label Fermi-type damping (F) for this specific function. In the TS(F) model\cite{tkatchenko_accurate_2009}, the parameter $d$ in Eq.\:\ref{eq:fermi_damping} is set to 20 and $R_\mathrm{0,AB}$ is determined from the vdW-radii of the free atoms $R^\mathrm{free}_{\mathrm{0,A/B}}$ according to
\begin{align}
\begin{split}
R_{\mathrm{0,AB}} &= R_{\mathrm{0,A}} + R_{\mathrm{0,B}},\\
R_{\mathrm{0,A/B}} &= {v_\mathrm{A/B}}^{1/3} R^\mathrm{free}_{\mathrm{0,A/B}}.
\label{eq:scale_r_ts}
\end{split}
\end{align}
Therefore, $R_\mathrm{0,AB}$ is an environment-dependent quantity in the TS model and not constant for a given element pair like in the DFT-D3 method. Since Eq.\:\ref{eq:Edisp(2)} is truncated after $n=6$ in the case of the TS model, the $s_{r,6}$ scaling factor is the only parameter that needs to be optimized for a given xc functional in the TS(F) model.
\\
In Fig.\:\ref{fig:damping-comparison} the approximate interaction potential and its negative gradient, i.e., the dispersion forces $F_\mathrm{disp}$, of two oxygen atoms are shown for DFT-D3(0) and TS(F) using the optimized parameters for the PBE xc functional of Grimme et al.\cite{grimme_consistent_2010} and Tkatchenko and Scheffler\cite{tkatchenko_accurate_2009}, respectively, as well as the $C_{6,\mathrm{AB}} / {r_\mathrm{AB}}^6$ term without damping. For visualization purposes, the environment-dependence of $C_{6/8,\mathrm{AB}}$ and $R_\mathrm{0,AB}$ has been omitted. This effect of the environment is very small in case of the TS model, while the DFT-D3 potential curve would exhibit a slight ``bump'' at short distances\cite{grimme_consistent_2010}, which is due to $C_{6,\mathrm{AB}}$ changing abruptly when the CN of the oxygen atoms becomes greater than zero and the interpolation switches from the free atom supporting point to the next, which is the OH radical with a CN of one.
\\
It can be seen that both energy curves, DFT-D3(0) and TS(F), follow the curve without damping closely in the long-range regime but start to deviate between 2.5\,\AA~to 3.5\,\AA~when they reach the scaled cutoffs (6.38~a$_0$ for TS and 4.7~a$_0$ for D3)  
after which they are damped to zero for short distances. The positions of the minima differ for the two methods, which is mainly due to the difference in parametrization of $f_\mathrm{damp}$. Additionally, DFT-D3(0) has an increased flexibility to adjust the potential to the xc functional due the additional $n=8$ term\cite{grimme_consistent_2010, schroder_reformulation_2015}, which results in different shapes of the potential wells of the two methods.
\\
In the lower plot of Fig. \ref{fig:damping-comparison}, the resulting dispersion forces of the potentials are shown. It can be seen that due to damping to zero, the forces become positive, i.e., repulsive, at short distances. This behavior can lead to larger interatomic distances when the equilibrium bond-length predicted by the uncorrected xc functional is lower than the scaled cutoff,\cite{grimme_effect_2011} which is incompatible with the definition of dispersion interactions that should be attractive at any range.
\\
This behavior of the forces for zero-damping is the motivation for the second ``damping-philosophy'', which argues that $E_\mathrm{disp}$ should not approach zero in the $r_{\mathrm{AB},\textbf{L}} \rightarrow 0$ limit, but instead should converge to a constant non-zero value.\cite{grimme_effect_2011} This form of damping is derived from a convergent multipole-expansion of $E_\mathrm{disp}$ as introduced by Koide\cite{koide_new_1976} and was first proposed by Becke and Johnson.\cite{becke_density-functional_2005,johnson_post-hartreefock_2005,johnson_post-hartree-fock_2006,grimme_effect_2011} 
\begin{equation}
    f^\mathrm{BJ}_{\mathrm{damp},n}\left(r_{\mathrm{AB},\textbf{L}}\right) = \frac{{r_{\mathrm{AB},\textbf{L}}^n}}{{r_{\mathrm{AB},\textbf{L}}^n}+(a_1 R_{0,\mathrm{AB}}+a_2)^n},
    \label{eq:BJ_damping}
\end{equation}
where $a_1$ and $a_2$ are functional-dependent parameters. $a_1$ has the same physical meaning as $s_{r,n}$ in Eq.\:\ref{eq:zero_damping} and Eq.\:\ref{eq:fermi_damping}~\cite{grimme_effect_2011} and $a_2$ is an additional parameter, which enables finer adjustment of the damping function to a given xc functional.\cite{grimme_effect_2011}
This form of damping was incorporated into DFT-D3 method in a later publication\cite{grimme_effect_2011} and will be called DFT-D3(BJ) in the following.
\\
In addition to the difference in damping between DFT-D3(0) and DFT-D3(BJ), $R_{0,\mathrm{AB}}$ is determined differently as well. In DFT-D3(BJ), $R_{0,\mathrm{AB}}$ is calculated from $C_{6,\mathrm{AB}}$ and $C_{8,\mathrm{AB}}$ according to
\begin{equation}
R_{0,\mathrm{AB}} = \sqrt{\frac{C_{8,\mathrm{AB}}}{C_{6,\mathrm{AB}}}}.
\end{equation}
However, since $C_{8,\mathrm{AB}}$ is calculated recursively from $C_{6,\mathrm{AB}}$ using element pair-dependent scaling coefficients, $R_{0,\mathrm{AB}}$ remains constant for a given pair.
\\
The resulting shape of the oxygen-oxygen (OO) interaction potential produced by DFT-D3(BJ) is displayed in Fig. \ref{fig:damping-comparison}. It can be seen that BJ-damping avoids the repulsive forces at short distances, which is found for DFT-D3(0), making the approach more consistent with the definition of dispersion interactions being always attractive. However, due to BJ-damping approaching a constant value, there is a significant contribution of $E_\mathrm{disp}$ even at short distances where correlation is approximately included in the xc functional. For the case of DFT-D3(BJ), Grimme et al.\cite{grimme_effect_2011} investigated the possibility of errors due to this double-counting, but did not observe any significant errors when combining the approach with standard, unmodified xc functionals.
\\
In the present work, we have implemented BJ-damping into the TS model to investigate if the behavior of this type of damping is consistent between different dispersion models. Also in this TS(BJ) case, $R_{0,\mathrm{AB}}$ is determined using Eq.\:\ref{eq:scale_r_ts}, i.e., the same way it is done for the TS(F) model.

\subsection{High-dimensional neural network potentials}

Machine learning potentials (MLPs)\cite{P4885,P5793,P6102,P6112,P6121} allow to significantly accelerate the time and length scales of \textit{ab initio} MD (AIMD) simulations by learning the potential energy surface (PES) of a system, as described by an electronic structure method, using machine learning algorithms. Using the learned PES, the forces, required to propagate the atomic positions forward in time, can be determined at a significantly reduced computational cost while the accuracy of the underlying electronic structure method is retained up to typical energy errors of about 1\,meV\,/\,atom and force errors of about 100\,meV\,/\,\AA.\cite{behler_four_2021}
\\
HDNNPs\cite{behler_generalized_2007, behler_four_2021} are a class of MLP, that approximate the potential energy $E$ of a system as a sum of environment-dependent atomic energy contributions $E^\mu_A$\cite{behler_generalized_2007,behler_four_2021},
\begin{equation}
E = \sum^{N_\mathrm{elem}}_{\mu=1}\sum^{N^\mu_\mathrm{atom}}_{A=1} E^\mu_A,
\label{eq:hdnnp_pot_e}
\end{equation}
where $N_\mathrm{elem}$ is the number of elements $\mu$ in the system, and $N^\mu_\mathrm{atom}$ is the number of atoms of the respective element.\cite{behler_four_2021}
\\
The local environment around a central atom A is defined by a cutoff function\cite{behler_atom-centered_2011},
\begin{equation}
    f_\mathrm{c}(r_\mathrm{AB})=\begin{cases}
0.5\cdot\left[\cos\left(\frac{\pi r_\mathrm{AB}}{R_c}\right)+1\right] & \text{if } r_\mathrm{AB}\leq R_\mathrm{c}\,,\\
0& \text{if } r_\mathrm{AB}> R_\mathrm{c}\,,
\end{cases}
\end{equation}
where $r_\mathrm{AB}$ is the distance between the central atom and a neighbor B and $R_\mathrm{c}$ is the cutoff radius.
Inside the cutoff sphere, the geometry is described by many-body descriptors called radial and angular atom-centered symmetry functions (ACSF)\cite{behler_atom-centered_2011},
\begin{align}
    \label{eq:ACSF_rad}
    G^{\mathrm{rad}}_\mathrm{A} &= \sum_\mathrm{B} \mathrm{e}^{-\eta(r_\mathrm{AB}-R_\mathrm{s})^2}\cdot f_\mathrm{c}(r_\mathrm{AB})\;, \\
    \begin{split}
        G_\mathrm{A}^{\mathrm{ang}} &= 2^{1-\zeta}\sum_\mathrm{B}\sum_\mathrm{C\neq B}(1+\lambda\cos\theta_\mathrm{ABC})^\zeta \\
        &\cdot\mathrm{e}^{-\eta \left(r_\mathrm{AB}^2+r_\mathrm{AC}^2+r_\mathrm{BC}^2\right)}\cdot f_\mathrm{c}(r_\mathrm{AB})f_\mathrm{c}(r_\mathrm{AC})f_\mathrm{c}(r_{\mathrm{BC}})\;.
        \label{eq:ACSF_ang}
    \end{split}
\end{align}
For each element in the system a set of radial functions employing Eq.\:\ref{eq:ACSF_rad} is used to describe the radial distribution of the neighboring atoms B around the central atom with parameters $\eta$ and $R_\mathrm{s}$ determining the shapes of the functions.
\\
Similarly, functions defined by Eq.\:\ref{eq:ACSF_ang} describe the angular distribution of the neighbors around the central atom. They depend on the angle $\theta_\mathrm{ABC}$ centered at A, which is formed with neighboring atoms B and C. Further, Eq.\:\ref{eq:ACSF_ang} depends on the cutoff functions of the interatomic distances $R_\mathrm{AB}$, $R_\mathrm{AC}$ and $R_\mathrm{BC}$ as well as on the parameters $\eta$-, $\zeta$- and $\lambda$. The $\zeta$-parameter can be used to control the distribution of the angles while the $\lambda$-parameter can have values of $+1$ and $-1$, which can be used to invert the cosine of the angle.\cite{behler_constructing_2015}
The sum over all neighbors in Eq.\:\ref{eq:ACSF_rad} and Eq.\:\ref{eq:ACSF_ang} ensures the required permutational invariance of the descriptor. Furthermore, relying on internal coordinates like distances and angles ensures rotational and translational invariance. More details about ACSFs and their construction can be found in Refs.~\citenum{behler_atom-centered_2011} and \citenum{P6548}.
The relation between the local atomic environment and contribution of the central atom to the total energy is learned using atomic feed-forward neural networks (FFNNs), which have the same architecture and weight parameters for all atoms of the same element. 

\section{Computational details}

\subsection{Density functional theory calculations}

All DFT calculations were performed using the Fritz-Haber-Institute ab initio molecular simulations (FHI-aims) code (version 221103)\cite{blum_ab_2009} using the PBE\cite{perdew_generalized_1996} and RPBE\cite{hammer_improved_1999} xc functionals. For calculations of periodic water boxes, the ``intermediate'' basis sets of numeric atom-centered orbitals were employed. The \textbf{k}-point grid was set up using a \textbf{k}-point density of at least 2.15\,$1/\mathrm{\AA}^{-1}$. The convergence criterion for the SCF cycle was set to $10^{-6}$\,eV for total energies, $10^{-3}$\,eV for the sum of eigenvalues and $5\cdot 10^{-5}\,\mathrm{e}/{a_0}^3$ for the electron density.
\\
The calculation of the interaction energies of the model complexes contained in the S22\cite{jurecka_benchmark_2006} and S66\cite{rezac_s66_2011, rezac_erratum_2014} data sets, and in the  BEGDB water cluster benchmark set\cite{rezac_quantum_2008, temelso_benchmark_2011} were performed using the ``tight'' FHI-aims basis sets instead with identical convergence criteria. To reduce the effect of the basis-set superposition error (BSSE) on the interaction energies, a Counterpoise correction as proposed by Boys and Bernardi\cite{boys_calculation_1970} was employed for the dimers contained in the datasets. For larger clusters beyond dimers, the generalized site-site function Counterpoise correction proposed by Wells and Wilson\cite{wells_van_1983} was employed.
\\
Additional details on the employed FHI-aims input parameters are given in the Supplementary Material in Sec.\:SI A. The performance of the different FHI-aims basis-sets and the effect of the Counterpoise correction on the interaction energies of the model complexes contained in both the S22 and S66 data sets is investigated in the Supplementary Material in Sec.\:SII D. The magnitude of the BSSE correction is on average below 5\,meV for the employed ``tight'' basis set. Further, a comparison between the RDF produced by RPBE using the ``intermediate'' and ``tight'' basis from previous work\cite{morawietz_how_2016} is shown in the Supplementary Material in Sec.\:SIII D. We find, that the RDFs obtained with both basis-sets are essentially identical and converged with respect to basis set size.

\subsection{Dispersion corrections}

Dispersion corrections according to the DFT-D3 model were determined using the DFT-D3 software developed by Grimme et al.\cite{grimme_consistent_2010, grimme_effect_2011} (version 3.1 from October 2015). The default parameters for the PBE and RPBE xc functionals were used for zero- as well as BJ-damping. Additionally, the default cutoff of 50.2\,\AA~was employed.
\\
For the TS dispersion corrections, a Hirshfeld analysis\cite{hirshfeld_bonded-atom_1977} was performed for each DFT calculation after convergence of the scf cycle using the FHI-aims code. The resulting atomic Hirshfeld volumes were used to correct for dispersion interactions using our own implementation of the TS model, which also offers the option to use BJ-damping as an alternative to Fermi damping. We use the reference polarizabilities $\alpha_{0, \mathrm{H}}^\mathrm{free} = 4.5\,{a_0^3}$ and $\alpha_{0, \mathrm{O}}^\mathrm{free} = 5.4\,{a_0^3}$ and the reference dispersion coefficients $C^\mathrm{free}_{6,\mathrm{HH}}=6.5\,E_\mathrm{H}\,{a_0^6}$ and $C^\mathrm{free}_{6,\mathrm{OO}}=15.6\,E_\mathrm{H}\,{a_0^6}$ for hydrogen and oxygen atoms, respectively, from the database of Chu and Dalgarno\cite{chu_linear_2004}. The vdW-radii of the free hydrogen and oxygen atoms are $R_{\mathrm{0,H}}^{\mathrm{free}}=3.1\,a_0$ and $R_{\mathrm{0,O}}^{\mathrm{free}}= 3.19\,a_0$.\cite{hermann_libmbd_2023} The parametrization of both damping functions for PBE and RPBE is discussed in Sec.\:\ref{sec:parametrization}. The cutoff for TS dispersion corrections was set to 50\,\AA.

\subsection{Training of high-dimensional neural network potentials}

The HDNNPs were trained using the RuNNer code.\cite{behler_constructing_2015, behler_first_2017}
Initial reference geometries, which include periodic liquid water and ice structures, were taken from previous works\cite{morawietz_how_2016, eckhoff_insights_2021} and the energies and forces were determined using DFT with the PBE and RPBE xc functionals. This initial reference data was further extended for both xc functionals individually using an active learning protocol.\cite{eckhoff_molecular_2019, eckhoff_high-dimensional_2021} The final PBE and RPBE datasets include 13,725 and 10,210 structures, respectively, which contain between eight and 128 water molecules.
\\
The two reference datasets were used to generate multiple datasets by applying dispersion corrections using either DFT-D3(0), DFT-D3(BJ), TS(F) or TS(BJ). For the TS models, two different parametrizations of each damping function were employed obtained using either the S22 and S66 benchmark sets as discussed in Sec.\:\ref{sec:parametrization}. Including the uncorrected PBE and RPBE data, this resulted in 14 different HDNNPs.
\\
For comparability, for all HDNNPs we use the same neural network architecture containing two hidden layers each consisting of 15 neurons, which we found to give accurate training results and stable potentials in all cases with energy root mean square errors (RMSEs) well below 1\;meV\,atom$^{-1}$. The oxygen environments are described by 32 radial (Eq.\:\ref{eq:ACSF_rad}) and 70 angular (Eq.\:\ref{eq:ACSF_ang}) ACSFs. For hydrogen 32 radial and 72 angular ACSFs were employed. A cutoff radius of $6.0\,\mathrm{\AA}$ was chosen. The ACSF parameters are summarized in the Supplementary Material in Sec.\:SI\:B\:2.
\\
90\% of the reference data was used for training while the remaining 10\% was used for validation. From the entire training data, $2.5\%$ of the force components were randomly chosen in each epoch, while all energies have been used throughout the optimization process of the neural network weights. The full RuNNer training settings can be found in the Supplementary Material in Sec.\:SI\:B\:1.
The RMSEs of the energies and force components of the 14 HDNNPs are summarized in Table\:\ref{tab:hdnnps}.

\begin{table}
\caption{Summary of the training results of the HDNNPs used for running the MD simulations of bulk water systems in the present work. The HDNNPs were trained on DFT bulk water reference datasets determined using the PBE and RPBE xc functionals and corrected for dispersion using DFT-D3 and TS models with different damping-functions. For DFT-D3, zero- (0) and BJ-damping and for TS, Fermi- (F) and BJ-damping was employed. The two damping functions of the TS models were parameterized for both xc functionals individually using both the S22\cite{jurecka_benchmark_2006} and the S66\cite{rezac_s66_2011, rezac_erratum_2014} benchmark sets as indicated by (22) and (66), as  discussed in Sec.\:\ref{sec:parametrization}. For the DFT-D3 models, the default parametrization of Grimme et al.\cite{grimme_consistent_2010, grimme_effect_2011} was employed. Root mean squared errors (RMSE) of the energies are given in meV\,atom$^{-1}$ and the RMSEs of the force components are in meV\,$a_0^{-1}$ for both the training and the test set.}
\label{tab:hdnnps}
\begin{ruledtabular}
\begin{tabular}{l c c c r}
\multirow{ 2}{*}{Method} & \multicolumn{2}{c}{RMSE (Energy)} & \multicolumn{2}{r}{RMSE (Forces)} \\ 
& train & test & train & test \\
\hline
PBE & 0.605  & 0.650 & 24.985  &  24.846\\
PBE-TS(F) (S22) &  0.578 & 0.644  & 25.398 & 25.497 \\
PBE-TS(F) (S66) & 0.587 & 0.645 & 25.266 & 25.524 \\
PBE-TS(BJ) (S22) & 0.600 & 0.642  &   26.306   &  26.030 \\
PBE-TS(BJ) (S66) & 0.585  &   0.653   &  25.970  &  26.252 \\
PBE-D3(0) &  0.651  &   0.695  &   27.729  &   27.457 \\
PBE-D3(BJ) & 0.574 & 0.639 & 24.468 & 24.687 \\
\hline
\hline
RPBE & 0.581 & 0.622  & 26.568 & 25.933\\
RPBE-TS(F) (S22) & 0.679 & 0.718 & 54.071  & 53.748 \\
RPBE-TS(F) (S66) &   0.659 & 0.768  & 37.830 & 37.321 \\
RPBE-TS(BJ) (S22) & 0.634 & 0.668 & 36.354 & 35.972 \\
RPBE-TS(BJ) (S66) & 0.603 & 0.641 & 32.167 & 31.698 \\
RPBE-D3(0) & 0.591  & 0.645  & 29.570 & 28.990 \\
RPBE-D3(BJ) & 0.589  &  0.637  & 28.126  & 27.633 \\
\end{tabular}
\end{ruledtabular}
\end{table}

\subsection{Molecular dynamics simulations}\label{sec:theory_md}

MD simulations with DFT accuracy were performed using the large-scale atomic/molecular massively parallel simulator (LAMMPS)\cite{plimpton_fast_1995, thompson_lammps_2022} in combination with the neural network potential package (n2p2)\cite{singraber_library-based_2019}. Since our focus is on the role of the damping function in dispersion corrections, classical simulations have been performed in all cases.
The RDFs were obtained in the canonical (\textit{NVT}) ensemble applying the Nosé-Hoover\cite{nose_molecular_1984, hoover_canonical_1985} thermostat with a coupling constant of 0.05\,ps. All MD simulations in the \textit{NVT} ensemble were performed in a cubic, periodic simulation cell with a box length of $24.83\;\mathrm{\AA}$ containing 512 water molecules, which corresponds to a density of $1\:\mathrm{g/cm^{3}}$. The trajectories were run with a time step of 0.5\,fs for 2.1\,ns. The first 100\,ps were used to equilibrate the system and discarded. Then, the RDFs were averaged over 2 ns using snapshots of the trajectories taken in 5\,fs time intervals. 
\\
Moreover, MD simulations in the isothermal-isobaric (\textit{NpT}) ensemble at $p=1 \:\mathrm{bar}$ were performed to obtain the density isobars. Here, a Nosé-Hoover barostat with a coupling constant of 0.5\,ps allowing isotropic changes of the simulations cell was employed in addition to the Nosé-Hoover thermostat. The densities were sampled every 10\,fs for 2\,ns following 100\,ps of equilibration. 
\\
The self-diffusion coefficients were obtained from MD simulations in the micro-canonical (\textit{NVE}) ensemble using a time step of 0.5\,fs for 100\,ps. To extrapolate the self-diffusion coefficient to an infinite system size, simulations for 64, 128, 216, 512 and 1024 water molecules have been performed. The starting configurations have been obtained from 100\,ps \textit{NVT} simulations of cubic simulation boxes with cell parameters resulting in a density of $1\:\mathrm{g/cm^{3}}$ for the respective number of water molecules. To ensure a sufficient sampling of the canonical ensemble, twenty different starting configurations have been generated for each system-size. For every \textit{NVE} simulation a self-diffusion coefficient was determined from the Green-Kubo relation between the self-diffusion coefficient and the integral over the velocity autocorrelation function (VACF).\cite{frenkel_2002} We sampled the velocities every 1\,fs and used every timestep as origin to calculate the VACF. The integration limit to determine the self-diffusion coefficient was 20\,ps. For each system size, the self-diffusion coefficient $D_L$ was determined by linear regression of the logarithm of the self-diffusion coefficients versus the inverse mean temperature of the \textit{NVE} simulations. For RPBE the simulations were carried out at 300~K, while an increased $T$ of 400~K has been used for the PBE functional due to the well-known low diffusion obtained with this functional.\cite{gillan_perspective_2016} Using the obtained $D_L$ values, a linear regression of the system size-specific self-diffusion coefficients versus the inverse simulation box length was performed. The self-diffusion coefficient of the infinite system $D_\infty$ was then determined from the $D_L$-intercept of the fitted function. Further information on the determination of the self-diffusion coefficients can be found in the Supplementary Material in Sec.\:SIV.

\section{Results and discussion}\label{sec:discussion}

\subsection{Optimization of damping function parameters for the Tkatchenko-Scheffler model}\label{sec:parametrization}

The free parameters of the Fermi-type damping function Eq.\:\ref{eq:fermi_damping} and BJ-damping function Eq.\:\ref{eq:BJ_damping} are usually optimized by minimizing the mean absolute error (MAE) of interaction energies predicted by a given xc functional with the TS model compared to benchmark results of a higher level of theory, often CCSD(T).\cite{tkatchenko_accurate_2009} We selected two different benchmark sets that are commonly employed for this purpose to probe if and how the choice of the benchmark set affects the optimal parameters.
\\
The first benchmark set is the S22  set of Jure\v{c}ka et al.\cite{jurecka_benchmark_2006}, which contains CCSD(T) interaction energies of twenty-two model complexes representing different NCIs, which are relevant in biological systems. The interactions can be separated into hydrogen bonded, dispersion stabilized and ``mixed'', where electrostatic and dispersion interactions are similar in magnitude.\cite{jurecka_benchmark_2006} Tkatchenko and Scheffler\cite{tkatchenko_accurate_2009} employed this benchmark set to parametrize the Fermi-type damping function for the PBE xc functional when they introduced their method. Caro reported optimized $s_{r,6}$ parameters for several xc functionals, among them RPBE and PBE\cite{caro_parametrization_2017}.
\\ 
While the S22 benchmark set of molecules is commonly employed, its scope is limited. Consequently, it has been argued that some interaction motifs are underrepresented\cite{rezac_s66_2011}, which would make the obtained parameters less transferable. Additionally, Grimme et al.\cite{grimme_effect_2011} argue that the limited size is insufficient to optimize BJ-damping, which has a more complex error surface resulting from the additional free parameter.
\\
The second benchmark set is the S66 data set of \v{R}ez\'{a}\v{c} et al.\cite{rezac_s66_2011, rezac_erratum_2014}, which represents an extension of S22. It includes the same interaction types, but consists of 66 biomolecular model complexes to achieve a more balanced composition. For this benchmark set, we make use of revised interaction energies reported by Schmitz and Hättig\cite{schmitz_accuracy_2017}, which are closer to the CBS limit than the original values.
\\
Using these two benchmark sets, we optimized the free parameters for both damping functions for the PBE and RPBE xc functionals. For the Fermi-type damping function, we screened $s_{r,6}$  in the range from 0.50 to 1.10 in steps of 0.01 and for BJ-damping we screened along $a_1$ in the range from 0.00 to 3.00 in steps of 0.01 and for $a_2$ from $0.00\,a_0$ to $6.50\,a_0$ in steps of $0.01\,a_0$. Here, $a_0$ is the Bohr radius. The optimal parameters were determined by finding the minimum of the respective error curve or surface. An overview over the optimized parameters and MAEs for the S22 and S66 benchmark sets is presented in Tables\:\ref{tab:S22_damp_parms} and \:\ref{tab:S66_damp_parms}, respectively. For comparison, we also included the MAEs for both xc functionals without dispersion correction and with DFT-D3(0) and DFT-D3(BJ) using the  parameters of Grimme et al.\cite{grimme_consistent_2010, grimme_effect_2011}. Additionally, we show the error curves and surfaces obtained in the optimization in the Supplementary Material in Sec.\:SII.
\\
\begin{table*}
\caption{Mean absolute errors (MAEs) between CCSD(T) interaction energies of model complexes contained in the S22\cite{jurecka_benchmark_2006} benchmark data set and DFT using the PBE and RPBE xc functionals. The interactions of the complexes can be separated into hydrogen bonded (HB), dispersion stabilized and ``mixed'', where electrostatic and dispersion interactions are similar in magnitude.\cite{jurecka_benchmark_2006} The ``Total'' MAEs take all complexes into account. The DFT interaction energies have been corrected for dispersion using different dispersion models. Boldface rows represent results obtained using models, whose damping function parameters --$s_{r,6}$ for the Fermi-type damping, and $a_1$ and $a_2$ for the BJ-damping function-- have been optimized on the same S22 benchmark data by minimizing the MAE.}
\label{tab:S22_damp_parms}
\begin{ruledtabular}
\begin{tabular}{ l  c  S[table-format=3.2]  S[table-format=3.2]  S[table-format=3.2]  rS[table-format=3.2]}
Method & Optimized parameters & \multicolumn{1}{c}{HB / meV} & \multicolumn{1}{c}{Dispersion / meV} & \multicolumn{1}{c}{Mixed / meV} & \multicolumn{1}{r}{Total / meV} \\
\hline
PBE & - & 45.55 & 209.62 & 87.07 & 118.42 \\
\textbf{PBE-TS(F) (S22)} & $\bm{s_{r,6} = 0.94}$ & \textbf{20.59} & \textbf{15.74} & \textbf{6.31} & \textbf{14.28} \\
PBE-TS(F) (S66) & $s_{r,6} = 0.99$ & 14.52 & 37.45 & 13.76 & 22.62 \\
\textbf{PBE-TS(BJ) (S22)} & $\bm{a_1 = 0.00}$\textbf{, }$\bm{a_2 = 5.90\;a_0}$ & \textbf{29.10} & \textbf{15.14} & \textbf{5.14} & \textbf{16.40}\\
PBE-TS(BJ) (S66) & $a_1 = 0.00$, $a_2 = 6.27\;a_0$ & 18.31 & 35.23 & 9.94 & 21.80\\
PBE-D3(0) & - & 26.37 & 39.51 & 9.82 & 25.88 \\
PBE-D3(BJ) & - & 28.57 & 33.82 & 9.54 & 24.42 \\
\hline
\hline
RPBE & - &  180.52 & 318.61 & 151.50 & 221.50 \\
\textbf{RPBE-TS(F) (S22)} & $\bm{s_{r,6} = 0.60}$ & \textbf{20.87} & \textbf{29.38} & \textbf{18.55} & \textbf{23.23} \\
RPBE-TS(F) (S66) & $s_{r,6} = 0.65$ & 32.54 & 52.37 & 13.91 & 33.82 \\
\textbf{RPBE-TS(BJ) (S22)} & $\bm{a_1 = 0.16}$\textbf{, }$\bm{a_2 = 2.95\;a_0}$ & \textbf{8.67} & \textbf{11.99} & \textbf{13.74} & \textbf{11.49}\\
RPBE-TS(BJ) (S66) & $a_1 = 0.67$, $a_2 = 0.01\;a_0$ & 11.06 & 54.95 & 8.29 & 26.14\\
RPBE-D3(0) & - & 34.73 & 50.23 & 13.84 & 33.72 \\
RPBE-D3(BJ) & - & 19.64 & 37.20 & 13.84 & 24.28\\
\end{tabular}
\end{ruledtabular}
\end{table*}

\begin{table*}
\caption{Mean absolute errors (MAEs) between CCSD(T) interaction energies of model complexes contained in the S66\cite{rezac_s66_2011, rezac_erratum_2014} benchmark data set and DFT using the PBE and RPBE xc functionals. The interactions of the complexes can be separated into hydrogen bonded (HB), dispersion stabilized and ``mixed'', where electrostatic and dispersion interactions are similar in magnitude.\cite{jurecka_benchmark_2006, rezac_s66_2011, rezac_erratum_2014} The ``total'' MAEs take all complexes into account. The DFT interaction energies have been corrected for dispersion using different dispersion models. Boldface rows represent results obtained using models, whose damping function parameters --$s_{r,6}$ for the Fermi-type damping, and $a_1$ and $a_2$ for the BJ-damping function-- have been optimized on the same S66 benchmark data by minimizing the MAE.}
\label{tab:S66_damp_parms}
\begin{ruledtabular}
\begin{tabular}{ l c  S[table-format=3.2]  S[table-format=3.2]  S[table-format=3.2]  rS[table-format=3.2]}
Method & Optimized parameters & \multicolumn{1}{c}{HB / meV} & \multicolumn{1}{c}{Dispersion / meV} & \multicolumn{1}{c}{Mixed / meV} & \multicolumn{1}{r}{Total / meV} \\
\hline
PBE & - & 39.52 & 149.94 & 84.20 & 91.54\\
PBE-TS(F) (S22) & $s_{r,6} = 0.94$ & 17.22 & 36.87 & 15.36 & 23.51 \\
\textbf{PBE-TS(F) (S66)} & $\bm{s_{r,6} = 0.99}$ & \textbf{11.28} & \textbf{19.82} & \textbf{6.43} & \textbf{12.79} \\
PBE-TS(BJ) (S22) & $a_1 = 0.00$, $a_2 = 5.90\;a_0$ & 19.99 & 23.91 & 11.63 & 18.82\\
\textbf{PBE-TS(BJ) (S66)} & $\bm{a_1 = 0.00}$\textbf{, }$\bm{a_2 = 6.27\;a_0}$ & \textbf{12.82} & \textbf{12.77} & \textbf{5.84} & \textbf{10.69}\\
PBE-D3(0) & - & 21.03 & 15.20 & 9.52 & 15.51 \\
PBE-D3(BJ) & - & 19.60 & 11.95 & 7.01 & 13.11\\
\hline
\hline
RPBE & - & 134.85 & 241.50 & 152.18 & 177.27 \\
RPBE-TS(F) (S22) & $s_{r,6} = 0.60$ & 21.62 & 41.45 & 20.64 & 28.23\\
\textbf{RPBE-TS(F) (S66)} & $\bm{s_{r,6} = 0.65}$ & \textbf{21.37} & \textbf{27.56} & \textbf{11.73} & \textbf{20.60}\\
RPBE-TS(BJ) (S22) & $a_1 = 0.16$, $a_2 = 2.95\;a_0$ & 13.25 & 55.41 & 24.50 & 31.35\\
\textbf{RPBE-TS(BJ) (S66)} & $\bm{a_1 = 0.67}$\textbf{, }$\bm{a_2 = 0.01\;a_0}$ & \textbf{8.17} & \textbf{33.52} & \textbf{7.27} & \textbf{16.73}\\
RPBE-D3(0) & - & 21.17 & 17.85 & 8.73 & 16.24 \\
RPBE-D3(BJ) & - & 17.67 & 9.34 & 8.86 & 12.10\\
\end{tabular}
\end{ruledtabular}
\end{table*}

\begin{figure*}
\centering
\begin{subfigure}{.49\textwidth}
    \centering
    \caption{}
    \label{fig:damping-results-OO}
    \includegraphics[scale=0.8]{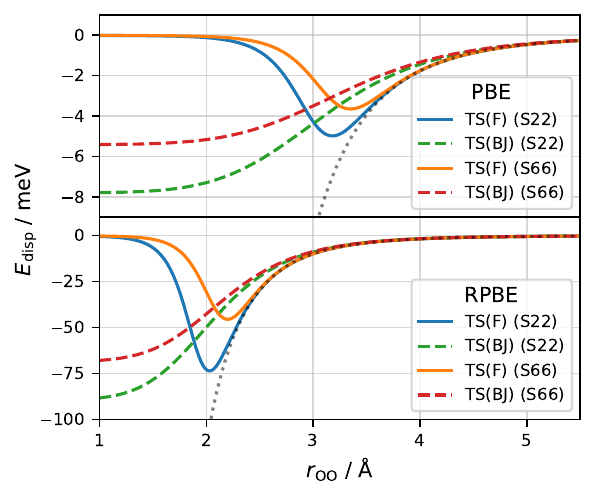}
\end{subfigure}
\begin{subfigure}{.49\textwidth}
    \centering
    \caption{}
    \label{fig:damping-results-OH}
    \includegraphics[scale=0.8]{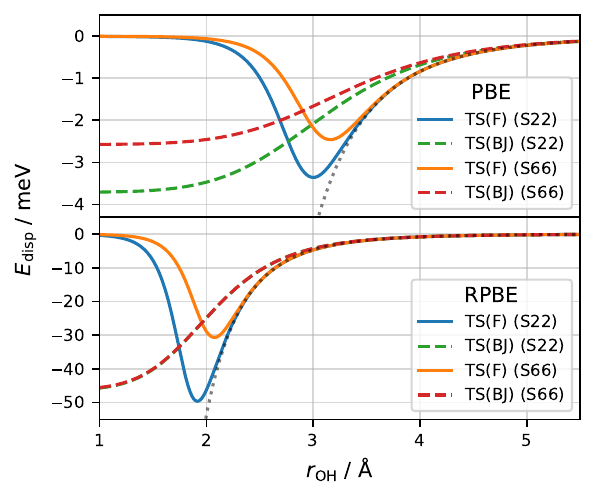}
\end{subfigure}
\caption{Approximate dispersion energies $E_\mathrm{disp}$ for OO (a) and OH (b) atom pairs as a function of the distance $r$ using the TS model with the parameters obtained in Sec.\:\ref{sec:parametrization}. %
The potential without damping is plotted as grey dotted line for comparison.}
\label{fig:damping-results}
\end{figure*}

First, when considering the results of PBE and RPBE without dispersion corrections, it can be seen that both functionals generally do not perform well with overall MAEs of 118.42\,meV and 221.50\,meV for S22 and 91.54\,meV and 177.27\,meV for S66, respectively. We show the energy differences for the individual complexes in the Supplementary Material in Fig.\:S1, Fig.\:S3 and Fig.\:S4. Here, it can be seen that both xc functionals generally underestimate interaction energies. A metric that is often referred to is the so-called ``chemical accuracy'', which corresponds to a MAE of 43\,meV (1\,kcal/mol) or below. Both xc functionals do not reach this accuracy in both benchmark sets. Looking at the individual interaction groups next, it can be seen that especially complexes that are predominately stabilized by dispersion interactions are described poorly with MAEs of 209.62\,meV and 318.61\,meV for S22, and 149.94\,meV and 241.50\,meV for S66, for PBE and RPBE, respectively. Both xc functionals perform better in describing hydrogen bonded or ``mixed'' complexes but never reach chemical accuracy. 
\\
Since errors are largest for dispersion bound complexes, it can be anticipated that predominantly errors in describing dispersion interactions would be reduced when optimizing the damping function parameters using these benchmarks sets. However, especially in the case of RPBE, the errors for the other NCIs are significant and necessarily influence the optimization of the parameters.
\\
The optimization yields different parameters for both xc functionals and benchmark sets. For the S22 benchmark set, we obtain an optimal $s_r = 0.94$ for PBE-TS(F) (S22), $s_r = 0.60$ for RPBE-TS(F) (S22), $a_1 = 0.00$, $a_2 = 5.90\,a_0$ for PBE-TS(BJ) (S22), and $a_1 = 0.67$, $a_2 = 0.01\;a_0$ for RPBE-TS(BJ) (S22). The obtained PBE-TS (S22) parameters are consistent with the values reported by Tkatchenko and Scheffler\cite{tkatchenko_accurate_2009} and Caro\cite{caro_parametrization_2017}. For RPBE-TS(F), Caro\cite{caro_parametrization_2017} reports an optimal $s_r = 0.59$, which is slightly smaller than our value of $0.60$. The small difference might be the result of using different DFT codes and basis sets. For the S66 benchmark set, we obtain an optimal $s_r = 0.99$ for PBE-TS(F) (S66) and $a_1 = 0.00$, $a_2 = 6.27\,a_0$  for PBE-TS(BJ) (S66). For RPBE-TS (S66), we obtain $s_r = 0.65$ and $a_1 = 0.16$, $a_2 = 2.95\,a_0$ for RPBE-TS(BJ) (S66).
\\ 
To visualize the effect of the different parameters, we show the oxygen-oxygen (OO) and oxygen-hydrogen (OH) dispersion interaction profiles for both functionals in Fig.\:\ref{fig:damping-results}. We note that $E_\mathrm{disp}$ depends on the Hirshfeld volumes of hydrogen and oxygen, which change depending on the  environment. As an estimate of the average OO and OH interaction profiles relevant for water, we thus set the Hirshfeld volumes to 0.67 for hydrogen and to 0.88 for oxygen. These values correspond to the average Hirshfeld volumes in the water cluster benchmark set, which will be discussed in Sec.\:\ref{sec:water_clusters}. This benchmark set includes thirty-eight water clusters, with up to ten water molecules and should represent the fundamental water interaction motifs providing a good estimate of the average Hirshfeld volume of both elements. We note, that the average volumes are consistent for both xc functionals.
\\
When comparing the TS OH dispersion interaction profiles of PBE and RPBE in Fig.\:\ref{fig:damping-results}, it can be seen that the correction generally starts to act at shorter distances in the case of RPBE. The minimum employing the Fermi-type damping function is at about 2\,\AA\:while it is at about 3\,\AA\:in the case of PBE. For TS(BJ)-damping, $E_\mathrm{disp}$ also converges to a value that is one order of magnitude larger for RPBE. This behavior reflects the benchmark results that RPBE underestimates NCIs and especially dispersion interactions more significantly than PBE. Looking for differences resulting from the benchmark set, it can be seen that S22 generally leads to a more attractive interaction potential. The minimum using the Fermi-type damping function is shifted towards shorter distances for both functionals and BJ-damping also converges to a lower value for PBE. For RPBE the profiles using BJ-damping align. However, in this case, the extent to which the two functions depend on the Hirshfeld volume is different. Due to the small $a_1=0.16$ and large $a_2=2.95\,a_0$ parameter obtained from S22, Eq.\:\ref{eq:BJ_damping} changes only slowly depending on the Hirshfeld volume. In contrast, with $a_1=0.67$ and $a_2=0.01\,a_0$ obtained from S66, the effect of the Hirshfeld volume on the interaction profile is more pronounced. Therefore, the order of the two potentials can change depending on the volume in this case, while it stays the same for the models employing the Fermi-type damping function. 
\\
In Fig.\:\ref{fig:damping-results}, we also show the OO interaction profile using the averaged oxygen Hirshfeld volume. The curves look generally very similar, but a notable difference is that the RPBE-TS(BJ) (S22) potential is significantly more attractive than RPBE-TS(BJ) (S66). In the case of PBE, it is also interesting that $a_1=0.00$ is optimal for both benchmark sets. Since $a_1$ is a global parameter, this means that the damping function is not only independent of the Hirshfeld volume and therefore the environment but also constant for every element pair. In this work, we only investigate water, therefore only three different element pairs can be realized, but it would be interesting to see how applicable an $a_1=0.0$ parameter is for systems containing a wider range of elements.
\\
Applying the TS model with these parameters drastically improves the MAE for both xc functionals. The optimization results for the S22 dataset are 14.28\,meV for PBE-TS(F) (S22), 16.40\,meV for PBE-TS(BJ) (S22), 23.23\,meV for RPBE-TS(F) (S22) and 11.49\,meV for RPBE-TS(BJ) (S22). They are additionally summarized in Table\:\ref{tab:S22_damp_parms}. For the S66 data set we obtain a MAE of 12.79\,meV for PBE-TS(F) (S66), 10.69\,meV for PBE-TS(BJ) (S66), 20.60\,meV for RPBE-TS(F) (S66) and 16.73\,meV for RPBE-TS(BJ) (S66). The S66 results are summarized in Table\:\ref{tab:S66_damp_parms}. Therefore, excellent results can be achieved with both types of damping function. This is consistent with results for DFT-D3\cite{grimme_effect_2011}, but has not been confirmed for the TS model before.
\\
In Tables\:\ref{tab:S22_damp_parms} and \:\ref{tab:S66_damp_parms} results for PBE and RPBE with DFT-D3(0) and DFT-D3(BJ) are shown for comparison. We note that the parameters for DFT-D3 have been optimized using a different database, which comprises multiple benchmark sets including the S22 dataset but based on interaction energies reported by Takatani et al.\cite{takatani_basis_2010}.  More details can be found in  Refs.\:\onlinecite{grimme_consistent_2010, grimme_effect_2011}.
\\
The MAEs of both xc functionals are equally improved when employing DFT-D3(0) or DFT-D3(BJ). PBE-D3(0) yields a MAE of 25.88\,meV on S22 and 15.51\,meV on S66 and PBE-D3(BJ) gives similar results with a MAE of 24.42\,meV for S22 and 13.11\,meV for S66. RPBE-D3(0) performs equally well with a MAE of 33.72\,meV on S22 and 16.24\,meV on S66. With RPBE-D3(BJ) a MAE of 24.28\,meV is obtained from S22 and 12.10\,meV on S66.
\\
Next, we have performed a simple form of cross-validation by using the parameters obtained from one benchmark set to predict the interaction energies of the other. We note, that there is some overlap between the complexes contained in the two benchmark sets, but the majority of systems is different. The obtained MAEs are also listed in Table\:\ref{tab:S22_damp_parms} and Table\:\ref{tab:S66_damp_parms} for S22 and S66, respectively. The MAEs of both benchmark sets are below the target chemical accuracy indicating that there is some tolerance of suitable parameters for both xc functionals and damping-functions. This can also be seen in the error curves and surfaces of the optimization, which are shown in the Supplementary Material in Fig.\:S2 and Fig.\:S5. Grimme et al.\cite{grimme_effect_2011} observed this behavior for the case of BJ-damping and therefore recommend more diverse datasets to determine the optimal set of parameters.\\
Especially in the case of the BJ-damping $a_1$-$a_2$ error surface, there is a large area associated with below chemical accuracy MAE values, which forms a straight path through the error surface. This indicates that there might be a linear relationship between the optimal $a_1$ and $a_2$ parameters. We investigate this further in the Supplementary Material in Sec.\:SII\:C and find that such a relationship can indeed be established for both benchmark datasets and xc functionals, which implies that also for BJ-damping a single damping-function parameter might be sufficient to achieve the desired chemical accuracy. However, the size of the S22 and S66 datasets might be too limited to draw final conclusions. Investigating this further is outside the scope of this study but could be very interesting for future work.\\ 
In the following sections, we present extended tests of the optimal S22 and S66 parameters for both damping-functions, focusing on water - using water clusters as well as dynamic properties of liquid water obtained from MD simulations.

\subsection{Water clusters}\label{sec:water_clusters}

\begin{figure*}[t]
    \centering
    \includegraphics[scale=0.8]{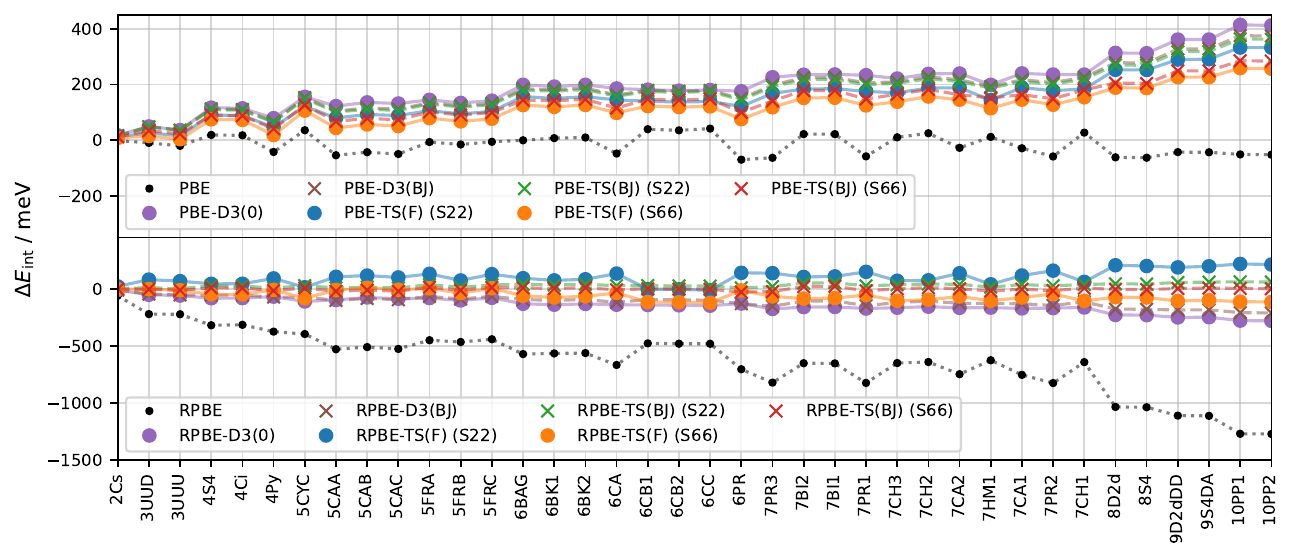}
    \caption{Results for the BEGDB water cluster benchmark set\cite{rezac_quantum_2008, temelso_benchmark_2011} for the PBE and RPBE xc functionals with different dispersion corrections. Labels on the \textit{x}-axis correspond to the names of structures in the benchmark set. The numbers at the beginning of each label corresponds to the number of water molecules contained in the respective cluster. Figures displaying the clusters can be found in Ref. \onlinecite{manna_conventional_2017}. The interaction energy differences $\Delta E_\mathrm{int}$ have been computed with respect to explicitly correlated CCSD(T) or composite MP2 interaction energy reference values of Manna et al.\cite{manna_conventional_2017} and have not been normalized by the number of molecules contained in a given cluster.}
    \label{fig:BEGDB_results}
\end{figure*}

As a first test of the performance of the TS models obtained in Sec.\:\ref{sec:parametrization}, we make use of the BEGDB water clusters benchmark set\cite{rezac_quantum_2008, temelso_benchmark_2011}. This benchmark set covers thirty-eight neutral water clusters, from dimers up to decamers in different configurations with geometries taken from Temelso et al.\cite{temelso_benchmark_2011}. We use the reference energies reported by Manna et al.\cite{manna_conventional_2017} that are interaction energies, obtained from explicitly correlated CCSD(T) or composite second-order Møller–Plesset perturbation theory (MP2) depending on the cluster size. As reference we have always used the highest level of theory available. We calculated the interaction energies $E_\mathrm{int}$ according to,
\begin{equation}
    E_\mathrm{int}=E_\mathrm{cluster}-\sum_i^{N_\mathrm{H_2O}} E_{\mathrm{H_2O,\:}i},
\end{equation}
where $N_\mathrm{H_2O}$ is the number of water molecules in the cluster, $E_{\mathrm{H_2O,\:}i}$ is the energy of water molecule $i$ in the basis of the full cluster and $E_\mathrm{cluster}$ is the energy of the cluster.
\\
The interaction energies have been determined for the structures as provided in the BEGDG set without further optimization using the PBE and RPBE functionals. Subsequently, we corrected the interaction energies for dispersion using the TS model with our optimized parameters for the Fermi-type and BJ-damping functions as well as DFT-D3 with the zero- and BJ-damping functions. Since the benchmark set contains clusters of different sizes, the MAE is dominated by the errors of the largest clusters. For a more balanced assessment of the accuracy, therefore we normalize the MAE values by the number of water molecules per cluster, which are reported as MAE$_\mathrm{rel}$ in Table\:\ref{tab:water_clusters}. Additionally, the interaction energy error, without normalization per molecule, $\Delta E_\mathrm{int} = E_\mathrm{int,\:DFT.} - E_\mathrm{int,\:ref}$ for the individual water clusters is shown in Fig.\:\ref{fig:BEGDB_results}.
\\
Focusing on the results of the plain xc functionals without dispersion correction first, it can be seen that the PBE interaction energies agree remarkably well with the reference energies while there is a strong underestimation for the RPBE functional. These results are consistent with previous studies comparing these two xc functionals for small water clusters.\cite{morawietz_density-functional_2013} When including dispersion corrections, the PBE functional yields a significant overestimation of the interaction strength. Further, it becomes apparent that the choice of benchmark set for optimization clearly influences the results, with the S22, S66 and DFT-D3 models providing distinct results that depend to a lesser extent on the choice of the damping function. Consistent with our analysis of damping parameters in Sec.\:\ref{sec:parametrization}, the TS models using parameters obtained from S22 are more attractive and therefore lead to a more pronounced overestimation than the parameters obtained from S66. Moreover, the results employing DFT-D3 with zero- and BJ-damping show the largest overestimation of interaction energies indicating that the parameters obtained by Grimme et al.\cite{grimme_consistent_2010, grimme_effect_2011} for DFT-D3 lead to a more attractive interaction potential in the case of PBE than those we obtained from S22 and S66 for the TS model.
\\
The picture is clearly different in the case of RPBE. Here, applying a dispersion correction significantly improves the interaction energies. Especially, RPBE-TS(BJ) (S66) with a MAE$_\mathrm{rel}$ of 1.83\;meV and RPBE-TS(BJ) (S22) with 4.45\;meV perform outstandingly. In the case of the TS models employing the Fermi-type damping function, both models perform worse than BJ-damping with  MAE$_\mathrm{rel}$ values of 16.48\;meV and 9.52\;meV respectively. As can be seen in Fig.\:\ref{fig:BEGDB_results}, RPBE-TS(F) (S22) slightly overestimates the interaction energies while RPBE-TS(F) (S66) leads to an underestimation. This order is also consistent with our analysis in Sec.\:\ref{sec:parametrization}. Employing DFT-D3 with RPBE also leads to better interaction energies compared to the plain xc functional. However, in the case of RPBE-D3 with zero- as well as BJ-damping, the interaction strength is underestimated the most indicating that the default parameters lead to a dispersion energy profile that is less attractive than the one we obtained from S22 and S66 for the TS model.
\\
Based on these results, an even worse description of water should be obtained when applying a dispersion correction to PBE while it should be much improved when combining any of the dispersion models with RPBE. However, this conclusion would be premature as we will see in the following sections. While some insights into the relative interaction strengths of the dispersion models can be gained from the water clusters interaction energy benchmark, there is no information about the behavior of these models in MD simulations of liquid water. Here, the gradient of the PES is more important that cannot be probed using equilibrium geometries of water clusters only. A promising strategy to solve this could be to also employ atomic forces, which provide valuable information on the shape of the PES, to parametrize and validate dispersion models. Currently, there is only a limited availability of fully converged forces from CCSD(T) but generation of such atomic force benchmark data could be highly valuable for future work.

\begin{table}
\caption{Mean absolute interaction energy errors of PBE and RPBE with and without dispersion corrections employing different damping functions for the BEGDB water cluster benchmark set\cite{rezac_quantum_2008, temelso_benchmark_2011} with respect to explicitly correlated CCSD(T) or composite MP2 interaction energies of Manna et al.\cite{manna_conventional_2017}. The MAE$_\mathrm{rel}$ has been normalized by the number of water molecules contained in each cluster.}
\label{tab:water_clusters}
\begin{ruledtabular}
\begin{tabular}{ l  rS[table-format=3.2]}
Method & \multicolumn{1}{r}{MAE$_\mathrm{rel}$ / meV} \\
\hline
PBE &  5.30 \\
PBE-TS(F) (S22) & 23.07 \\
PBE-TS(F) (S66) & 17.52 \\
PBE-TS(BJ) (S22) & 27.31 \\
PBE-TS(BJ) (S66) & 21.25 \\
PBE-D3(0) & 30.54 \\
PBE-D3(BJ) & 28.12 \\
\hline
\hline
RPBE & 98.31 \\
RPBE-TS(F) (S22) & 16.48 \\
RPBE-TS(F) (S66) & 9.52 \\
RPBE-TS(BJ) (S22) & 4.45 \\
RPBE-TS(BJ) (S66) & 1.83 \\
RPBE-D3(0) & 22.29 \\
RPBE-D3(BJ) & 17.62 \\
\end{tabular}
\end{ruledtabular}
\end{table}

\subsection{Liquid water}\label{sec:liquid}

\subsubsection{Radial distribution functions}\label{sec:RDF}

\begin{figure*}
\centering
\begin{subfigure}{.49\textwidth}
    \centering
    \caption{}
    \label{fig:RDF-comp-OO}
    \includegraphics[scale=0.8]{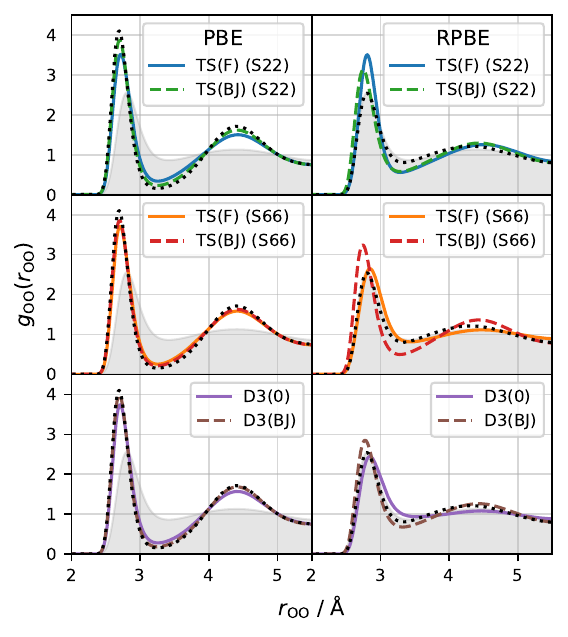}
\end{subfigure}
\begin{subfigure}{.49\textwidth}
    \centering
    \caption{}
    \label{fig:RDF-comp-OH}
    \includegraphics[scale=0.8]{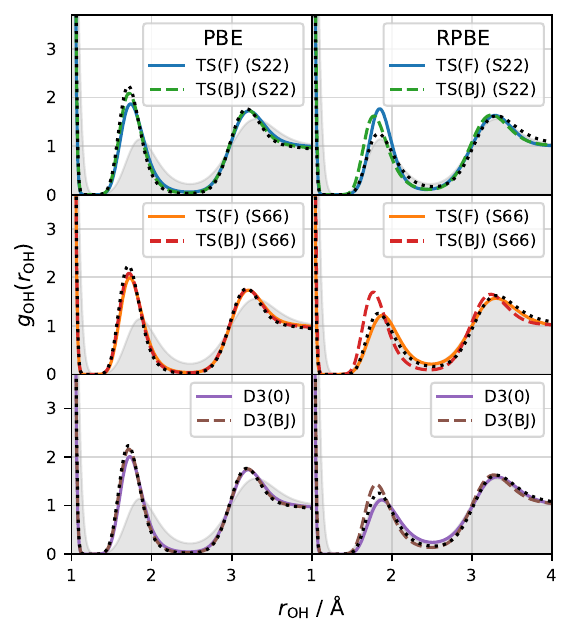}
\end{subfigure}
\caption{OO (a) and OH (b) RDFs produced by the PBE and RPBE functionals in HDNNP-driven MD simulations in the $NVT$ ensmeble at a temperature of 300\,K and density of 1.0\,g/cm$^3$ with different dispersion corrections. In light grey reference RDFs from CCMD obtained at 298.15\,K reported by Daru et al.\cite{daru_coupled_2022} are shown. Additionally, the black dotted lines correspond to RDFs of the respective xc functional without dispersion correction.}
\label{fig:RDF-comp}
\end{figure*}

Radial distribution functions provide valuable information about the structure of a liquid and allow assessing the strength of the H-bond network of water.\cite{gillan_perspective_2016} We obtain RDFs of liquid water from MD simulations driven by HDNNPs, which were trained on PBE and RPBE reference data corrected for dispersion using the TS and DFT-D3 models with the investigated damping functions. We focus on the comparison among the different dispersion models and assess the quality using data from Daru et al.\cite{daru_coupled_2022}, who reported RDFs of liquid water at 298.15\,K obtained from coupled cluster molecular dynamics (CCMD)
that are in excellent agreement with results from neutron scattering and x-ray diffraction experiments. We note that we have performed classical MD simulations, while NQEs have been included in CCMD. However, NQEs were shown to have only a small effect on the OO RDF while the impact on the OH RDF is somewhat more pronounced.\cite{marsalek_quantum_2017, cheng_ab_2019, torres_using_2021} 
\\
First, we focus on the OO RDFs, which are shown in Fig.\:\ref{fig:RDF-comp-OO}, while characteristic distances $r_\mathrm{OO}$ and associated values $g_\mathrm{OO}$ are summarized in Table\;\ref{tab:MD_results}. The data confirms the well-known property that plain PBE produces an over-structured liquid with peak positions of the first maximum and minimum being at too short distances and the associated height and depth being too pronounced. RPBE on the other hand, produces a structure that is very close to the CCMD data. These observations are in good agreement with previous studies.\cite{torres_using_2021, gillan_perspective_2016, morawietz_how_2016, sakong_structure_2016}
\\
Applying dispersion corrections has a significant effect on the OO RDF produced by both functionals. In case of PBE, they lead to a very slight softening of the structure regardless of the parameters or functional form of the damping function. This is surprising given the water clusters benchmark results in Sec.\:\ref{sec:parametrization}, where we found that applying any of the investigated dispersion models to PBE results in an overestimation of interaction energies, which could lead to the expectation of increased structuring. Nevertheless, the height of the first and second maximum is decreased while a shallower first minimum is obtained. This implies that a larger number of water molecules can enter interstitial sites, which decreases the number of molecules in the first and second solvation shell. Softening is more pronounced for both TS and DFT-D3 when damping to zero. Additionally, only in case of zero damping the position of the first maximum is shifted towards slightly larger distances indicating that it could be related to the repulsive gradient present for this type of damping.\cite{grimme_effect_2011}
\\ 
In the case of RPBE, the effect of applying a dispersion correction is significantly more pronounced and can be markedly different compared to PBE. The larger effect compared to PBE can be rationalized by considering the dispersion energy profiles in Fig.\:\ref{fig:damping-results} showing that the contribution of the correction can be up to an order of magnitude larger at short distances for RPBE. Moreover, applying a dispersion correction to RPBE can also enhance structuring of the liquid significantly. This behavior is observed for all dispersion models employing BJ-damping and the TS Fermi-damping model using the parameters obtained from the S22 data set. For these models, the height of the first maximum is significantly increased while the first minimum is further depleted. At the same time, the second maximum is shifted to larger distances and slightly increased. Similar observations regarding RPBE-D3(BJ) have been reported by Sakong et al.\cite{sakong_structure_2016}. Caro\cite{caro_parametrization_2017} found that the slightly smaller $s_{r,6}=0.59$ parameter obtained from the S22 benchmark set for RPBE-TS(F) also produces an enhanced RDF. In contrast to these results, RPBE-TS(F) (S66) leads to only a very mildly enhanced and RPBE-D3(0) to a slightly less structured OO RDF. Additionally, both shift the position of the first maximum to larger distances, which can be ascribed to the repulsive nature of the employed damping functions.
\\
For the OH RDFs presented in Fig.\:\ref{fig:RDF-comp-OH} we focus on the second maximum at around 1.75\;\AA, which corresponds to the intermolecular H-bond.\cite{gillan_perspective_2016} The peak positions $r^\mathrm{HB,\:max}_\mathrm{OH}$ and the associated values $g^\mathrm{HB,\:max}_\mathrm{OH}$ are summarized in Table\:\ref{tab:MD_results}. Here, for both xc functionals similar results to those found in the OO RDF can be observed. The uncorrected PBE functional produces also a significantly over-structured OH RDF while the RPBE results compare quite favorably with CCMD. In case of PBE, the peak height is decreased and its position is shifted towards larger distances indicating a more flexible H-bond when applying a dispersion correction. For the OH RDF, the effect is also more pronounced for dispersion models that damp to zero. For RPBE, the models employing the BJ-damping function and the TS model optimized on S22 employing the Fermi-type damping function enhance the H-bond peak significantly while RPBE-D3(0) and RPBE-TS(F) (S66) show a slight softening instead, which brings the RDF closer towards  CCMD. For both xc functionals, the Fermi-type or zero-damping function, respectively, lead to a more pronounced elongation of the H-bond compared to BJ-damping, which can be seen from the shift of the H-bond peak towards larger distances. Here, we also assume that this is the result of the negative gradient at short distances of Fermi- and zero-damping.
\\
To explain the contrasting effects of the different dispersion models on the RDF observed in RPBE-based MD simulations, we first consider both RPBE-TS(F) models that employ the Fermi-type damping function. In Fig.\:\ref{fig:damping-results} the differences between the dispersion energy profiles of these two models can be seen. The S22 profile is more attractive and the minimum is located at a distance that is about 0.2\,\AA\:shorter. In Sec.\:\ref{sec:water_clusters}, we could already identify a difference with RPBE-TS(F) (S22) overestimating water cluster interaction energies while RPBE-TS(F) (S66) leads to an underestimation. The RDFs in Fig.\:\ref{fig:RDF-comp} follow this trend showing a drastically over-structured liquid in the case of RPBE-TS(F) (S22) while the structure is only mildly enhanced in the case of RPBE-TS(F) (S66). These results suggest that there may be a very sharp threshold up to which the dispersion correction can contribute, as otherwise there seems to be significant double-counting of interactions, which results in an over-structured liquid. The location of this threshold seems to be around 2\,\AA\:in the case of RPBE according to Fig.\:\ref{fig:damping-results}. This might also explain the behavior of the models employing BJ-damping. Consulting the energy profiles in Fig.\:\ref{fig:damping-results} again, it can be seen that, while the gradient of the BJ-damping models is smaller compared to the ones employing the Fermi-type damping function, the interactions remain attractive up to an even shorter distance than both Fermi-type damping models. This results in a still over-structured but less enhanced RDF due to the smaller gradient with the first maximum shifted towards shorter distances. 
\\
Of the three considered BJ-damping models, RPBE-D3(BJ) shows the least amount of over-structuring. This is consistent with our observations for the water cluster benchmark in Fig.\:\ref{fig:BEGDB_results}. Here, RPBE-D3(BJ) was found to significantly underestimate interaction energies compared to the RPBE-TS(BJ) models indicating that the model provides less attractive interactions and therefore less double-counting is observed. What remains at odds with our reasoning so far is that the two RPBE-TS(BJ) damping models show no signs of double counting in the water clusters benchmark in Sec.\:\ref{sec:water_clusters}. This might be due to the fact that this benchmark only includes equilibrium structures. When employing BJ-damping, the dispersion correction still contributes significantly at short distances, which leads to double-counting in the case of RPBE. However, these distances might not be sufficiently represented when only considering equilibrium cluster geometries.

\begin{table*}
\caption{Equilibrium densities $\rho_\mathrm{eq}$, diffusion coefficients extrapolated to the infinite system size $D_\infty$, and peak positions of OO and OH RDFs obtained from different HDNNP-based MD simulations of liquid water using the PBE and RPBE xc functionals employing different dispersion corrections. The simulation temperature $T$ is given in K. The equilibrium density $\rho_\mathrm{eq}$ in g/cm$^3$ was obtained in the $NpT$-ensemble. Diffusion coefficients $D_\infty$ are in $\mathrm{\AA}^2$/ps and were extrapolated from system-size specific $D_L$ values at a density of 1\,g/cm$^3$ from $NVE$ simulations, which is shown in Fig.\:\ref{fig:D_extrapolation}. RDFs were obtained in the $NVT$ ensemble at a density of 1\,g/cm$^3$ and are shown in Fig.\:\ref{fig:RDF-comp}. 
The positions, in  \AA, $r^\mathrm{1,max}_\mathrm{OO}$, $r^\mathrm{1,min}_\mathrm{OO}$ and $r^\mathrm{2,max}_\mathrm{OO}$ correspond to the first maximum $g^\mathrm{1,max}_\mathrm{OO}$, first minimum $g^\mathrm{1,min}_\mathrm{OO}$ and second maximum $g^\mathrm{2,max}_\mathrm{OO}$ of the OO RDF respectively. The positions, in \AA, $r^\mathrm{HB,max}_\mathrm{OO}$ correspond to the second maximum $g^\mathrm{HB,max}_\mathrm{OH}$ of the OH RDF, i.e., a H-bond}.
\label{tab:MD_results}
\begin{ruledtabular}
\begin{tabular}{ l c  S[table-format=2.3]  S[table-format=2.5]  S[table-format=2.3]  S[table-format=2.3]  S[table-format=2.3]  S[table-format=2.3]  S[table-format=2.3]  S[table-format=2.3]  S[table-format=2.3]  rS[table-format=2.3]}
Method & $T$ & \multicolumn{1}{c}{$\rho_\mathrm{eq}$} & \multicolumn{1}{c}{$D_\infty$} & \multicolumn{1}{c}{$r^\mathrm{1,max}_\mathrm{OO}$} & \multicolumn{1}{c}{$g^\mathrm{1,max}_\mathrm{OO}$} & \multicolumn{1}{c}{$r^\mathrm{1,min}_\mathrm{OO}$} & \multicolumn{1}{c}{$g^\mathrm{1,min}_\mathrm{OO}$} & \multicolumn{1}{c}{$r^\mathrm{2,max}_\mathrm{OO}$} & \multicolumn{1}{c}{$g^\mathrm{2,max}_\mathrm{OO}$} & \multicolumn{1}{c}{$r^\mathrm{HB,\:max}_\mathrm{OH}$} &	 \multicolumn{1}{r}{$g^\mathrm{HB,\:max}_\mathrm{OH}$}\\
\hline
PBE                 &    300  &   0.88  &  0.004 \pm 0.005  &  2.70  &  4.10  &  3.26  &  0.16  &  4.41  &  1.71 & 1.70  &  2.23 \\
PBE-TS(F) (S22)        &    300  &   0.99  &  0.02 \pm 0.01  &  2.71  &  3.52  &  3.26  &  0.34  &  4.42  &  1.50 & 1.74  &  1.87 \\
PBE-TS(F) (S66)        &    300  &   0.96  &  0.009 \pm 0.005  &  2.71  &  3.71  &  3.26  &  0.26  &  4.42  &  1.59 & 1.74  &  1.99 \\
PBE-TS(BJ) (S22)    &    300  &   0.96  &  0.005 \pm 0.002  &  2.70  &  3.87  &  3.26  &  0.23  &  4.42  &  1.62 & 1.72  &  2.09 \\
PBE-TS(BJ) (S66)    &    300  &   0.95  &  0.007 \pm 0.002  &  2.70  &  3.85  &  3.26  &  0.22  &  4.41  &  1.63 & 1.72  &  2.09 \\
PBE-D3(0)           &    300  &   0.97  &  0.011 \pm 0.004  &  2.71  &  3.74  &  3.26  &  0.28  &  4.41  &  1.57 & 1.74  &  2.01 \\
PBE-D3(BJ)          &    300  &   0.95  &  0.005 \pm 0.004  &  2.70  &  3.99  &  3.24  &  0.19  &  4.41  &  1.69 & 1.72  &  2.16 \\
\hline
\hline
PBE                 &    400   &  0.79  &  0.33 \pm 0.03  &  2.74  &  2.59  &  3.32  &  0.72  &  4.41  &  1.22 & 1.77  &  1.29 \\
PBE-TS(F) (S22)        &    400   &  0.99  &  0.49 \pm 0.03  &  2.76  &  2.35  &  3.35  &  0.90  &  4.41  &  1.11 & 1.78  &  1.15 \\
PBE-TS(F) (S66)        &    400   &  0.96  &  0.43 \pm 0.03  &  2.74  &  2.42  &  3.30  &  0.84  &  4.38  &  1.15 & 1.78  &  1.19 \\
PBE-TS(BJ) (S22)    &    400   &  0.97  &  0.42 \pm 0.02  &  2.74  &  2.53  &  3.35  &  0.80  &  4.41  &  1.16 & 1.77  &  1.25 \\
PBE-TS(BJ) (S66)    &    400   &  0.95  &  0.41 \pm 0.02  &  2.74  &  2.52  &  3.34  &  0.79  &  4.46  &  1.17 & 1.77  &  1.24 \\
PBE-D3(0)           &    400   &  0.99  &  0.48 \pm 0.03  &  2.76  &  2.42  &  3.38  &  0.86  &  4.38  &  1.12 & 1.78  &  1.18 \\
PBE-D3(BJ)          &    400   &  0.95  &  0.37 \pm 0.03  &  2.74  &  2.57  &  3.34  &  0.76  &  4.42  &  1.19 & 1.77  &  1.28 \\
\hline
\hline
RPBE                &    300   &  0.63  &  0.157 \pm 0.007  &  2.81  &  2.54  &  3.34 &   0.81  &  4.31 &   1.21 & 1.83 &   1.26 \\
RPBE-TS(F) (S22)       &    300   &  0.96  &  0.07 \pm 0.02  &  2.81  &  3.51  &  3.29 &   0.56  &  4.50 &   1.25 & 1.85 &   1.77 \\
RPBE-TS(F) (S66)       &    300   &  0.96  &  0.24 \pm 0.02  &  2.84  &  2.65  &  3.43 &   0.81  &  4.47 &   1.12 & 1.88 &   1.22 \\
RPBE-TS(BJ) (S22)   &    300   &  0.98  &  0.087 \pm 0.009  &  2.74  &  3.13  &  3.29 &   0.58  &  4.44 &   1.29 & 1.77 &   1.62 \\
RPBE-TS(BJ) (S66)   &    300   &  0.95  &  0.06 \pm 0.02  &  2.74  &  3.24  &  3.29 &   0.50  &  4.42 &   1.37 & 1.77 &   1.70 \\
RPBE-D3(0)          &    300   &  0.90  &  0.28 \pm 0.02  &  2.84  &  2.44  &  3.53 &   0.94  &  4.47 &   1.08 & 1.88 &   1.11 \\
RPBE-D3(BJ)         &    300   &  0.90  &  0.12 \pm 0.01  &  2.78  &  2.85  &  3.32 &   0.68  &  4.42 &   1.26 & 1.80 &   1.43 \\
\hline
\hline
CCMD\cite{daru_coupled_2022} &	298.15 & \multicolumn{1}{c}{-} & \multicolumn{1}{c}{0.234(9)\footnote{Value taken from Ref.\:\onlinecite{stolte_nuclear_2024}.}} & 2.80 & 2.56 & 3.42 & 0.87 & 4.40 & 1.13 & 1.86 & 1.14 \\
\end{tabular}
\end{ruledtabular}
\end{table*}

\subsubsection{Self-diffusion coefficient}

\begin{figure*}
\centering
\begin{subfigure}{.49\textwidth}
    \centering
     \caption{}
    \label{fig:D_PBE_400K}
    \includegraphics[scale=0.8]{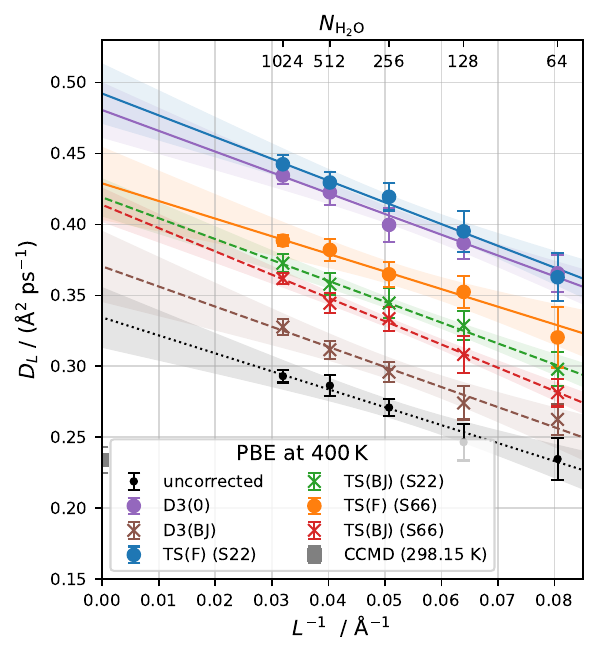}
\end{subfigure}
\begin{subfigure}{.49\textwidth}
    \centering
    \caption{}
    \label{fig:D_RPBE_300K}
    \includegraphics[scale=0.8]{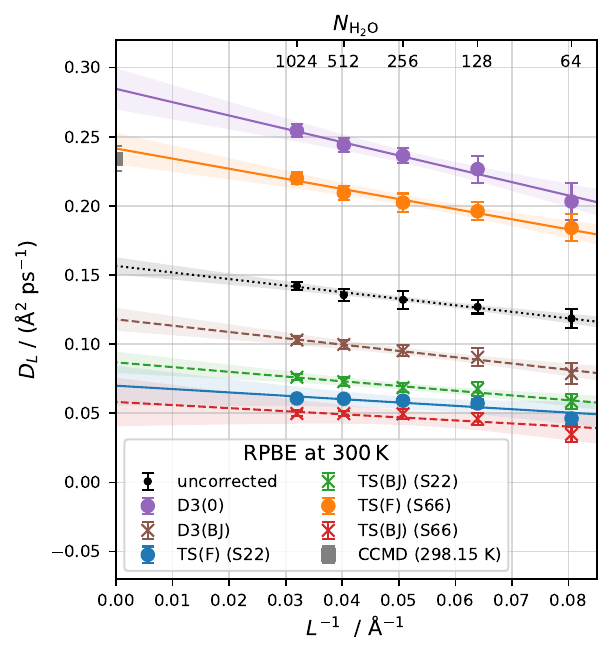}
\end{subfigure}
\caption{System-size specific diffusion coefficients $D_L$ obtained from HDNNP-driven MD simulations of PBE at 400\;K (a) and RPBE at 300\,K (b) at a density of 1\,g/cm$^3$ with different dispersion corrections plotted against the inverse length of the simulation box $L$ to extrapolate the value to an infinite system size. Error bars correspond to the 95\% confidence interval of the interpolation of the $NVE$ diffusion coefficients to 300\,K and 400\,K for RPBE and PBE, respectively. Colored lines correspond to the extrapolation functions obtained from linear fits with the area of matching color representing the 95\% confidence interval. The extrapolated CCMD value reported by Stolte et al.\cite{stolte_nuclear_2024}, was obtained at 298.15\,K and is shown in light grey for comparison.}
\label{fig:D_extrapolation}
\end{figure*}

 The self-diffusion of water molecules is the result of rearrangements in the H-bond network that require an exchange of H-bond partners.\cite{gomez_water_2022} Therefore, the predicted self-diffusion coefficient $D$ of water is highly dependent on the strength of the H-bonding network and its dynamics\cite{gillan_perspective_2016, gomez_water_2022, stolte_nuclear_2024} making it an excellent property for assessing the influence of the  dispersion models. 
\\
We obtain self-diffusion coefficients from MD simulations in the $NVE$-ensemble driven by HDNNPs, for PBE and RPBE with our optimized TS-models and DFT-D3 with zero- and BJ-damping. Starting geometries have been generated from $NVT$ simulations at 300\,K and a density of 1 g/cm$^3$ for both xc functionals. In the case of PBE, the rate of self-diffusion at 300\,K is very low making it difficult to obtain statistically meaningful data for different dispersion corrections. Therefore, for PBE $D$ was additionally determined at 400\,K, which was shown to give good agreement with the experimental RDF at 300\,K.\cite{grossman_towards_2004, schwegler_towards_2004}
\\
The self-diffusion coefficient is subject to finite-size effects\cite{stolte_nuclear_2024}. Therefore, we run MD simulations at multiple system sizes and subsequently extrapolate the obtained values to the diffusion coefficient corresponding to an infinite system size $D_\infty$. System-size specific diffusion coefficients $D_L$ extrapolated to the infinite system are shown in Fig.\:\ref{fig:D_extrapolation} for PBE at 400\,K and RPBE at 300\,K. The results for PBE at 300\,K are shown in the Supplementary Material in Fig.\:S10. Additionally, all $D_\infty$ values are summarized in Table\:\ref{tab:MD_results}.
\\
In the case of PBE, applying any of the investigated dispersion models increases $D$. 
When comparing the rate of self-diffusion for PBE among the different dispersion models in Fig.\:\ref{fig:D_PBE_400K}, it can be seen that models damping to zero show significantly faster diffusion. This in good agreement with the OH RDF results in Sec.\:\ref{sec:RDF} where this also reduced the over-structuring of the OH RDF more significantly compared to BJ-damping, which we traced back to the repulsive short-range forces. A less structured RDF indicates a weaker H-bond network that allows for more frequent switches of H-bond partners resulting in faster translational motion. For comparison, we also determined the OO and OH RDFs for PBE with and without dispersion corrections at 400\,K, which we show in the Supplementary Material in Fig.\:S6. In general, these RDFs show the same trend as the ones obtained at 300\,K albeit with a generally much softer profile.
\\
In the case of RPBE, %
including a dispersion correction can increase the rate of self-diffusion compared to the uncorrected xc functional similar to the case of PBE, but all RPBE models employing BJ-damping and the RPBE-TS(F) (S22) model exhibit reduced diffusion. This observation is consistent with the behavior seen in Fig.\:\ref{fig:RDF-comp-OO} where these models also lead to a more structured OH RDF indicating a more stable H-bond, which would decrease self-diffusion according to the proposed H-bond jump mechanism by Gomez et al.\cite{gomez_water_2022}. In accordance with this mechanism and the results obtained for PBE with dispersion corrections, the two models, RPBE-D3(0) and RPBE-TS(F) (S66), which reduce the structuring of the OH RDF compared to the plain xc functional indicating weaker H-bonding, also show an increase in the rate of self-diffusion.

\subsubsection{Density isobar}

\begin{figure*}
\centering
\begin{subfigure}{.49\textwidth}
    \centering
     \caption{}
    \label{fig:isobars}
    \includegraphics[scale=0.8]{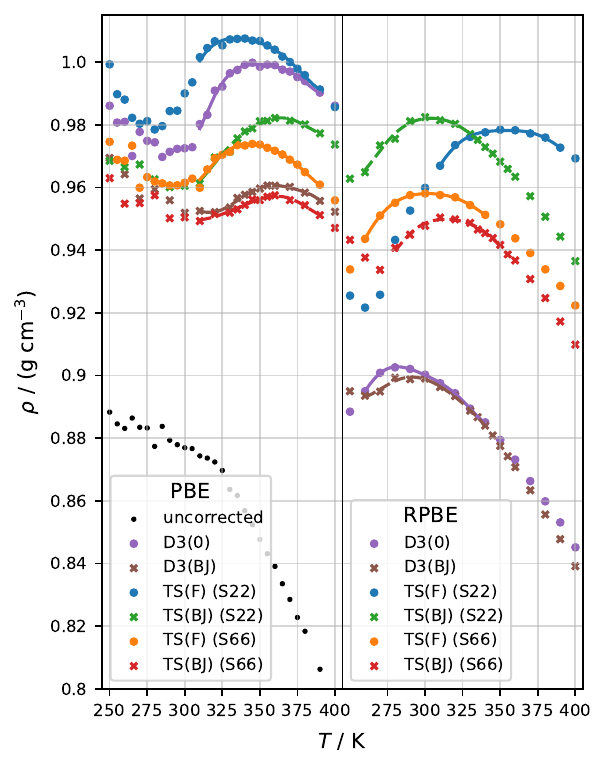}
\end{subfigure}
\begin{subfigure}{.49\textwidth}
    \centering
    \caption{}
    \label{fig:nearest_neighbors}
    \includegraphics[scale=0.8]{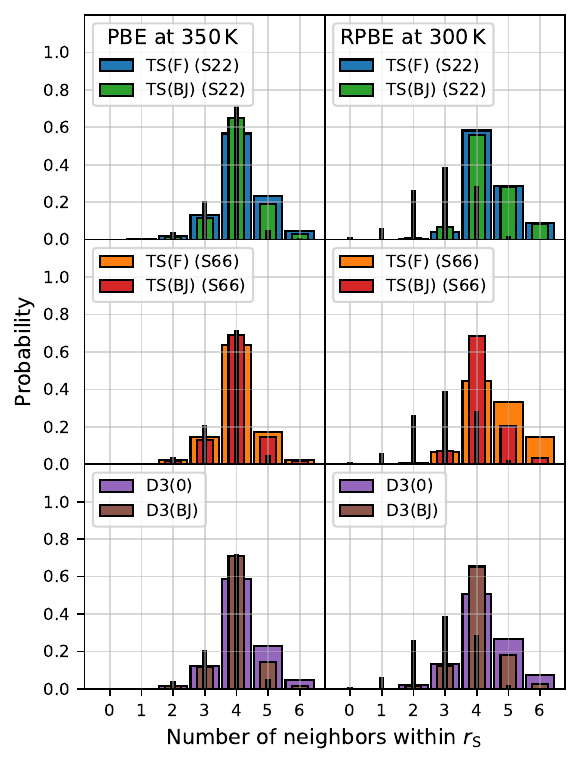}
\end{subfigure}
\caption{(a) Density isobars of liquid water obtained from HDNNP-driven MD simulations employing the PBE and RPBE functionals in combination with different dispersion corrections in the \textit{NpT}-ensemble at a pressure of $1\,\text{bar}$. The densities produced by the uncorrected RPBE xc functional are severely underestimated and below the range of the plot. The solid lines correspond to a sixth-order polynomial fit of the density isobars around the density maximum. Additionally, (b) shows the probability of a water molecule having a specific number of neighbors within the Stillinger radius $r_\mathrm{S}$ determined from a MD trajectory run in the $NpT$-ensemble at a temperature of 350\,K for PBE and 300\,K for RPBE. We set $r_\mathrm{S}$ to 3.3\,\AA\:for PBE and 3.4\,\AA\:for RPBE based on the approximate location of the first minimum in the OO RDF in Fig.\:\ref{fig:RDF-comp-OO}. The histogram of the respective xc functional without dispersion correction is shown in black in each panel.}
\label{fig:isobar_overview}
\end{figure*}

The density isobar of water features a maximum at about 277\;K\cite{mallamace_anomalous_2007} resulting from a balance between two competing effects.\cite{jedlovszky_molecular_2000} Upon heating, thermal fluctuations weaken the tetrahedral H-bonding network, which leads to an increased disorder and allows for an increase in the population of water molecules in interstitial sites\cite{jedlovszky_molecular_2000, distasio_individual_2014} increasing the density. This compensates the opposite effect of thermal expansion between neighboring molecules.\cite{jedlovszky_molecular_2000, morawietz_how_2016} At about 277\,K there is an occupation maximum of the interstitial sites, which leads to the appearance of the density maximum.\cite{jedlovszky_molecular_2000, morawietz_how_2016} Therefore, to produce the correct density isobar, the relative strength of these two competing effects needs to be accurately described. Earlier work by Morawietz et al.\cite{morawietz_how_2016} showed that including dispersion is necessary to predict the density maximum of water when applying RPBE. The uncorrected RPBE functional significantly underestimates the strength of the H-bonding network and therefore massively overestimates thermal expansion. As a consequence, there is no pronounced occupation of interstitial sites and the density decreases monotonically. Correcting for dispersion increases the strength of the H-bonding network, which restores the balance resulting in the appearance of the density maximum.\cite{morawietz_how_2016} Very recently, de Hijes et al.\cite{montero_de_hijes_density_2024} observed that the density isobar of water predicted by several xc functionals is significantly affected by the choice of damping function. Therefore, here we investigate this more closely to assess how the different models perform and if the observed behavior correlates with our results discussed above.
\\
We compute the density isobars from MD simulations in the $NpT$-ensemble driven by the HDNNPs representing the different xc functional and dispersion model combinations. The simulations contain 512 water molecules in a temperature range from 250\,K to 400\,K, which we scanned in intervals of 10\,K. In some cases, we did an additional scan in intervals of 5\,K around the density maximum. The resulting density isobars are shown in Fig.\:\ref{fig:isobars}. To determine the temperature of the density maximum $T_\mathrm{max}$, we performed a sixth-order polynomial fit of the data. The resulting $T_\mathrm{max}$ values and the associated densities $\rho_{T_\mathrm{max}}$ are summarized in Table\:\ref{tab:density}. Additionally, we included the equilibrium density at 300\,K $\rho_\mathrm{eq}$ in Table\:\ref{tab:MD_results}.
\\
Focusing on PBE first, it can be seen that uncorrected PBE fails to produce a density maximum. The equilibrium density at 300\,K is too low at 0.88\,g/cm$^3$ and decreases monotonically with increasing temperature. Adding a dispersion correction increases the equilibrium density significantly and results in a pronounced density maximum for all models. There seems to be a correlation between the employed benchmark set, damping function type and obtained density. Comparing the S22, S66 and DFT-D3 models, it can be seen that employing the Fermi-type or zero-damping function leads to a higher density and a density maximum at a lower temperature compared to the BJ-damping function.
\\
For RPBE, the results are similar, but the effect is more pronounced. The equilibrium density at 300\,K produced by the uncorrected functional is very low, only about 0.63\,g/cm$^3$, and decreases monotonically with increasing temperature. The density isobar features densities so low that we did not include them in Fig.\:\ref{fig:isobars}. Adding a dispersion correction increases the equilibrium density at 300\,K drastically to a much more reasonable value of 0.90\,g/cm$^3$ in the case of both DFT-D3 models and above 0.95\,g/cm$^3$ in the case of the TS models. For the latter, we also see a strong correlation to the benchmark data set used for optimization with both RPBE-TS (S22) models yielding the highest density, followed by both RPBE-TS (S66) models and both RPBE-D3 models, which produce the lowest density. This order was already observed for the water clusters in Sec.\:\ref{sec:water_clusters} where we found that S22-optimized TS models are most attractive, overestimating the interaction strength, followed by S66 and DFT-D3. Damping to zero also seems to result in a higher density for the RPBE-TS(F) (S66) and RPBE-D3 models with RPBE-TS(F) (S22) being an exception having an equilibrium density lower than RPBE-TS(BJ) (S22). As for PBE, all dispersion models generate a pronounced density maximum, which when employing the Fermi-type or zero-damping function is shifted towards lower temperatures compared to BJ-damping. The exception is RPBE-TS(F) (S22), for which the density maximum is at a much higher temperature of 355\,K compared to 304\,K obtained from RPBE-TS(BJ) (S22). The RPBE-TS(F) (S22) model stood out in all prior sections and seems to suffer from the highest degree of double-counting. Therefore, thermal expansion is drastically underestimated leading to a rise in density up 355\,K after which it starts to decline.
\\
The higher density resulting from damping to zero was also observed by de Hijes et al.\cite{montero_de_hijes_density_2024}, which they argue is due to the zero-damping function yielding stronger dispersion interactions compared to the BJ-damping function in DFT-D3 for the xc functionals they have assessed. However, this is not fully compatible with our results of the two previous sections, where we found that employing BJ-damping leads to an enhanced RDF and lower self-diffusion indicating the opposite behavior. As described above, the density of water can increase in two ways, due to contraction and due to occupation of interstitial sites. For both the zero- and Fermi-type damping functions, we observed that the maxima in the RDFs in Fig.\:\ref{fig:RDF-comp} are shifted towards larger distances indicating an expansion instead. However, compared to BJ-damping there seems to be an increased occupation of interstitial sites based on generally higher $g^\mathrm{1,min}_\mathrm{OO}$ values. 
\\
To assess this further, we identified the number of nearest neighbors around a given water molecule based on the Stillinger cluster condition.\cite{stillinger_rigorous_1963} Here, two molecules are considered to be clustered when they are within the Stillinger radius $r_\mathrm{S}$, which corresponds to the position of the first minimum of the RDF.\cite{wedekind_what_2007} We use the OO RDF in Fig.\:\ref{fig:RDF-comp-OO} and approximate a value of 3.3\,\AA\:for PBE and 3.4\,\AA\:for RPBE. Using this condition, we determine the probability of a water molecule having between zero and six neighbors based on a similar analysis of Jedlovszky et al.\cite{jedlovszky_molecular_2000}. Generally, the maximum of the probability can be expected to be at four due to the tetrahedral H-bonding network in water. An increased occupation of interstitial sites can be detected by a decrease of four-fold coordinated water molecules and an increase in higher-than-four-fold coordinated molecules.\cite{jedlovszky_molecular_2000} We performed this analysis on the 2\,ns $NpT$-trajectories obtained at 350\,K and 300\,K for PBE and RPBE, respectively, using snapshots every 1\,ps. These temperatures correspond to the average $T_\mathrm{max}$ values of the respective xc functional excluding the RPBE-TS(F) (S22) outlier. The resulting histograms are shown in Fig.\:\ref{fig:nearest_neighbors}. It can be seen that without dispersion corrections both xc functionals do not produce a considerable occupation of interstitial sites. The majority of water molecules is four-fold coordinated and there are almost none with five or six neighbors. However, there is a significant amount of molecules that have less than four neighbors indicating an increased neighbor distance consistent with the severely underestimated density. 
\\
When applying a dispersion correction, the distribution is different. The majority of water molecules is still four-fold coordinated but there is also a significant number of interstitial molecules, which have five or six neighbors regardless of the dispersion model showing the importance of long-range attractive dispersion interactions for restoring the balance between thermal expansion and the strength of the local tetrahedral H-bonding network. Comparing the distribution between the two types of damping, it can be seen that the relative number of tetrahedrally-coordinated water molecules is generally lower when damping to zero compared to BJ-damping, indicating that there is a larger occupation of interstitial sites. Following the conclusions of previous sections, we assume this increase to be due to the repulsive behavior associated with this type of damping, which weakens the local tetrahedral network. This leads to a larger degree of disorder and allows for an increase in the population of water molecules in interstitial sites, which increases the density. This can also explain the generally lower $T_\mathrm{max}$ values observed when damping to zero. The repulsive gradient amplifies perturbations of the H-bonding network, induced by thermal fluctuations, which allows for an increased population of water molecules in interstitial sites at lower temperatures compared to BJ-damping. Therefore, we suggest that the observed density difference between zero-damping and BJ-damping is not due to a difference in interaction strength, but due to the different short-range behavior of the two types of damping.

\begin{table}
\caption{Temperature of the density maximum $T_{\mathrm{max}}$ and corresponding density $\rho_{\mathrm{max}}$ obtained from the density isobars of water in Fig. \ref{fig:isobars}.}
\label{tab:density}
\begin{ruledtabular}
\begin{tabular}{ lcr }
Method & $T_{\rm{max}}\:/\:\rm{K}$ & $\rho_{T_{\rm{max}}}\:/\:\rm{(g/cm^3)}$\\
\hline
PBE & - & -\\
PBE-TS(F) (S22) & 342 & 0.97\\
PBE-TS(F) (S66) & 334 & 1.01\\
PBE-TS(BJ) (S22) & 364 & 0.98\\
PBE-TS(BJ) (S66) & 361 & 0.96\\
PBE-D3(0) & 347 & 1.00\\
PBE-D3(BJ) & 359 & 0.96\\
\hline
\hline
RPBE & - & - \\
RPBE-TS(F) (S22) & 355 & 0.98\\
RPBE-TS(F) (S66) & 302 & 0.96\\
RPBE-TS(BJ) (S22) & 304 & 0.98\\
RPBE-TS(BJ) (S66) & 314 & 0.95\\
RPBE-D3(0) & 282 & 0.90\\
RPBE-D3(BJ) & 291 & 0.90\\
\end{tabular}
\end{ruledtabular}
\end{table}

\section{Conclusions}\label{sec:conclusions}

The choice of the damping function in dispersion-corrected DFT can have a significant effect on the predicted properties of water. The strength of dispersion corrections in the intermediate range is determined by the reference data used for optimizing the parameters of the damping function. Therefore, zero- and BJ-damping perform similarly in benchmarks focusing predominantly on interaction energies of equilibrium geometries, e.g., of optimized water clusters. However, for non-equilibrium geometries as encountered in MD simulations of liquid water artifacts resulting from double counting, which result in an increased over-structuring of the liquid, can be present, in particular when employing BJ-damping. While for both forms of damping a softening of the structure, an acceleration of self-diffusion, a higher equilibrium density at ambient conditions and the appearance of a density maximum can be observed, the effect is generally more pronounced when employing zero-damping. 
\\
These findings can be explained by the different short-range behavior of the two types of damping. While the attractive short-range behavior of BJ-damping is more likely to lead to an overestimation of the strength of the H-bonding network, the repulsive short-range forces of zero-damping artificially weaken this network. This repulsive short-range behavior can compensate for deficiencies in the PBE and RPBE functionals in applications to liquid water. In case of BJ-damping such a compensation is not found due to its strictly attractive interaction profile. Consequently, the extent to which applying a dispersion correction improves the description of liquid water strongly depends on the employed damping function and needs to be reevaluated.
\\
The impact of both types of damping is consistent between the investigated DFT-D3 and TS dispersion models. Therefore, we expect that our results are transferable to other dispersion corrections that similarly apply a BJ- or zero-damping function, such as XDM\cite{becke_exchange-hole_2005, becke_exchange-hole_2007}, DFT-D4\cite{caldeweyher_generally_2019} or many-body dispersion (MBD)\cite{tkatchenko_accurate_2012,hermann_libmbd_2023}. For instance, we compared RDFs of liquid water obtained from DFT-D3 to the more recently introduced DFT-D4, which uses BJ-damping exclusively. We found that DFT-D4 models produce RDFs that are similar to those obtained using DFT-D3(BJ) (see Supplementary Material in Sec.\:SIII\:C). Therefore, the improved description of water, which can be obtained for PBE and RPBE when employing a zero-damping function, cannot be achieved with current implementations of DFT-D4 due to its exclusive use of BJ-damping, which is also the case for XDM.\cite{becke_exchange-hole_2005, becke_exchange-hole_2007} This finding has implications beyond water studied here as an important model system, because uncertainties related to the choice and parameterization of the damping function are likely to occur for many different types of systems and should not be exclusively ascribed to the employed dispersion model or xc functional.

\section*{Supplementary Material}

See Supplementary Material for FHI-aims settings (SI\:A), RuNNer settings (SI\:B\:1), ACSF parameters (SI\:B\:2), additional details on the optimization of the damping function parameters for the TS model (SII\:A and SII\:B), investigation into the relationship between the $a_1$ and $a_2$ BJ-damping parameters (SII\:C), effect of choice of basis-set and Counterpoise correction on interaction energies of the S22 and S66 benchmark datasets (SII\:D), water RDFs for PBE at 300\,K (SIII\:A), water RDFs at the equilibrium density at ambient conditions for all investigated PBE and RPBE models (SIII\:B), comparison between water RDFs obtained using DFT-D3 and DFT-D4 for PBE and RPBE (SIII\:C), RDF basis-set convergence test (SIII\:D), additional information on our workflow for determining the self-diffusion coefficient (SIV\:A) and self-diffusion coefficients at 300\;K for all investigated PBE models (SIV\:B).

\begin{acknowledgments}
The authors are grateful for funding by the Deutsche Forschungsgemeinschaft (DFG, German Research Foundation) - (217133147/SFB 1073, project C03) and under Germany’s Excellence Strategy – EXC 2033–(390677874 – RESOLV). We gratefully acknowledge computing time provided by the Paderborn Center for Parallel Computing (PC$^2$). Discussions on the determination of the diffusion coefficient of water with Nore Stolte are gratefully acknowledged.
\end{acknowledgments}

\section*{Author Declarations}

The authors have no conflicts to disclose.

\section*{Data Availability Statement}
The data that support the findings of this study are available from the corresponding author upon reasonable request.

\section*{References}
\bibliography{literature}

\clearpage

\section*{Supplementary Material - Impact of the damping function in dispersion-corrected density functional theory on the properties of liquid water}

\renewcommand{\thefigure}{S\arabic{figure}}
\renewcommand{\thetable}{S\arabic{table}}
\renewcommand{\thesection}{S\Roman{section}}
\setcounter{figure}{0} 
\setcounter{table}{0} 
\setcounter{section}{0} 

\section{Computational details}

\subsection{FHI-aims settings}

\subsubsection{non-periodic}
\noindent
{\scriptsize
relativistic   atomic\_zora scalar\\
charge         0.\\
xc              pbe / rpbe\\
sc\_accuracy\_rho            5.0E-6\\
sc\_accuracy\_eev            1.0E-3\\
sc\_accuracy\_etot           1.0E-6\\
sc\_accuracy\_forces         not\_checked\\
sc\_iter\_limit              100\\
sc\_init\_iter               50\\
KS\_method                  elpa\\
RI\_method                  LVL\_fast\\
density\_update\_method      density\_matrix\\
packed\_matrix\_format       index\\
preconditioner kerker 2.0\\
mixer pulay\\
n\_max\_pulay 12\\
charge\_mix\_param           0.2\\
occupation\_type gaussian   0.01\\
vdw\_correction\_hirshfeld\\
output                     hirshfeld\_always}\\

\subsubsection{periodic}
\noindent
{\scriptsize
relativistic   atomic\_zora scalar\\
charge         0.\\
xc              pbe / rpbe\\
sc\_accuracy\_rho            5.0E-6\\
sc\_accuracy\_eev            1.0E-3\\
sc\_accuracy\_etot           1.0E-6\\
sc\_accuracy\_forces         not\_checked\\
sc\_iter\_limit              1000\\
sc\_init\_iter               150\\
k\_grid\_density 2.05\\
KS\_method                  elpa\\
RI\_method                  LVL\_fast\\
density\_update\_method      density\_matrix\\
packed\_matrix\_format       index\\
preconditioner kerker 2.0\\
mixer pulay\\
n\_max\_pulay 12\\
charge\_mix\_param           0.02\\
occupation\_type gaussian   0.1\\
vdw\_correction\_hirshfeld\\
output                     hirshfeld\_always}\\

\subsection{High-dimensional neural network potentials}

\subsubsection{RuNNer settings}
\noindent
{\scriptsize
runner\_mode                    1 / 2 \\
nnp\_gen                        2 \\
test\_fraction                  0.1 \\
remove\_atom\_energies \\
atom\_energy H                  -0.5052583648142086 \\
atom\_energy O                  -75.16774963320022 \\
use\_short\_nn \\
nn\_type\_short                  1 \\
elements                       H O \\
number\_of\_elements             2 \\
bond\_threshold                 0.40000000 \\
random\_seed                    (changed for each fit)\\
random\_number\_type             6 \\
weights\_max                    1.00000000 \\
weights\_min                    -1.00000000 \\
epochs                         30 \\
nguyen\_widrow\_weights\_short \\
repeated\_energy\_update \\
use\_short\_forces \\
short\_force\_fraction           0.02500000 \\
precondition\_weights                       \\                
optmode\_short\_energy           1\\
optmode\_short\_force            1\\
points\_in\_memory               1000\\
scale\_symmetry\_functions\\
cutoff\_type                    2\\
global\_activation\_short        t t l\\
global\_hidden\_layers\_short     2\\
global\_nodes\_short             15 15\\
kalman\_lambda\_short            0.98000000\\
kalman\_nue\_short               0.99870000\\
short\_energy\_error\_threshold   0.80000000\\
short\_energy\_fraction          1.00000000\\
short\_force\_error\_threshold    0.80000000\\
mix\_all\_points \\
write\_weights\_epoch            1\\
center\_symmetry\_functions\\
calculate\_forces\\
}

\subsubsection{Atom-centered symmetry functions}
The radial atom-centered symmetry function\cite{behler_atom-centered_2011} (ACSF) parameters used to train the high-dimensional neural network potentials\cite{behler_generalized_2007, behler_constructing_2015, behler_first_2017, behler_four_2021} (HDNNPs) are summarized in Table\:\ref{tab:rad_ACSF}. The angular ACSF parameters are summarized in Tables\:\ref{tab:ang_ACSF1}, \ref{tab:ang_ACSF2} and \ref{tab:ang_ACSF3}.

\begin{table}
\caption{Summary of radial ACSF parameters for element pairs AB - HH, OH and OO.}
\label{tab:rad_ACSF}
\begin{ruledtabular}
\begin{tabular}{l c c c r}
A & B & $\eta\:/\:a_{0}^{-2}$ & $R_s\:/\:a_0$ & $R_c\:/\:a_0$\\
\hline
H &  H &     0.001   &    0.0   &   11.33835676  \\ 
H &  H &     0.010   &    0.0   &   11.33835676  \\ 
H &  H &     0.030   &    0.0   &   11.33835676  \\ 
H &  H &     0.060   &    0.0   &   11.33835676  \\ 
H &  H &     0.150   &    1.9   &   11.33835676  \\ 
H &  H &     0.300   &    1.9   &   11.33835676  \\ 
H &  H &     0.600   &    1.9   &   11.33835676  \\ 
H &  H &     1.500   &    1.9   &   11.33835676  \\ 
H &  O &     0.001   &    0.0   &   11.33835676  \\ 
H &  O &     0.010   &    0.0   &   11.33835676  \\ 
H &  O &     0.030   &    0.0   &   11.33835676  \\ 
H &  O &     0.060   &    0.0   &   11.33835676  \\ 
H &  O &     0.150   &    0.9   &   11.33835676  \\ 
H &  O &     0.300   &    0.9   &   11.33835676  \\ 
H &  O &     0.600   &    0.9   &   11.33835676  \\ 
H &  O &     1.500   &    0.9   &   11.33835676  \\ 
O &  H &     0.001   &    0.0   &   11.33835676  \\ 
O &  H &     0.010   &    0.0   &   11.33835676  \\ 
O &  H &     0.030   &    0.0   &   11.33835676  \\ 
O &  H &     0.060   &    0.0   &   11.33835676  \\ 
O &  H &     0.150   &    0.9   &   11.33835676  \\ 
O &  H &     0.300   &    0.9   &   11.33835676  \\ 
O &  H &     0.600   &    0.9   &   11.33835676  \\ 
O &  H &     1.500   &    0.9   &   11.33835676  \\ 
O &  O &     0.001   &    0.0   &   11.33835676  \\ 
O &  O &     0.010   &    0.0   &   11.33835676  \\ 
O &  O &     0.030   &    0.0   &   11.33835676  \\ 
O &  O &     0.060   &    0.0   &   11.33835676  \\ 
O &  O &     0.150   &    4.0   &   11.33835676  \\ 
O &  O &     0.300   &    4.0   &   11.33835676  \\ 
O &  O &     0.600   &    4.0   &   11.33835676  \\ 
O &  O &     1.500   &    4.0   &   11.33835676  \\
\end{tabular}
\end{ruledtabular}
\end{table}

\begin{table}
\caption{Summary of the angular ACSF parameters for element triplets ABC - H-H-H and H-H-O.}
\label{tab:ang_ACSF1}
\begin{ruledtabular}
\begin{tabular}{l c c c c c r}
A & B & C & $\eta\:/\:a_{0}^{-2}$ & $\lambda$ & $\xi$ & $R_c\:/\:a_0$\\
\hline
H  & H & H   &   0.00000000   &   -1    &   1   &   11.33835676   \\ 
H  & H & H   &   0.00000000   &   -1    &   2   &   11.33835676   \\ 
H  & H & H   &   0.00000000   &   -1    &   4   &   11.33835676   \\ 
H  & H & H   &   0.00000000   &   -1    &   8   &   11.33835676   \\ 
H  & H & H   &   0.00000000   &   -1    &  16   &   11.33835676   \\ 
H  & H & H   &   0.00000000   &   -1    &  32   &   11.33835676   \\ 
H  & H & H   &   0.04718483   &   -1    &   1   &   11.33835676   \\ 
H  & H & H   &   0.04718483   &   -1    &   2   &   11.33835676   \\ 
H  & H & H   &   0.04718483   &   -1    &   4   &   11.33835676   \\ 
H  & H & H   &   0.04718483   &   -1    &   8   &   11.33835676   \\ 
H  & H & H   &   0.04718483   &   -1    &  16   &   11.33835676   \\ 
H  & H & H   &   0.04718483   &   -1    &  32   &   11.33835676   \\ 
H  & H & H   &   0.00000000   &    1    &   1   &   11.33835676   \\ 
H  & H & H   &   0.00000000   &    1    &   2   &   11.33835676   \\ 
H  & H & H   &   0.00000000   &    1    &   4   &   11.33835676   \\ 
H  & H & H   &   0.00000000   &    1    &   8   &   11.33835676   \\ 
H  & H & H   &   0.00000000   &    1    &  16   &   11.33835676   \\ 
H  & H & H   &   0.00000000   &    1    &  32   &   11.33835676   \\ 
H  & H & H   &   0.04718483   &    1    &   1   &   11.33835676   \\ 
H  & H & H   &   0.04718483   &    1    &   2   &   11.33835676   \\ 
H  & H & H   &   0.04718483   &    1    &   4   &   11.33835676   \\ 
H  & H & H   &   0.04718483   &    1    &   8   &   11.33835676   \\ 
H  & H & H   &   0.04718483   &    1    &  16   &   11.33835676   \\ 
H  & H & H   &   0.04718483   &    1    &  32   &   11.33835676   \\ 
H  & H & O   &   0.00000000   &   -1    &   1   &   11.33835676   \\ 
H  & H & O   &   0.00000000   &   -1    &   2   &   11.33835676   \\ 
H  & H & O   &   0.00000000   &   -1    &   4   &   11.33835676   \\ 
H  & H & O   &   0.00000000   &   -1    &   8   &   11.33835676   \\ 
H  & H & O   &   0.00000000   &   -1    &  16   &   11.33835676   \\ 
H  & H & O   &   0.00000000   &   -1    &  32   &   11.33835676   \\ 
H  & H & O   &   0.06057924   &   -1    &   1   &   11.33835676   \\ 
H  & H & O   &   0.06057924   &   -1    &   2   &   11.33835676   \\ 
H  & H & O   &   0.06057924   &   -1    &   4   &   11.33835676   \\ 
H  & H & O   &   0.06057924   &   -1    &   8   &   11.33835676   \\ 
H  & H & O   &   0.06057924   &   -1    &  16   &   11.33835676   \\ 
H  & H & O   &   0.06057924   &   -1    &  32   &   11.33835676   \\ 
H  & H & O   &   0.00000000   &    1    &   1   &   11.33835676   \\ 
H  & H & O   &   0.00000000   &    1    &   2   &   11.33835676   \\ 
H  & H & O   &   0.00000000   &    1    &   4   &   11.33835676   \\ 
H  & H & O   &   0.00000000   &    1    &   8   &   11.33835676   \\ 
H  & H & O   &   0.00000000   &    1    &  16   &   11.33835676   \\ 
H  & H & O   &   0.00000000   &    1    &  32   &   11.33835676   \\ 
H  & H & O   &   0.06057924   &    1    &   1   &   11.33835676   \\ 
H  & H & O   &   0.06057924   &    1    &   2   &   11.33835676   \\ 
H  & H & O   &   0.06057924   &    1    &   4   &   11.33835676   \\ 
H  & H & O   &   0.06057924   &    1    &   8   &   11.33835676   \\ 
H  & H & O   &   0.06057924   &    1    &  16   &   11.33835676   \\ 
H  & H & O   &   0.06057924   &    1    &  32   &   11.33835676   \\
\end{tabular}
\end{ruledtabular}
\end{table}

\begin{table}
\caption{Summary of the angular ACSF parameters for element triplets ABC - H-O-O and O-H-H.}
\label{tab:ang_ACSF2}
\begin{ruledtabular}
\begin{tabular}{l c c c c c r}
A & B & C & $\eta\:/\:a_{0}^{-2}$ & $\lambda$ & $\xi$ & $R_c\:/\:a_0$\\
\hline
H  & O & O   &   0.00000000   &   -1    &   1   &   11.33835676   \\ 
H  & O & O   &   0.00000000   &   -1    &   2   &   11.33835676   \\ 
H  & O & O   &   0.00000000   &   -1    &   4   &   11.33835676   \\ 
H  & O & O   &   0.00000000   &   -1    &   8   &   11.33835676   \\ 
H  & O & O   &   0.00000000   &   -1    &  16   &   11.33835676   \\ 
H  & O & O   &   0.00000000   &   -1    &  32   &   11.33835676   \\ 
H  & O & O   &   0.04724594   &   -1    &   1   &   11.33835676   \\ 
H  & O & O   &   0.04724594   &   -1    &   2   &   11.33835676   \\ 
H  & O & O   &   0.04724594   &   -1    &   4   &   11.33835676   \\ 
H  & O & O   &   0.04724594   &   -1    &   8   &   11.33835676   \\ 
H  & O & O   &   0.04724594   &   -1    &  16   &   11.33835676   \\ 
H  & O & O   &   0.04724594   &   -1    &  32   &   11.33835676   \\ 
H  & O & O   &   0.00000000   &    1    &   1   &   11.33835676   \\ 
H  & O & O   &   0.00000000   &    1    &   2   &   11.33835676   \\ 
H  & O & O   &   0.00000000   &    1    &   4   &   11.33835676   \\ 
H  & O & O   &   0.00000000   &    1    &   8   &   11.33835676   \\ 
H  & O & O   &   0.00000000   &    1    &  16   &   11.33835676   \\ 
H  & O & O   &   0.00000000   &    1    &  32   &   11.33835676   \\ 
H  & O & O   &   0.04724594   &    1    &   1   &   11.33835676   \\ 
H  & O & O   &   0.04724594   &    1    &   2   &   11.33835676   \\ 
H  & O & O   &   0.04724594   &    1    &   4   &   11.33835676   \\ 
H  & O & O   &   0.04724594   &    1    &   8   &   11.33835676   \\ 
H  & O & O   &   0.04724594   &    1    &  16   &   11.33835676   \\ 
H  & O & O   &   0.04724594   &    1    &  32   &   11.33835676   \\ 
O  & H & H   &   0.00000000   &   -1    &   1   &   11.33835676   \\ 
O  & H & H   &   0.00000000   &   -1    &   2   &   11.33835676   \\ 
O  & H & H   &   0.00000000   &   -1    &   4   &   11.33835676   \\ 
O  & H & H   &   0.00000000   &   -1    &   8   &   11.33835676   \\ 
O  & H & H   &   0.00000000   &   -1    &  16   &   11.33835676   \\ 
O  & H & H   &   0.00000000   &   -1    &  32   &   11.33835676   \\ 
O  & H & H   &   0.06057924   &   -1    &   1   &   11.33835676   \\ 
O  & H & H   &   0.06057924   &   -1    &   2   &   11.33835676   \\ 
O  & H & H   &   0.06057924   &   -1    &   4   &   11.33835676   \\ 
O  & H & H   &   0.06057924   &   -1    &   8   &   11.33835676   \\ 
O  & H & H   &   0.06057924   &   -1    &  16   &   11.33835676   \\ 
O  & H & H   &   0.06057924   &   -1    &  32   &   11.33835676   \\ 
O  & H & H   &   0.00000000   &    1    &   1   &   11.33835676   \\ 
O  & H & H   &   0.00000000   &    1    &   2   &   11.33835676   \\ 
O  & H & H   &   0.00000000   &    1    &   4   &   11.33835676   \\ 
O  & H & H   &   0.00000000   &    1    &   8   &   11.33835676   \\ 
O  & H & H   &   0.00000000   &    1    &  16   &   11.33835676   \\ 
O  & H & H   &   0.00000000   &    1    &  32   &   11.33835676   \\ 
O  & H & H   &   0.06057924   &    1    &   1   &   11.33835676   \\ 
O  & H & H   &   0.06057924   &    1    &   2   &   11.33835676   \\ 
O  & H & H   &   0.06057924   &    1    &   4   &   11.33835676   \\ 
O  & H & H   &   0.06057924   &    1    &   8   &   11.33835676   \\ 
O  & H & H   &   0.06057924   &    1    &  16   &   11.33835676   \\ 
O  & H & H   &   0.06057924   &    1    &  32   &   11.33835676   \\ 
\end{tabular}
\end{ruledtabular}
\end{table}

\begin{table}
\caption{Summary of the angular ACSF parameters for element triplets ABC - O-H-O and O-O-O.}
\label{tab:ang_ACSF3}
\begin{ruledtabular}
\begin{tabular}{l c c c c c r}
A & B & C & $\eta\:/\:a_{0}^{-2}$ & $\lambda$ & $\xi$ & $R_c\:/\:a_0$\\
\hline
O  & H & O   &   0.00000000   &   -1    &   1   &   11.33835676   \\ 
O  & H & O   &   0.00000000   &   -1    &   2   &   11.33835676   \\ 
O  & H & O   &   0.00000000   &   -1    &   4   &   11.33835676   \\ 
O  & H & O   &   0.00000000   &   -1    &   8   &   11.33835676   \\ 
O  & H & O   &   0.00000000   &   -1    &  16   &   11.33835676   \\ 
O  & H & O   &   0.00000000   &   -1    &  32   &   11.33835676   \\ 
O  & H & O   &   0.04724594   &   -1    &   1   &   11.33835676   \\ 
O  & H & O   &   0.04724594   &   -1    &   2   &   11.33835676   \\ 
O  & H & O   &   0.04724594   &   -1    &   4   &   11.33835676   \\ 
O  & H & O   &   0.04724594   &   -1    &   8   &   11.33835676   \\ 
O  & H & O   &   0.04724594   &   -1    &  16   &   11.33835676   \\ 
O  & H & O   &   0.04724594   &   -1    &  32   &   11.33835676   \\ 
O  & H & O   &   0.00000000   &    1    &   1   &   11.33835676   \\ 
O  & H & O   &   0.00000000   &    1    &   2   &   11.33835676   \\ 
O  & H & O   &   0.00000000   &    1    &   4   &   11.33835676   \\ 
O  & H & O   &   0.00000000   &    1    &   8   &   11.33835676   \\ 
O  & H & O   &   0.00000000   &    1    &  16   &   11.33835676   \\ 
O  & H & O   &   0.00000000   &    1    &  32   &   11.33835676   \\ 
O  & H & O   &   0.04724594   &    1    &   1   &   11.33835676   \\ 
O  & H & O   &   0.04724594   &    1    &   2   &   11.33835676   \\ 
O  & H & O   &   0.04724594   &    1    &   4   &   11.33835676   \\ 
O  & H & O   &   0.04724594   &    1    &   8   &   11.33835676   \\ 
O  & H & O   &   0.04724594   &    1    &  16   &   11.33835676   \\ 
O  & H & O   &   0.04724594   &    1    &  32   &   11.33835676   \\ 
O  & O & O   &   0.00000000   &   -1    &   1   &   11.33835676   \\ 
O  & O & O   &   0.00000000   &   -1    &   2   &   11.33835676   \\ 
O  & O & O   &   0.00000000   &   -1    &   4   &   11.33835676   \\ 
O  & O & O   &   0.00000000   &   -1    &   8   &   11.33835676   \\ 
O  & O & O   &   0.00000000   &   -1    &  16   &   11.33835676   \\ 
O  & O & O   &   0.00000000   &   -1    &  32   &   11.33835676   \\ 
O  & O & O   &   0.00718491   &   -1    &   1   &   11.33835676   \\ 
O  & O & O   &   0.00718491   &   -1    &   2   &   11.33835676   \\ 
O  & O & O   &   0.00718491   &   -1    &   4   &   11.33835676   \\ 
O  & O & O   &   0.00718491   &   -1    &   8   &   11.33835676   \\ 
O  & O & O   &   0.00000000   &    1    &   1   &   11.33835676   \\ 
O  & O & O   &   0.00000000   &    1    &   2   &   11.33835676   \\ 
O  & O & O   &   0.00000000   &    1    &   4   &   11.33835676   \\ 
O  & O & O   &   0.00000000   &    1    &   8   &   11.33835676   \\ 
O  & O & O   &   0.00000000   &    1    &  16   &   11.33835676   \\ 
O  & O & O   &   0.00000000   &    1    &  32   &   11.33835676   \\ 
O  & O & O   &   0.00718491   &    1    &   1   &   11.33835676   \\ 
O  & O & O   &   0.00718491   &    1    &   2   &   11.33835676   \\ 
O  & O & O   &   0.00718491   &    1    &   4   &   11.33835676   \\ 
O  & O & O   &   0.00718491   &    1    &   8   &   11.33835676   \\ 
O  & O & O   &   0.00718491   &    1    &  16   &   11.33835676   \\ 
O  & O & O   &   0.00718491   &    1    &  32   &   11.33835676   \\ 
\end{tabular}
\end{ruledtabular}
\end{table}

\section{Optimization of damping function parameters for the Tkatchenko-Scheffler model}

\subsection{S22}

\begin{figure*}
\centering
\begin{subfigure}{.49\textwidth}
    \centering
    \includegraphics[scale=0.8]{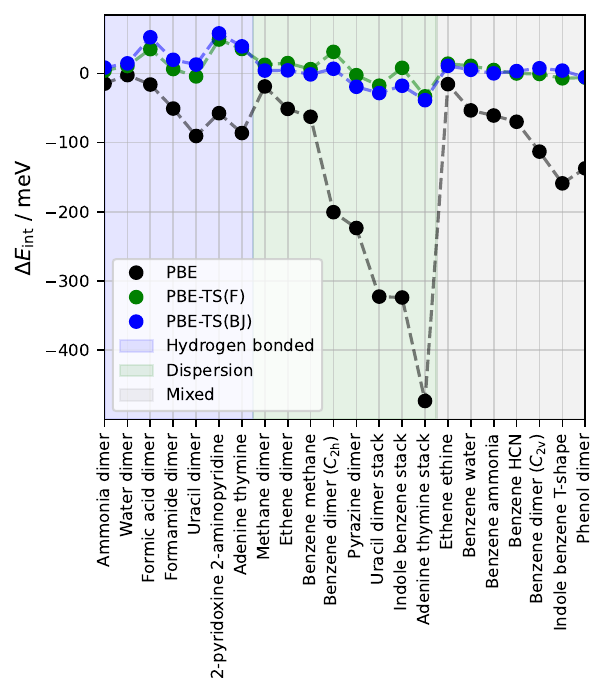}
    \caption{}
    \label{fig:s22_pbe}
\end{subfigure}
\begin{subfigure}{.49\textwidth}
    \centering
    \includegraphics[scale=0.8]{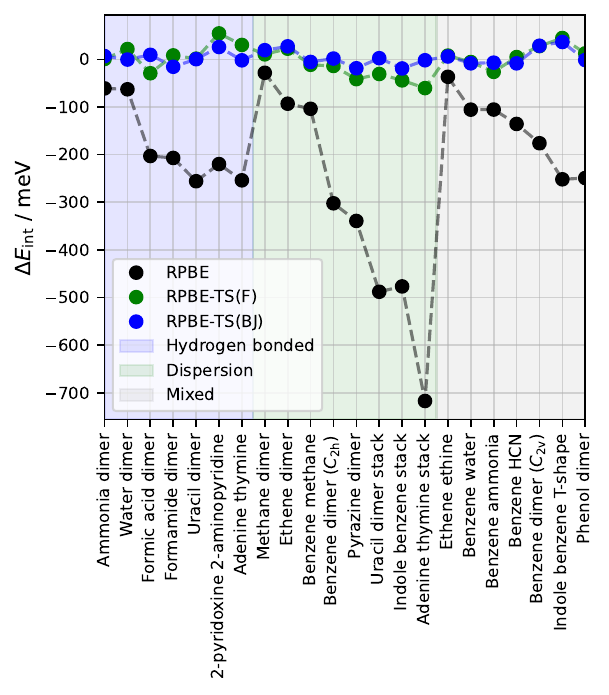}
    \caption{}
    \label{fig:s22_rpbe}
\end{subfigure}
\caption{Energy differences $\Delta E_\mathrm{int}$ between the PBE (a) and RPBE (b) xc functional, with and without TS dispersion correction, and CCSD(T) interaction energies of model complexes contained in the S22\cite{jurecka_benchmark_2006} dataset.}
\label{fig:s22}
\end{figure*}

\begin{figure*}
\centering
\begin{subfigure}{.49\textwidth}
    \centering
    \includegraphics[scale=0.8]{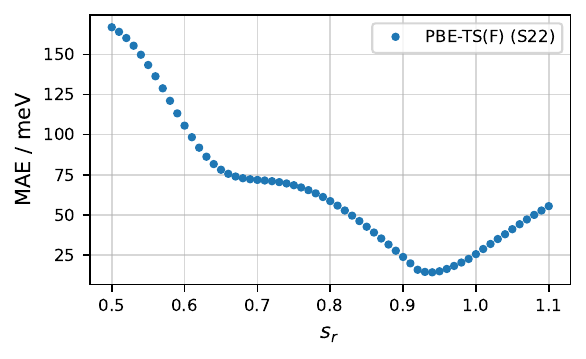}
    \caption{}
    \label{fig:s22_pbe_sr}
\end{subfigure}
\begin{subfigure}{.49\textwidth}
    \centering
    \includegraphics[scale=0.8]{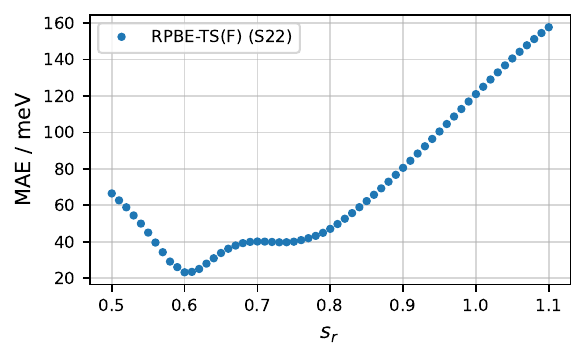}
    \caption{}
    \label{fig:s22_rpbe_sr}
\end{subfigure}
\begin{subfigure}{.49\textwidth}
    \centering
    \includegraphics[scale=0.8]{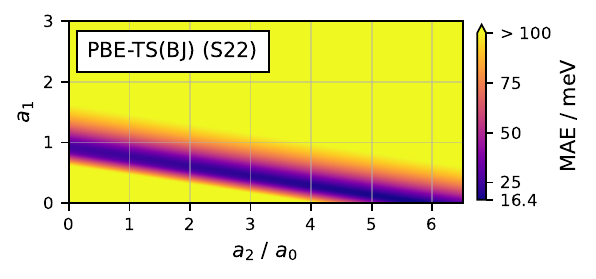}
    \caption{}
    \label{fig:s22_pbe_a1a2}
\end{subfigure}
\begin{subfigure}{.49\textwidth}
    \centering
    \includegraphics[scale=0.8]{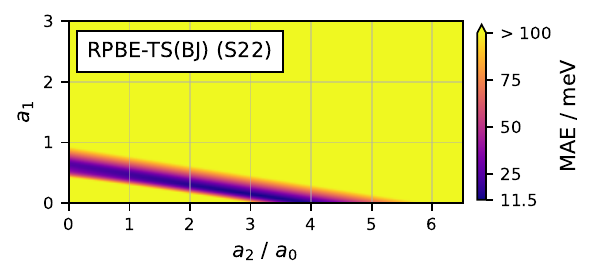}
    \caption{}
    \label{fig:s22_rpbe_a1a2}
\end{subfigure}
\caption{MAE of PBE-TS and RPBE-TS interaction energies with respect to CCSD(T) values contained in the S22\cite{jurecka_benchmark_2006} dataset as a function of the Fermi-damping function parameter $s_r$ is shown in panels (a) and (b) respectively. The MAE of PBE-TS(BJ) and RPBE-TS(BJ) as a function of the $a_1$ and $a_2$ parameters of the BJ-damping function is shown in panels (c) and (d) respectively.}
\label{fig:s22_param_opt}
\end{figure*}

The energy differences $\Delta E_\mathrm{int}$ between the PBE\cite{perdew_generalized_1996} and RPBE\cite{hammer_improved_1999} exchange-correlation (xc) functionals and the coupled cluster theory using singles, doubles and perturbative triples (CCSD(T)), interaction energies contained in the S22 benchmark set of Jurečka et al.\cite{jurecka_benchmark_2006}, are shown in Fig.\:\ref{fig:s22}. The mean absolute error (MAE) curves and surfaces obtained from the optimization of the Fermi- ($s_r$) and Becke-Johnson (BJ)-damping function parameters ($a_1,\:a_2$) for the Tkatchenko Scheffler\cite{tkatchenko_accurate_2009} (TS) model applied to correct the two xc functionals for long-range dispersion are shown in Fig.\:\ref{fig:s22_param_opt}. The $\Delta E_\mathrm{int}$ values of the optimized PBE-TS (S22),  PBE-TS(BJ) (S22), RPBE-TS (S22) and RPBE-TS(BJ) (S22) models are shown in Fig.\:\ref{fig:s22} as well.

\subsection{S66}

\begin{figure*}
    \centering
    \includegraphics[width=\textwidth]{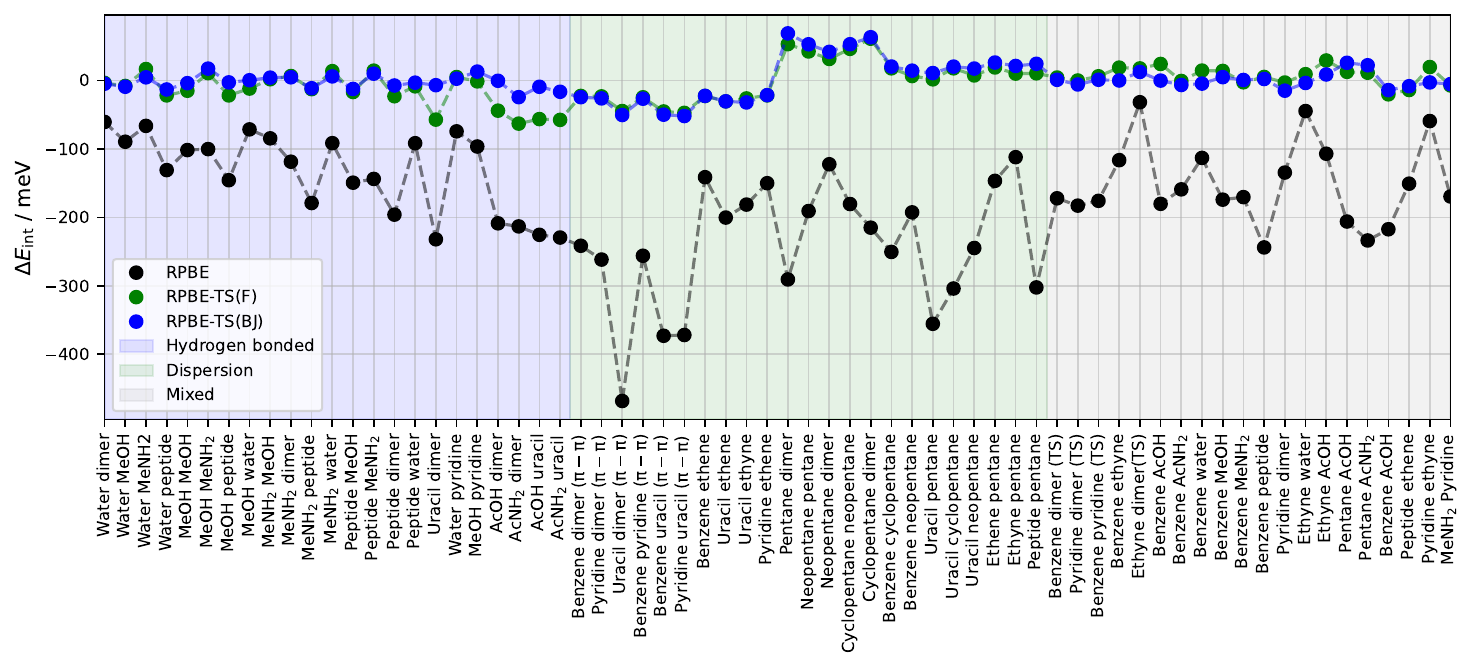}
    \caption{Energy differences $\Delta E_\mathrm{int}$ between RPBE, with and without TS dispersion correction, and explicitly correlated CCSD(T) interaction energies reported by Schmitz and Hättig\cite{schmitz_accuracy_2017} of model complexes contained in the S66\cite{rezac_s66_2011, rezac_erratum_2014} dataset.}
    \label{fig:s66}
\end{figure*}

\begin{figure*}
\centering
\begin{subfigure}{.49\textwidth}
    \centering
    \includegraphics[scale=0.8]{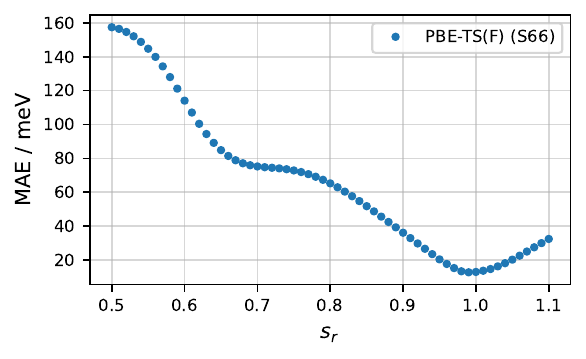}
    \caption{}
    \label{fig:s66_pbe_sr}
\end{subfigure}
\begin{subfigure}{.49\textwidth}
    \centering
    \includegraphics[scale=0.8]{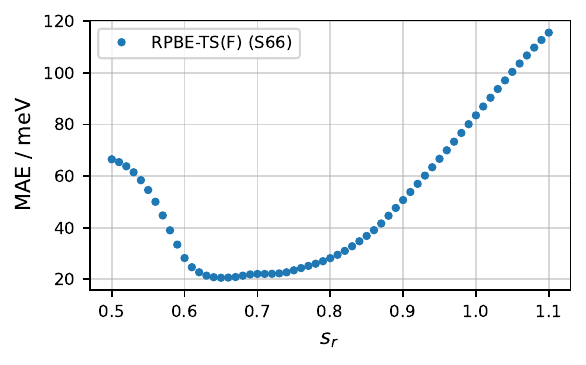}
    \caption{}
    \label{fig:s66_rpbe_sr}
\end{subfigure}
\begin{subfigure}{.49\textwidth}
    \centering
    \includegraphics[scale=0.8]{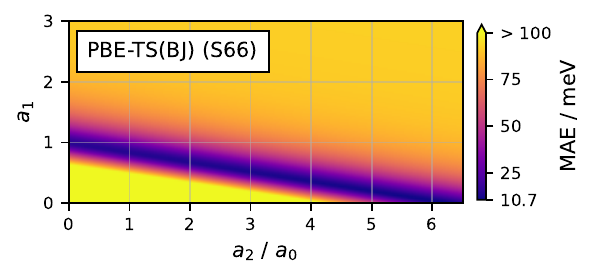}
    \caption{}
    \label{fig:s66_pbe_a1a2}
\end{subfigure}
\begin{subfigure}{.49\textwidth}
    \centering
    \includegraphics[scale=0.8]{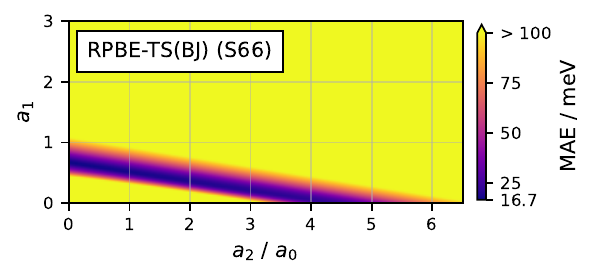}
    \caption{}
    \label{fig:s66_rpbe_a1a2}
\end{subfigure}
\caption{MAE of PBE-TS and RPBE-TS interaction energies with respect to explicitly correlated CCSD(T) interaction energies reported by Schmitz and Hättig\cite{schmitz_accuracy_2017} of model complexes contained in the S66\cite{rezac_s66_2011, rezac_erratum_2014} dataset as a function of the Fermi-damping function parameter $s_r$ are shown in panels (a) and (b) respectively. The MAE as a function of the $a_1$ and $a_2$ parameters of the BJ-damping function is shown in panels (c) and (d) for PBE-TS(BJ) and RPBE-TS(BJ) respectively.}
\label{fig:s66_param_opt}
\end{figure*}

The energy differences $\Delta E_\mathrm{int}$ between PBE and RPBE and explicitly correlated CCSD(T) interaction energies reported by Schmitz and Hättig\cite{schmitz_accuracy_2017} of model complexes contained in the S66\cite{rezac_s66_2011, rezac_erratum_2014} dataset are shown in Fig.\:\ref{fig:s66}. The MAE curves and surfaces obtained from the optimization of the Fermi- ($s_r$) and BJ-damping function parameters ($a_1,\:a_2$) for the TS model applied to correct for long-range dispersion are shown in Fig.\:\ref{fig:s66_param_opt}. The $\Delta E_\mathrm{int}$ values of the optimized PBE-TS (S66),  PBE-TS(BJ) (S66), RPBE-TS (S66) and RPBE-TS(BJ) (S66) models are shown in Fig.\:\ref{fig:s66} as well.

\begin{figure*}
    \centering
    \includegraphics[width=\textwidth]{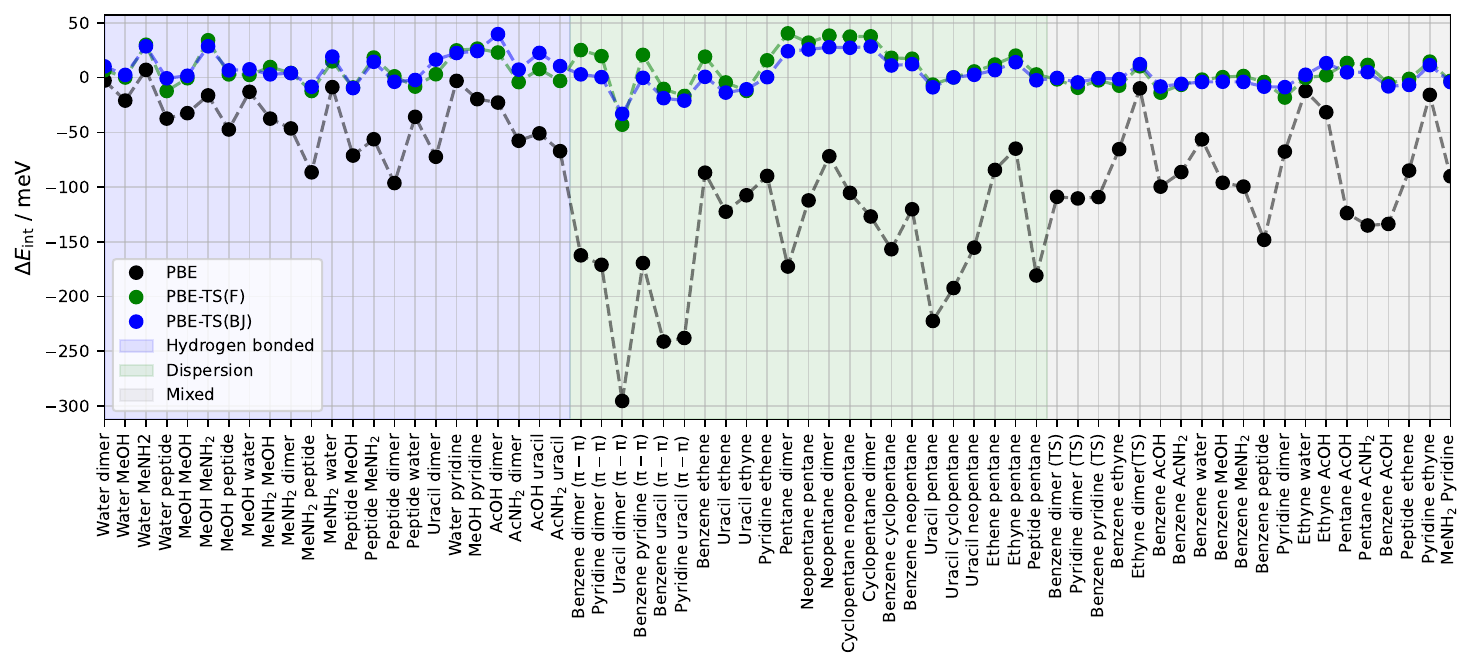}
    \caption{Energy differences $\Delta E_\mathrm{int}$ between PBE, with and without TS dispersion correction, and explicitly correlated CCSD(T) interaction energies reported by Schmitz and Hättig\cite{schmitz_accuracy_2017} of model complexes contained in the S66\cite{rezac_s66_2011, rezac_erratum_2014} dataset.}
    \label{fig:}
\end{figure*}

\subsection{Linear relation between BJ-damping parameters}

\begin{figure*}
\centering
\begin{subfigure}{.49\textwidth}
    \centering
    \includegraphics[scale=0.8]{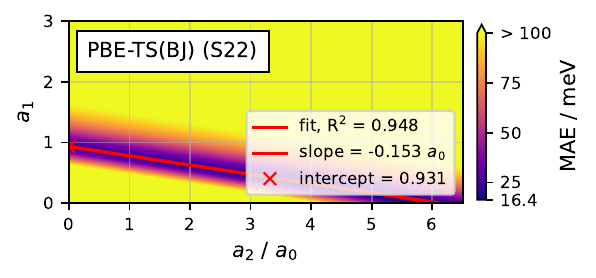}
    \caption{}
    \label{fig:s22_pbe_a1_a2_fit}
\end{subfigure}
\begin{subfigure}{.49\textwidth}
    \centering
    \includegraphics[scale=0.8]{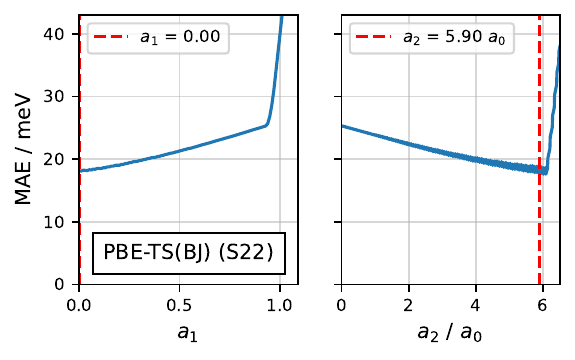}
    \caption{}
    \label{fig:s22_pbe_a1_a2_scan}
\end{subfigure}
\begin{subfigure}{.49\textwidth}
    \centering
    \includegraphics[scale=0.8]{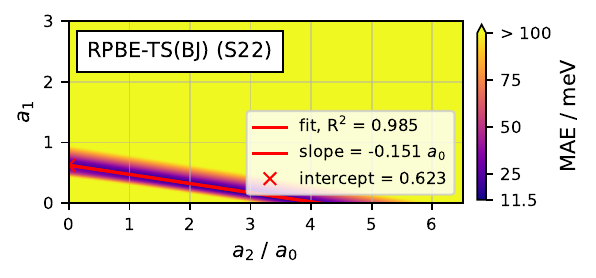}
    \caption{}
    \label{fig:s22_rpbe_a1_a2_fit}
\end{subfigure}
\begin{subfigure}{.49\textwidth}
    \centering
    \includegraphics[scale=0.8]{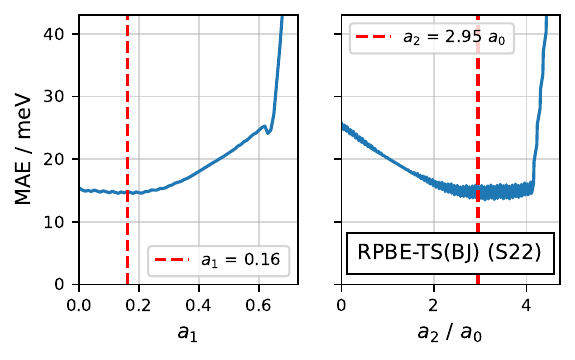}
    \caption{}
    \label{fig:s22_rpbe_a1_a2_scan}
\end{subfigure}
\caption{MAEs of PBE-TS(BJ) and RPBE-TS(BJ) interaction energies with respect to CCSD(T) interaction energies of model complexes contained in the S22\cite{jurecka_benchmark_2006} dataset as a function of the $a_1$ and $a_2$ parameters of the BJ-damping function are shown in panels (a) and (c) for PBE-TS(BJ) and RPBE-TS(BJ) respectively. The $a_1$-$a_2$ pairs, which yield a MAE below chemical accuracy (43\,meV), were used to fit a linear relation between $a_1$ and $a_2$ ($a_2 = m\cdot a_1 + b$, where $m$ is the slope and $b$ is the intercept). The resulting fit line is drawn in red in (a) and (c) on top of the associated error surface. Using the linear relation, BJ-damping can be made single-parametric by replacing one of the two parameters as a function of the other. In panels (b) and (d), we show the resulting MAEs for PBE-TS(BJ) and RPBE-TS(BJ) respectively. Additionally, the optimal $a_1$ and $a_2$ parameter, which we obtained from a grid search of the $a_1$-$a_2$ error surface, is drawn as a red dashed line.}
\label{fig:s22_bj_a1_a2_lin_rel}
\end{figure*}

\begin{figure*}
\centering
\begin{subfigure}{.49\textwidth}
    \centering
    \includegraphics[scale=0.8]{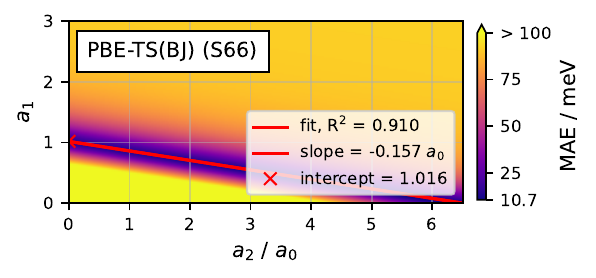}
    \caption{}
    \label{fig:s66_pbe_a1_a2_fit}
\end{subfigure}
\begin{subfigure}{.49\textwidth}
    \centering
    \includegraphics[scale=0.8]{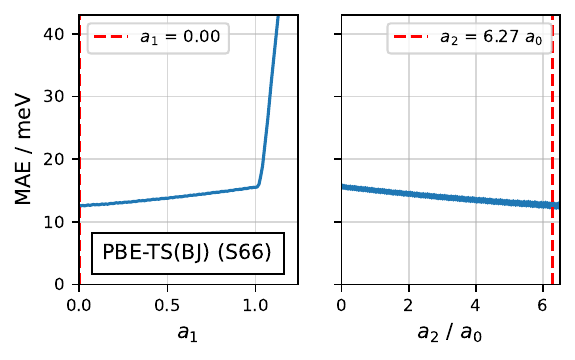}
    \caption{}
    \label{fig:s66_pbe_a1_a2_scan}
\end{subfigure}
\begin{subfigure}{.49\textwidth}
    \centering
    \includegraphics[scale=0.8]{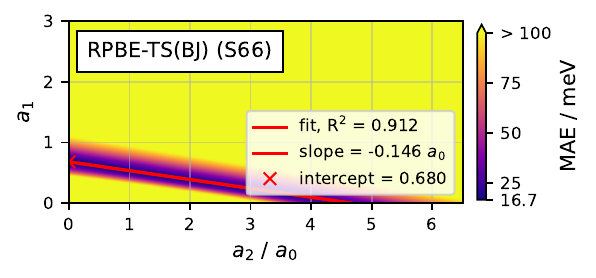}
    \caption{}
    \label{fig:s66_rpbe_a1_a2_fit}
\end{subfigure}
\begin{subfigure}{.49\textwidth}
    \centering
    \includegraphics[scale=0.8]{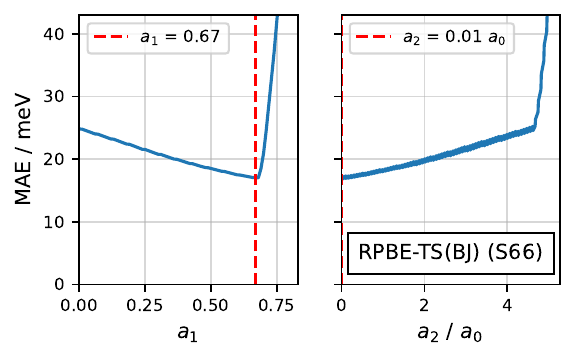}
    \caption{}
    \label{fig:s66_rpbe_a1_a2_scan}
\end{subfigure}
\caption{MAEs of PBE-TS and RPBE-TS interaction energies with respect to explicitly correlated CCSD(T) interaction energies reported by Schmitz and Hättig\cite{schmitz_accuracy_2017} of model complexes contained in the S66\cite{rezac_s66_2011, rezac_erratum_2014} dataset as a function of the $a_1$ and $a_2$ parameters of the BJ-damping function are shown in panels (a) and (c) for PBE-TS(BJ) and RPBE-TS(BJ) respectively. The $a_1$-$a_2$ pairs, which yield a MAE below chemical accuracy (43\,meV), were used to fit a linear relation between $a_1$ and $a_2$ ($a_2 = m\cdot a_1 + b$, where $m$ is the slope and $b$ is the intercept). The resulting fit line is drawn in red in (a) and (c) on top of the associated error surface. Using the linear relation, BJ-damping can be made single-parametric by replacing one of the two parameters as a function of the other. In panels (b) and (d), we show the resulting MAEs for PBE-TS(BJ) and RPBE-TS(BJ) respectively. Additionally, the optimal $a_1$ and $a_2$ parameter, which we obtained from a grid search of the $a_1$-$a_2$ error surface, is drawn as a red dashed line.}
\label{fig:s66_bj_a1_a2_lin_rel}
\end{figure*}

On both the S22 and S66 MAE error surfaces obtained from the BJ-damping parameter optimization, shown in Figs. \ref{fig:s22_param_opt} and \ref{fig:s66_param_opt} respectively, there is a large area of $a_1$-$a_2$ pairs visible, which achieve a below chemical accuracy (43\,meV) MAE. This area forms a straight path through the error surface indicating that there might be a linear relationship between the optimal $a_1$ and $a_2$ parameter.\\
To investigate this further, we used the below chemical accuracy $a_1$-$a_2$ pairs and the associated MAE values as weights to fit a linear relation between $a_1$ and $a_2$ for both the S22 and S66 data,
\begin{equation}
    a_2 = m\cdot a_1 + b,
    \label{eq:a2_fit}
\end{equation}
where $m$ is the slope and $b$ is the intercept of the line fit. The resulting fit parameters and lines are presented in Figs. \ref{fig:s22_bj_a1_a2_lin_rel} and \ref{fig:s66_bj_a1_a2_lin_rel} for S22 and S66 respectively. For both functionals and datasets, a good model fit could be achieved with $R^2$ values higher than 0.9. It is interesting that the slope of the fit is about $-0.15\,a_0$ for both functionals and datasets. The intercept differs between the RPBE and PBE data, but the difference between the S22 and S66 intercept of the respective functional is relatively small indicating that there might be a universal relationship between $a_1$ and $a_2$ for a given functional.\\
Using these linear relations, BJ-damping can be made single-parametric by replacing one of the two parameters as a function of the other. Also in Figs. \ref{fig:s22_bj_a1_a2_lin_rel} and \ref{fig:s66_bj_a1_a2_lin_rel}, we show the resulting MAE curves as a function of either $a_1$ or $a_2$. Here, it can be seen that there is little loss in accuracy when using this approach. This indicates that, similarly to the Fermi- or zero-damping function, also for BJ-damping a single damping function parameter might be sufficient.

\subsection{Basis-set and Counterpoise correction}
\label{sec:basis_and_counterpoise}

\begin{figure*}
\centering
\begin{subfigure}{.49\textwidth}
    \centering
    \includegraphics[scale=0.8]{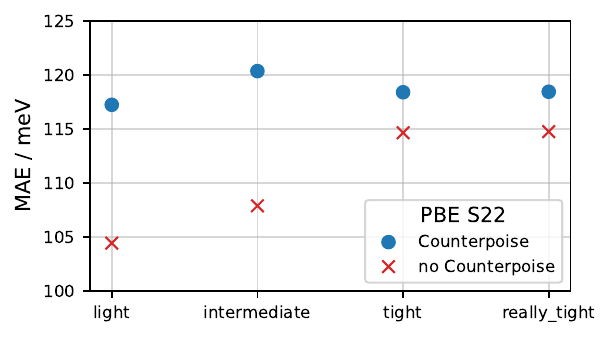}
    \caption{}
    \label{fig:s22_pbe_bsse}
\end{subfigure}
\begin{subfigure}{.49\textwidth}
    \centering
    \includegraphics[scale=0.8]{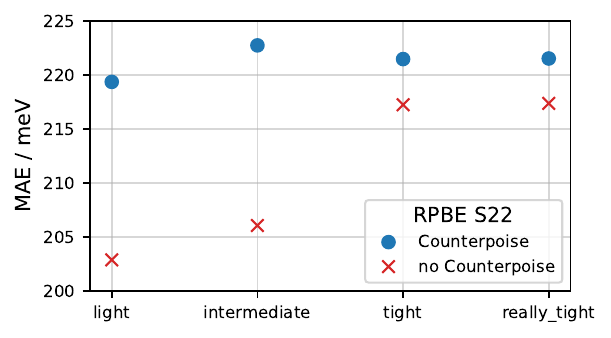}
    \caption{}
    \label{fig:s22_rpbe_bsse}
\end{subfigure}
\begin{subfigure}{.49\textwidth}
    \centering
    \includegraphics[scale=0.8]{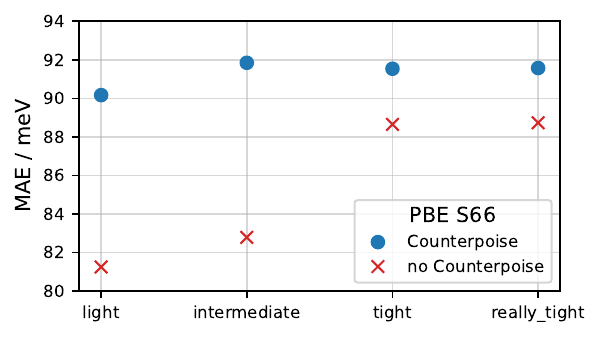}
    \caption{}
    \label{fig:s66_pbe_bsse}
\end{subfigure}
\begin{subfigure}{.49\textwidth}
    \centering
    \includegraphics[scale=0.8]{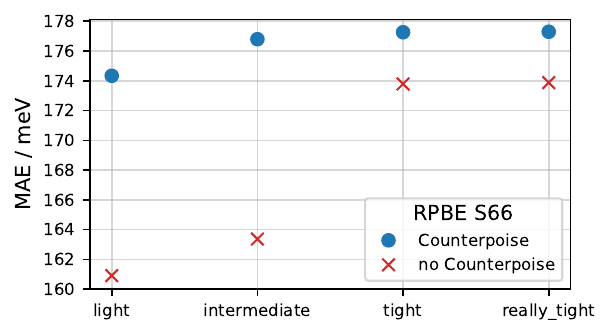}
    \caption{}
    \label{fig:s66_rpbe_bsse}
\end{subfigure}
\caption{MAE of PBE and RPBE interaction energies of the S22\cite{jurecka_benchmark_2006} ((a) and (b)) and S66\cite{rezac_s66_2011, rezac_erratum_2014, schmitz_accuracy_2017} ((c) and (d)) benchmark sets obtained using different FHI-aims\cite{blum_ab_2009} species settings with and without Counterpoise correction. The species defaults provided by FHI-aims are ``light'', ``intermediate'', ``tight'', and ``really\_tight'', in increasing accuracy, and define the basis functions and numerical settings, e.g., integration grids and the Hartree potential, of the DFT calculation. Generally, the number of basis functions increases by going up a tier; however, the ``really\_tight'' default uses the same basis-sets as the ``tight'' preset but improves numerical aspects of the calculation. The values presented in the main text correspond to the ``tight'' settings with Counterpoise correction.}
\label{fig:basis_set_Counter_Poise}
\end{figure*}

The FHI-aims code\cite{blum_ab_2009} provides four predefined settings (species defaults), which define the basis functions and numerical settings, e.g., integration grids and the Hartree potential, of the DFT calculation. These are named ``light'', ``intermediate'', ``tight'', and ``really\_tight''. In Fig.\;\ref{fig:basis_set_Counter_Poise}, we show the MAE of the PBE and RPBE interaction energies with respect to CCSD(T) interaction energies of the S22\cite{jurecka_benchmark_2006} and explicitly correlated CCSD(T) interaction energies reported by Schmitz and Hättig\cite{schmitz_accuracy_2017} of the S66\cite{rezac_s66_2011, rezac_erratum_2014} datasets using the different FHI-aims settings for the PBE and RPBE functionals. Additionally, we show the interaction energy MAEs of each FHI-aims setting with Counterpoise correction as proposed by Boys and Bernardi\cite{boys_calculation_1970}, which aims to reduce the effect of the basis-set superposition error (BSSE) on the interaction energies for incomplete basis-sets. Here, it can be seen that the MAE compared to CCSD(T) is generally lower for lower tier settings, i.e., ``light'', which employ fewer basis functions, without Counterpoise correction. This behavior can be explained by an error compensation of the DFT interaction energy error and the BSSE. PBE and RPBE generally underestimate interaction energies compared to the CCSD(T) reference due missing interactions, i.e., dispersion interactions. The BSSE artificially increases interaction energies due to the stabilization of both monomers in the dimer calculation provided by the extended basis-set size compared to the respective monomer calculation\cite{kestner_1999}. These two effects cancel each other and this leads to the observed lower MAE for smaller basis-sets without BSSE correction. When the damping function parameters of a dispersion correction would be parametrized using a small basis-set, this error compensation would be transferred onto the dispersion model leading to a model that provides weaker dispersion interactions. Consequently, the applicability and transferability of a such a dispersion model to DFT calculations employing larger basis-sets would be limited since dispersion interactions would be underestimated in this case. Additionally, the transferabilty of this error compensation to systems outside the benchmark dataset is questionable. A method to mitigate the BSSE is the Counterpoise correction where the DFT calculation of each monomer is performed in the basis of the dimer so that the stabilization due to the extended basis also applies to the monomer calculation. This decreases the monomer energy compared to the smaller basis and therefore leads to a lowering of the interaction energy compared to the calculation without Counterpoise correction.\\
When comparing the MAEs of the different FHI-aims settings in Fig.\;\ref{fig:basis_set_Counter_Poise}, it can be seen that when using the Counterpoise correction the MAEs of the different settings agree withing a few meV. With increasing number of basis functions, the difference between the Counterpoise corrected and uncorrected MAEs decreases. In the case of the ``tight'' and ``really\_tight'' settings, the difference is about 5\,meV for both functionals and datasets. The reason for the similar performance of the ``tight'' and ``really\_tight'' settings is that they employ the same basis functions and therefore no further decrease in BSSE is to be expected.\\
The parametrization of the TS dispersion models presented in the main text were performed using the ``tight'' settings with Counterpoise correction. Based on the data in Fig.\;\ref{fig:basis_set_Counter_Poise}, the BSSE should not affect the parametrization of the damping function parameters in a significant way resulting in parameters that should be transferable to different similar performing basis-sets.

\section{Radial distribution functions}

\subsection{PBE at 400\,K}

\begin{figure*}
\centering
\begin{subfigure}{.49\textwidth}
    \centering
    \includegraphics[scale=0.8]{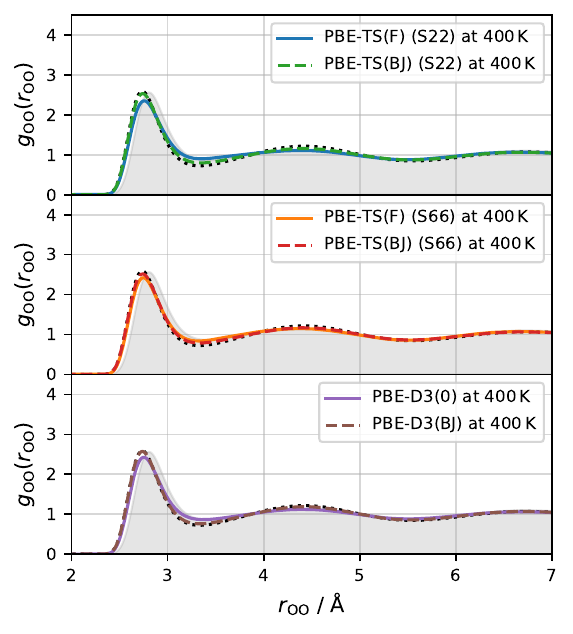}
    \caption{}
    \label{fig:PBE-400K-OO}
\end{subfigure}
\begin{subfigure}{.49\textwidth}
    \centering
    \includegraphics[scale=0.8]{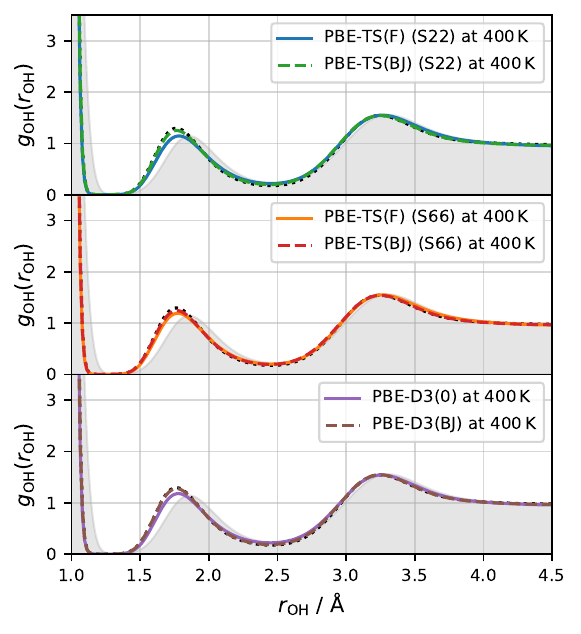}
    \caption{}
    \label{fig:PBE-400K-OH}
\end{subfigure}
\caption{OO (a) and OH (b) RDFs produced by PBE in HDNNP driven MD simulations at a temperature of 400\,K and density of 1.0\,g/cm$^3$ with different dispersion corrections. In light grey reference RDFs from CCMD obtained at 298.15\,K reported by Daru et al.\cite{daru_coupled_2022} are shown. Additionally, the black dotted lines correspond to results from the respective xc functional without dispersion correction.}
\label{fig:RDF-PBE-400K}
\end{figure*}

Radial distribution functions (RDFs) of liquid water from classical MD simulations driven by HDNNPs, which were trained on PBE reference data corrected for dispersion using TS-models and DFT-D3\cite{grimme_consistent_2010, grimme_effect_2011} with zero- and BJ-damping are shown in Fig.\:\ref{fig:RDF-PBE-400K}. The simulations were conducted in the canonical ($NVT$)-ensemble using a periodic simulation box of 512 water molecules at a density of 1\;g/cm$^3$ and a temperature of 400\,K. The oxygen-oxygen (OO) and oxygen-hydrogen (OH) RDFs are compared to RDFs of liquid water at 298.15\,K obtained from coupled cluster molecular dynamics (CCMD), which accounts for nuclear quantum effects (NQEs), reported by Daru et al.\cite{daru_coupled_2022}, in Fig.\:\ref{fig:PBE-400K-OO} and Fig.\:\ref{fig:PBE-400K-OH} respectively.

\subsection{Equilibrium density at ambient conditions}

\begin{figure*}
\centering
\begin{subfigure}{.49\textwidth}
    \centering
    \includegraphics[scale=0.8]{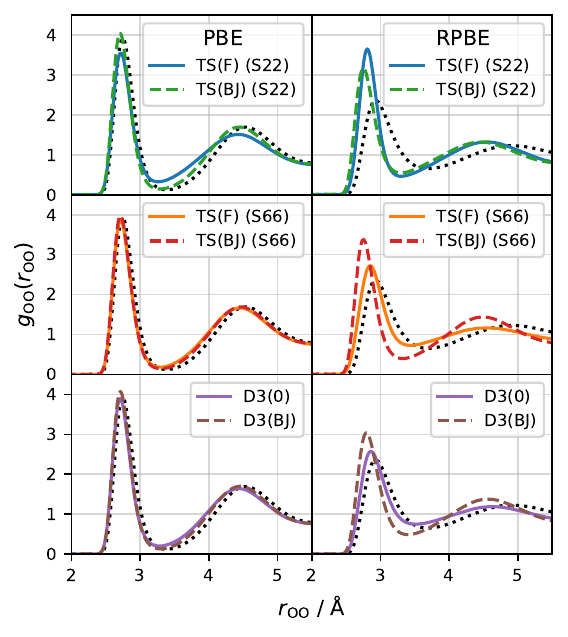}
    \caption{}
    \label{fig:equ-OO}
\end{subfigure}
\begin{subfigure}{.49\textwidth}
    \centering
    \includegraphics[scale=0.8]{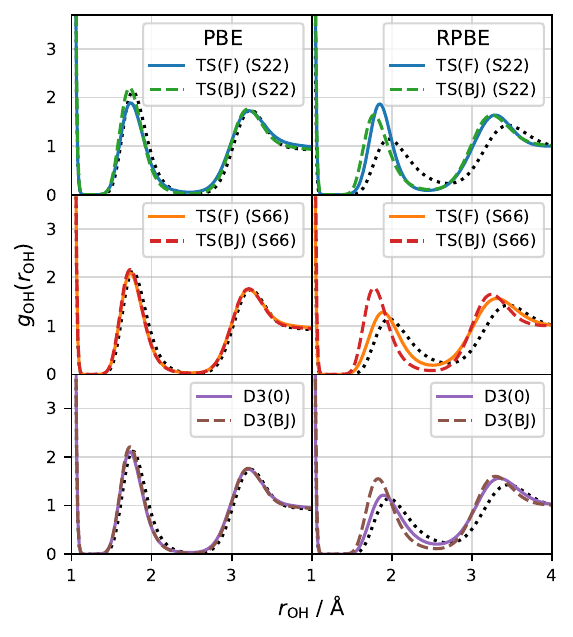}
    \caption{}
    \label{fig:equ-OH}
\end{subfigure}
\caption{OO (a) and OH (b) RDFs produced by PBE and RPBE in HDNNP driven MD simulations at a temperature of 300\,K with different dispersion corrections. The MD simulations were conducted at the equilibrium density at ambient conditions of each method. The black dotted lines correspond to results from the respective xc functional without dispersion correction.}
\label{fig:RDF-equ}
\end{figure*}

Equilibrium density RDFs of liquid water from classical MD simulations driven by HDNNPs, which were trained on PBE and RPBE reference data corrected for dispersion using TS-models and DFT-D3 with zero- and BJ-damping are shown in Fig.\:\ref{fig:RDF-equ}. The simulations were conducted in the $NVT$-ensemble using a periodic simulation box of 512 water molecules at a temperature of 300\,K and the equilibrium density at this temperature of the respective method. The equilibrium densities were obtained from MD simulations in the isothermal–isobaric ($NpT$)-ensemble at a temperature of 300\,K and pressure of 1\,bar. The equilibrium densities $\rho_\mathrm{eq}$ of each method are summarized in the main text.

\subsection{DFT-D4}

\begin{figure*}
\centering
\begin{subfigure}{.49\textwidth}
    \centering
    \includegraphics[scale=0.8]{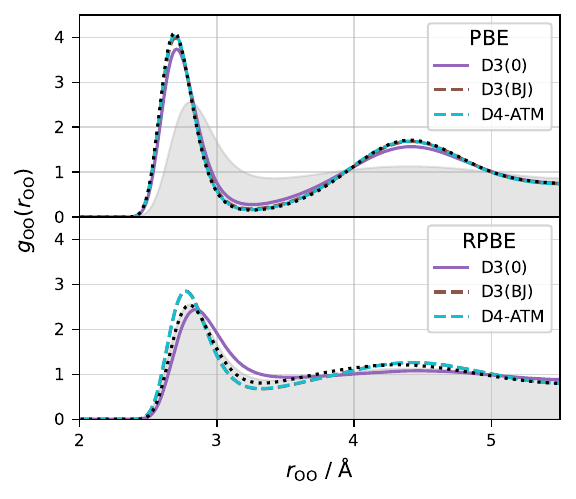}
    \caption{}
    \label{fig:D4-OO}
\end{subfigure}
\begin{subfigure}{.49\textwidth}
    \centering
    \includegraphics[scale=0.8]{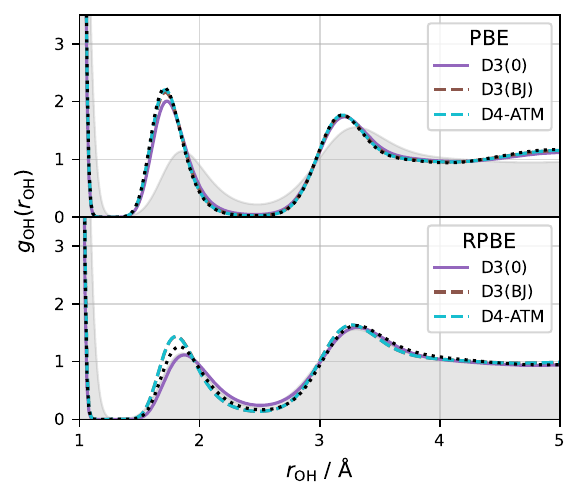}
    \caption{}
    \label{fig:D4-OH}
\end{subfigure}
\caption{OO (a) and OH (b) RDFs produced by PBE and RPBE in HDNNP driven MD simulations of 512 water molecules at a temperature of 300\,K and density of 1.0\,g/cm$^3$ corrected for dispersion using DFT-D3(BJ) employing BJ-damping, DFT-D3(0) employing zero-damping and DFT-D4-ATM employing BJ-damping. In light grey reference RDFs from CCMD obtained at 298.15\,K reported by Daru et al.\cite{daru_coupled_2022} are shown. The black dotted lines correspond to results from the respective xc functional without dispersion correction.}
\label{fig:RDF-D4}
\end{figure*}

To compare the DFT-D3 results to its successor DFT-D4\cite{caldeweyher_generally_2019}, we have additionally trained HDNNPs on the PBE and RPBE reference dataset corrected for dispersion using the DFT-D4 software developed by Grimme et al.\cite{caldeweyher_generally_2019} (version 3.7.0), with the default settings and parameters for both functionals. The fitting results are summarized in Table\:\ref{tab:d4-fit-results}. By default DFT-D4 includes three-body Axilrod-Teller\cite{axilrod_interaction_1943}-Muto (ATM) interactions, which we did not include in our DFT-D3 calculations. Additionally, BJ-damping is exclusively employed in DFT-D4-ATM. We compare the DFT-D4-ATM RDFs at 300\,K and a density of 1\,g/cm$^{-3}$ to those obtained by the HDNNPs trained on reference data corrected using DFT-D3 with both types of damping in Fig.\:\ref{fig:RDF-D4}. Here, it can be seen that the produced RDFs are a close match for the DFT-D3(BJ) RDFs for both xc functionals. Therefore, the overstructuring observed for RPBE-D3(BJ) is also observed when employing RPBE-D4-ATM. In the case of PBE-D4-ATM, the softening of the RDF is also less pronounced compared to PBE-D3(0).

\begin{table}
\caption{Summary of the fitting results of the HDNNPs for the DFT-D4-ATM models. RMSEs of the energies are in meV\,atom$^{-1}$ and the RMSEs of the force components are in meV\,$a_0^{-1}$.}
\label{tab:d4-fit-results}
\begin{ruledtabular}
\begin{tabular}{l c c c r}
\multirow{ 2}{*}{Method} & \multicolumn{2}{c}{RMSE (Energy)} & \multicolumn{2}{r}{RMSE (Forces)} \\ 
& train & test & train & test \\
\hline
PBE-D4-ATM &  0.588  &  0.645  & 25.470  & 25.550 \\
\hline
\hline
RPBE-D4-ATM & 0.589  & 0.649 & 28.400 & 27.934 \\
\end{tabular}
\end{ruledtabular}
\end{table}

\subsection{Effect of basis-set}

\begin{figure*}
\centering
\begin{subfigure}{.49\textwidth}
    \centering
    \includegraphics[scale=0.8]{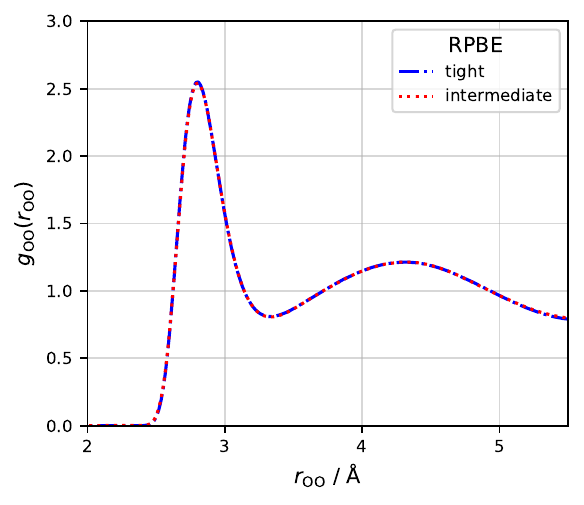}
    \caption{}
    \label{fig:basis-OO}
\end{subfigure}
\begin{subfigure}{.49\textwidth}
    \centering
    \includegraphics[scale=0.8]{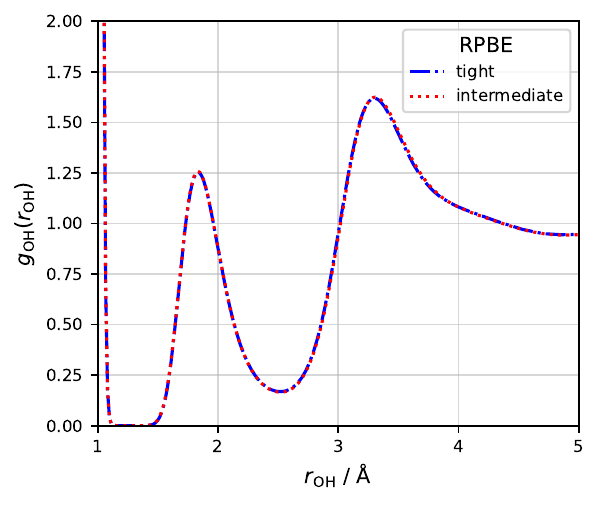}
    \caption{}
    \label{fig:basis-OH}
\end{subfigure}
\caption{Comparison of OO (a) and OH (b) RDFs produced by RPBE using different basis-sets in HDNNP driven MD simulations of 512 water molecules at a temperature of 300\,K and density of 1.0\,g/cm$^3$. The red dotted lines correspond to the results presented in the main text in Fig. 4, which are based on the ``intermediate'' species defaults provided by the FHI-aims code.\cite{blum_ab_2009} The blue dash-dotted lines correspond to RDFs obtained using a HDNNP published by Morawietz et al.\cite{morawietz_how_2016}, which has been trained on bulk water RPBE energies and forces obtained using the ``tight'' (``\textit{tier} 2'') species defaults. The associated peak positions of both the OO and OH RDFs are summarized in Tab.\;\ref{tab:RDF_basis}.}
\label{fig:basis_comp}
\end{figure*}

The DFT calculations of the interaction energies of the model complexes contained in the S22\cite{jurecka_benchmark_2006} and S66\cite{rezac_s66_2011, rezac_erratum_2014} data sets for the parametrization of the damping function parameters of the TS dispersion models were performed using the large ``tight'' FHI-aims basis-sets and Counterpoise correction to reduce the influence of the basis-set incompleteness error (BSIE) and basis-set superposition error (BSSE) on the obtained parameters. In \textit{ab initio} MD simulations, often smaller basis-sets are employed to reduce the computational costs and allow for longer simulation times necessary to achieve statistical significant results. Similarly, we have performed the DFT calculations of the reference data for the bulk water HDNNPs using the smaller ``intermediate'' basis-sets. This could potentially introduce inconsistencies between the energies and forces provided by the dispersion model and DFT basis. In Sec.\:\ref{sec:basis_and_counterpoise}, we investigated the effect of the choice of basis-set and Counterpoise correction on the interaction energies of the model complexes contained in the S22 and S66 benchmark datasets. Here we found that smaller basis-sets without BSSE correction tend to produce stronger interaction energies than larger basis-sets. Parametrizing the damping function using a small basis-sets would therefore result in a less attractive dispersion model, which would lead to an underestimation of dispersion interactions when combined with DFT calculations employing a larger basis-set. Conversely, employing a dispersion model parametrized using a large basis-set and pairing it with a smaller basis-set DFT calculation could lead to an overestimation of dispersion interactions.\\
In this study, we have employed the smaller ``intermediate'' basis-sets for MD simulations only. MD trajectories are primarily governed by the forces and therefore energy offsets due to a smaller basis-set are less critical as long as these offsets maintain the overall shape of the PES. Therefore, to estimate the effect of the change in basis on MD simulations of water, we have performed a 2\,ns simulation of 512 water molecules in the $NVT$ ensemble at 300\,K and a density of 1\,g/cm$^3$ using a HDNNP published by Morawietz et al.\cite{morawietz_how_2016}, which has been trained on bulk water RPBE energies and forces obtained using the ``tight'' (``\textit{tier} 2'') basis-sets. In Fig.\;\ref{fig:basis_comp}, we compare the resulting OO and OH water RDFs to the ones produced by the dispersion uncorrected RPBE HDNNP presented in the main text. The associated peak positions are summarized in Table\;\ref{tab:RDF_basis}. Here, it can be seen that the RDFs produced by the ``intermediate'' basis is in almost perfect agreement with the ``tight'' basis result. Therefore, it can be assumed that the overall shape of the ``tight'' basis PES is maintained when employing the smaller ``intermediate'' basis for MD simulations of water. Consequently, the application of the dispersion models parametrized on the `tight'' basis-set data should not introduce inconsistencies in MD results obtained using the ``intermediate'' basis-set.

\begin{table*}
\caption{Peak positions of the OO and OH RDFs presented in Fig.\;\ref{fig:basis_comp}. The RDFs were obtained from MD simulations of 512 water molecules in the $NVT$ ensemble at a density of 1\,g/cm$^3$. The positions, in  \AA, $r^\mathrm{1,max}_\mathrm{OO}$, $r^\mathrm{1,min}_\mathrm{OO}$ and $r^\mathrm{2,max}_\mathrm{OO}$ correspond to the first maximum $g^\mathrm{1,max}_\mathrm{OO}$, first minimum $g^\mathrm{1,min}_\mathrm{OO}$ and second maximum $g^\mathrm{2,max}_\mathrm{OO}$ of the OO RDF respectively. The positions, in \AA, $r^\mathrm{HB,max}_\mathrm{OO}$ correspond to the second maximum $g^\mathrm{HB,max}_\mathrm{OH}$ of the OH RDF, i.e., a H-bond}.
\label{tab:RDF_basis}
\begin{ruledtabular}
\begin{tabular}{ l c  S[table-format=2.3]  S[table-format=2.5]  S[table-format=2.3]  S[table-format=2.3]  S[table-format=2.3]  S[table-format=2.3]  S[table-format=2.3]  S[table-format=2.3]}
Species default & $T$ & \multicolumn{1}{c}{$r^\mathrm{1,max}_\mathrm{OO}$} & \multicolumn{1}{c}{$g^\mathrm{1,max}_\mathrm{OO}$} & \multicolumn{1}{c}{$r^\mathrm{1,min}_\mathrm{OO}$} & \multicolumn{1}{c}{$g^\mathrm{1,min}_\mathrm{OO}$} & \multicolumn{1}{c}{$r^\mathrm{2,max}_\mathrm{OO}$} & \multicolumn{1}{c}{$g^\mathrm{2,max}_\mathrm{OO}$} & \multicolumn{1}{c}{$r^\mathrm{HB,\:max}_\mathrm{OH}$} &	 \multicolumn{1}{r}{$g^\mathrm{HB,\:max}_\mathrm{OH}$}\\
\hline
intermediate                 &    300  &  2.81  &  2.54  &  3.34 &   0.81  &  4.31 &   1.21 & 1.83 &   1.26 \\
tight\cite{morawietz_how_2016}              &    300  &  2.81  &  2.55  &  3.34 &   0.81  &  4.33 &   1.21 & 1.83 &   1.25 \\
\end{tabular}
\end{ruledtabular}
\end{table*}

\section{Self-diffusion coefficient}

\subsection{Method}

\begin{figure*}
\centering
\begin{subfigure}{.49\textwidth}
    \centering
    \includegraphics[scale=0.8]{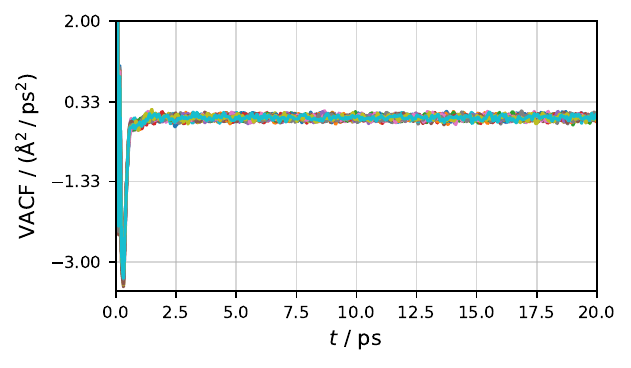}
    \caption{}
    \label{fig:VACF}
\end{subfigure}
\begin{subfigure}{.49\textwidth}
    \centering
    \includegraphics[scale=0.8]{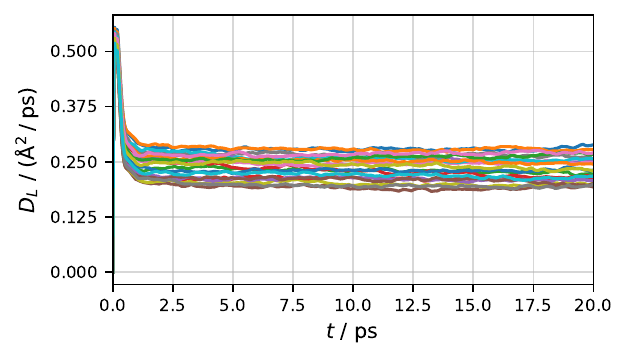}
    \caption{}
    \label{fig:D_cumtrapz}
\end{subfigure}
\begin{subfigure}{.49\textwidth}
    \centering
    \includegraphics[scale=0.8]{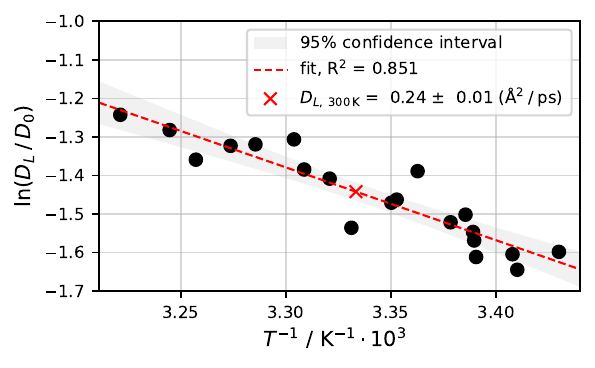}
    \caption{}
    \label{fig:D_int}
\end{subfigure}
\begin{subfigure}{.49\textwidth}
    \centering
    \includegraphics[scale=0.8]{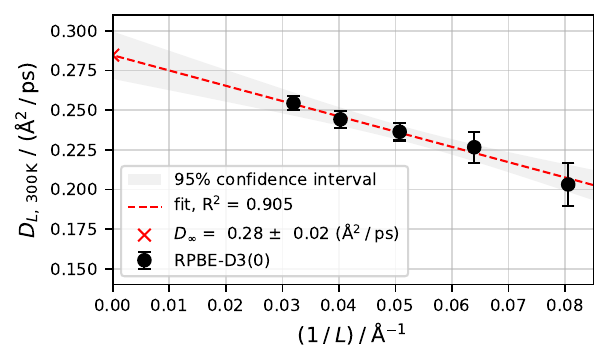}
    \caption{}
    \label{fig:D_extra}
\end{subfigure}
\caption{Workflow for determining the self-diffusion coefficient of the infinite system $D_\infty$. Here, exemplary shown for data from MD simulations in the microcanonical ($NVE$) ensemble of 256 water molecules at a temperature of 300\,K and a density of 1.0\,g/cm$^3$, driven by a HDNNP trained on RPBE-D3(0) data (panels (a) to (c)). In (a), the VACFs obtained from twenty $NVE$ simulations are shown. All $NVE$ simulations were started from different initial configuration, which were sampled from $NVT$ simulations at 300\,K. Next, the system-size dependent self-diffusion coefficient $D_L$ is obtained using the Kubo-Green relation between the integral of the VACF and self-diffusion coefficient.\cite{frenkel_2002} In (b), $D_L$ values obtained from the cumulative integral of the VACFs shown in (a) up to a given time $t$ are shown. In (c), the natural logarithm of the self-diffusion coefficients obtained using an upper integration limit of 20\,ps is plotted against the inverse mean temperature $T$ of the associated $NVE$ trajectory. Using a linear fit of the data, the system-size specific self-diffusion coefficient at the target temperature $D_{L,\:300\,\mathrm{K}}$ is determined by interpolation. The uncertainty is estimated from the 95\% confidence interval of the fit evaluated at the target temperature of 300\,K. This procedure was additionally repeated for systems containing 64, 128, 512 and 1024 water molecules. All obtained $D_{L,\:300\,\mathrm{K}}$ values are plotted against the associated inverse simulation box length $L$ in (d). To determine $D_\infty$ from the \textit{y}-intercept, a linear fit of the data is performed using the uncertainty of each $D_{L,\:300\,\mathrm{K}}$ value as fitting weight. The uncertainty of $D_\infty$ is estimated using the 95\% confidence interval of the fit evaluated at the \textit{y}-intercept.}
\label{fig:D_L}
\end{figure*}

In Fig.\:\ref{fig:D_L}, we visualize our workflow for determining the self-diffusion coefficient of the infinite system $D_\infty$ exemplary using our RPBE-D3(0) data .

\subsection{PBE at 300\,K}

\begin{figure}
\centering
\includegraphics[scale=0.8]{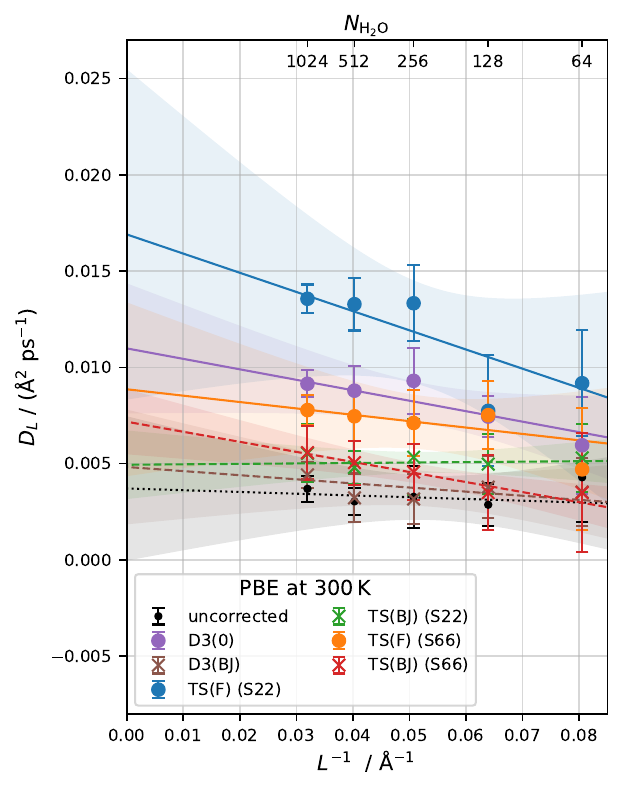}
\caption{System-size specific diffusion coefficients $D_L$ obtained from HDNNP driven MD simulations of PBE at 300\;K and density of 1\,g/cm$^3$ with different dispersion corrections plotted against the inverse length of the simulation box $L$ to extrapolate the value to that of the infinite system $D_\infty$. Error bars correspond to the 95\% confidence interval of the interpolation of the $NVE$ diffusion coefficients to 300\,K. Colored lines correspond to the extrapolation function obtained from a linear fit with the area of matching color representing the 95\% confidence interval.}
\label{fig:PBE-D-300K}
\end{figure}

In Fig.\:\ref{fig:PBE-D-300K}, we show the system-size specific diffusion coefficients $D_L$ extrapolated to the infinite system for PBE at 300\,K and a density of 1\,g/cm$^3$. The rate of self-diffusion is very low when applying PBE at 300\,K. Additionally, there is only a minor change in the rate when applying a dispersion correction. Therefore, additional or longer MD simulations would be necessary to converge the values so that there is no overlap among the estimated uncertainties for the self-diffusion coefficients produced by the investigated methods and work out differences resulting from the choice of damping function. We got around this issue, in the main text, by redetermining the self-diffusion coefficient at 400\,K, where the rate is generally an order of magnitude larger, which gives an improved separation among the results and associated uncertainties of the different methods. 

\end{document}